\pdfoutput=1
\documentclass[useAMS,usenatbib]{mn2e}
\pdfoutput=1
\usepackage{enumerate}
\usepackage{amsmath}
\usepackage{graphicx}
\usepackage{color}
\usepackage{hyperref}
 
 \providecommand{\adsurl}[1]{\href{#1}{ADS}}
  
\usepackage{afterpage}
\usepackage{amsfonts}
\newcommand{\tickYes}{\checkmark}
\usepackage{pifont}
\newcommand{\tickNo}{\hspace{1pt}\ding{55}}
\usepackage{ulem}
\title{Statistical Classification Techniques for Photometric Supernova Typing}
\author[Newling et al.]{J. Newling$^{1,2}$\thanks{E-mail:
james.newling@gmail.com}, M. Varughese$^{3}$,  B. Bassett$^{1,2,4}$, H. Campbell$^{5}$, R. Hlozek$^{6}$, \newauthor M. Kunz$^{7}$, H. Lampeitl$^{5}$, B. Martin$^{2,9,10}$, R. Nichol$^{5}$, D. Parkinson$^{8}$, M. Smith$^{1,10}$\\
$^{1}$Department of Mathematics and Applied Mathematics, University of Cape Town, Rondebosch 7701, South Africa\\
$^{2}$African Institute for Mathematical Sciences, 6-8 Melrose Road, Muizenberg,  7945, South Africa\\
$^{3}$Department of Statistical Sciences, University of Cape Town, Rondebosch 7701, South Africa\\
$^{4}$South African Astronomical Observatory, PO Box 9, Observatory 7935, South Africa\\
$^{5}$Institute of Cosmology and Gravitation, University of Portsmouth, Portsmouth PO1 3FX, United Kingdom\\
$^{6}$Department of Astrophysics, Oxford University, Oxford OX1 3RH, United Kingdom\\
$^{7}$D\'{e}partement de Physique Th\'{e}orique, Universit\'{e} de Gen\`{e}ve, Gen\`{e}ve CH1211, Switzerland\\
$^{8}$Astronomy Centre, University of Sussex, Brighton BN1 9QH, United Kingdom\\
$^{9}$Department of Astronomy, University of Cape Town, Rondebosch 7701, South Africa\\
$^{10}$Astrophysics, Cosmology and Gravity Centre, University of Cape Town, Rondebosch 7701, South Africa}
\begin{document}

\date{Submitted to arxiv, \today}

\pagerange{\pageref{firstpage}--19} \pubyear{2010}
\maketitle

\label{firstpage}

\begin{abstract}
Future photometric supernova surveys will produce vastly more candidates than can be followed up spectroscopically, highlighting the need for effective classification methods based on lightcurves alone. Here we introduce boosting and kernel density estimation techniques which have minimal astrophysical input and compare their performance on 20,000 simulated Dark Energy Survey lightcurves. We demonstrate that these methods are comparable to the best template fitting methods currently used, and in particular do not require the redshift of the host galaxy or candidate. However both methods require a training sample that is representative of the full population, so typical spectroscopic supernova subsamples will lead to poor performance. To enable the full potential of such blind methods, we recommend that representative training samples should be used and so specific attention should be given to their creation in the design phase of future photometric surveys. 
\end{abstract}


\section{Introduction}

Type Ia Supernovae (SNeIa) provided the first widely accepted evidence for cosmic acceleration in the late 1990's
\citep{sn1,sn2}. Based on small numbers of predominantly spectroscopically-confirmed SNeIa, those results have been
confirmed by independent analyses \citep{bao1,bao2,clusters,lensing,ISW,bao3,CMB} and by a series of steadily improving SNeIa
surveys. These modern SNeIa surveys have acquired about an order of magnitude more SNeIa than those early offerings, now covering redshifts out to $z \sim 1.5$
\citep{kait,snfactory,SNLS,hst, 
sdss,csp}. In addition, these surveys now have excellent lightcurve
coverage with rolling search strategies and multi-frequency lightcurve data with significantly better control of
photometric errors due to the use of a single telescope to acquire the data in each major survey.

The next generation of SNeIa surveys will be integrated into major photometric surveys, such as the Dark Energy Survey (DES)
 \citep{des}, PanSTARRS \citep{panstarrs}, SkyMapper \citep{SkyMapper} and LSST \citep{lsst}. These next generation surveys promise to catalyse a new revolution in SNIa
research due to the sheer number of high-quality SNIa candidates that will be discovered: tens of thousands and
perhaps millions of good SNIa candidates over the decade 2013-2023. Spectroscopic followup will probably be
limited to a very narrow subset of these candidates and so finding ways to best choose the followup subset to utilise the photometric data is a key challenge
in SN cosmology for the coming decade.

In this paper we are interested in methods that can be used to accurately identify SNeIa from their lightcurves
alone, that is, their variation in brightness in different colour bands as a function of time. This is a departure
from traditional studies of SNeIa where all SNe used in cosmological parameter estimation studies have had their type confirmed via one or more spectra.

There are two ways that one can imagine using photometric candidates. The first approach is to use all
the SNe, irrespective of how likely they are to actually be a SNIa.
This is the approach exemplified
by the BEAMS formalism, which accounts for the contamination from non-Ia SN data using the appropriate Bayesian
framework \citep{beams}. The more conservative approach is to try to classify the candidates into Ia,
Ibc or II supernovae, and then only use those objects that are believed to be SNeIa above some threshold
of confidence.

The origin of this paper was the Supernova Photometric Classification Challenge (SNPCC) run by \citet{sntc}.
The
SNPCC provided a simulated spectroscopic training data sample of approximately 1000 known supernovae. The
challenge was then to predict the types of approximately $20\,000$ other objects from their lightcurves alone. The challenge is now over, and the results from the different contributors are summarized in \citet{sntc_results}.

In this
paper we present the details of a number of approaches to this problem, and their successes and failures. 
In Section~\ref{sec:methods} we discuss methods we have implemented to go from multi-band
lightcurves to probabilities while in Section~\ref{sec:results} we discuss the performance of the methods in the SNPCC. In particular we highlight how a non-representative training sample negatively affects the performance of the different algorithms. Finally 
we conclude with recommendations for the future.

\section{The Lightcurve Data}\label{data}

\subsection{The Supernova Challenge Data}\label{sec:SNPCC}
The data used in this paper consists of $\sim 20\,000$ simulated SN lightcurves with associated SN types released after the SNPCC\footnote{These post-SNPCC lightcurves are available at http://sdssdp62.fnal.gov/sdsssn/SIMGEN\_PUBLIC/}. The SNPCC data \footnote{These competition lightcurves are available from http://www.hep.anl.gov/SNchallenge/} are only relevant in our discussion of competition scores. Our reason for using the post-data is that it has numerous improvements and bug-fixes and is a more accurate simulation. The simulation was based on a DES-like survey, consisting of 5 SN fields, each of 3 deg$^2$, such that $10\%$ of the total survey time is allocated to the SN survey. The SNPCC dataset consists of a mixture of SN types (Ia, II, Ib, Ic), sampled randomly with proportions given by their expected rates as a function of redshift. 

Each simulated SN consists of flux measurements in the \textit{griz} filters  \citep{1996AJ111.1748F}
and includes information about the sky-noise, point spread function, and atmospheric conditions  that are anticipated for the DES site. Distances were calculated assuming a standard $\Lambda$CDM cosmology ($\Omega_M = 0.3, \Omega_\Lambda = 0.7$ and $w = -1$), with anomalous scatter around the Hubble diagram drawn from a Gaussian distribution with $\sigma_m = 0.09$ and applied coherently to each passband. The SNPCC data includes two selection criteria. Each object is required to have at least one observation with a signal-to-noise ratio (S/N) above 5 in any filter, and must also have at least 5 observations after explosion. A complete summary of the SNPCC is given in \citet{sntc,sntc_results}.

We took part in two of the SNPCC challenges. In the first (\texttt{+HOSTZ}) challenge, participants were provided with photometric host galaxy redshift estimates, based on simulated galaxies analysed using the methods discussed in \citet{2008ApJ674768O} 
 and asked to return the type of each SN candidate. In the second (\texttt{-HOSTZ}) challenge, no redshift estimates for simulated SNe were provided. Both challenges are considered in this paper, but with emphasis on the \texttt{+HOSTZ} challenge. We did not attempt to distinguish between non-Ia sub-types (such as type II and type Ib/c SNe).

\begin{figure}
	\begin{center}
		\includegraphics[width=3.5in]{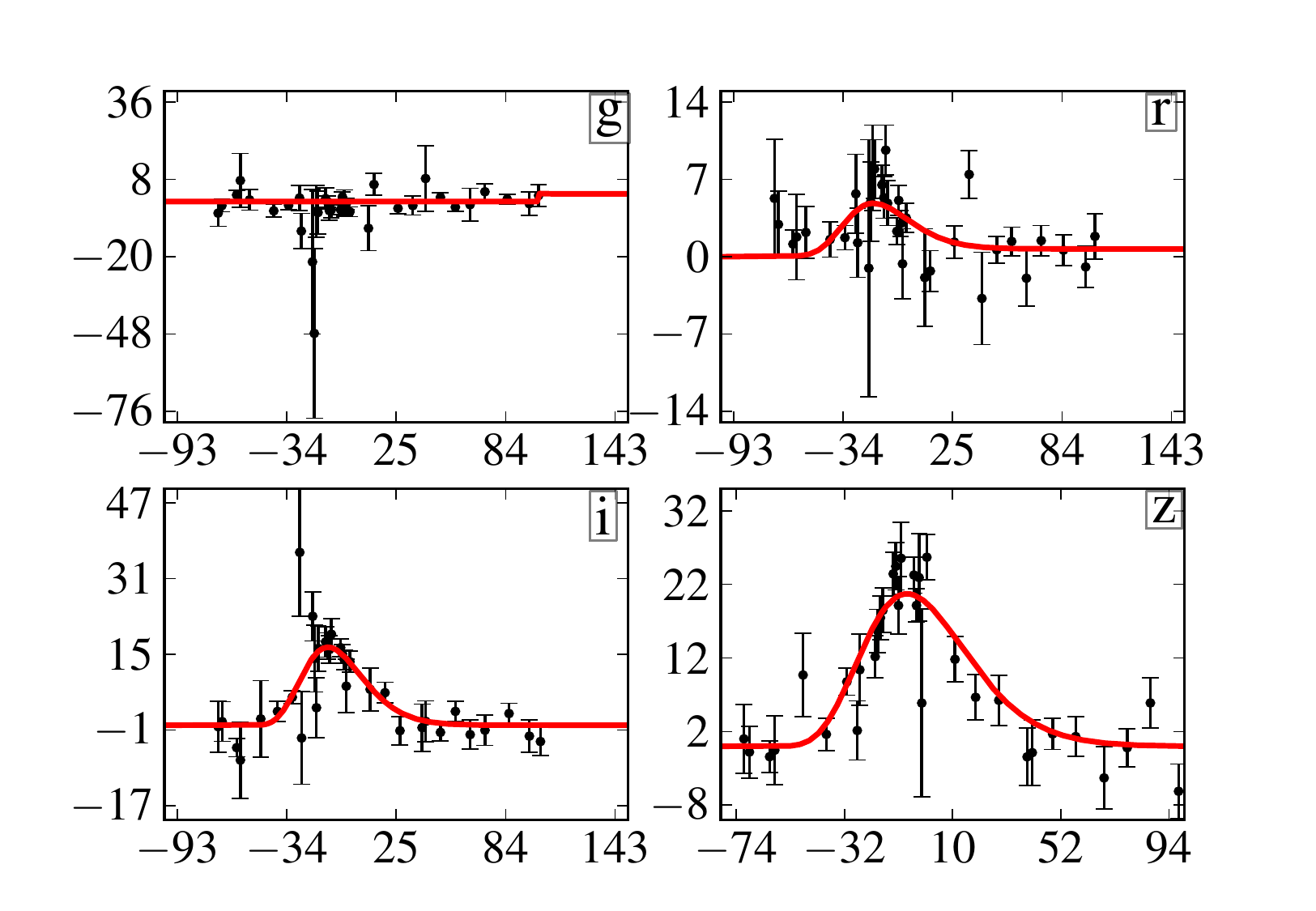}
		
		\includegraphics[width=3.5in]{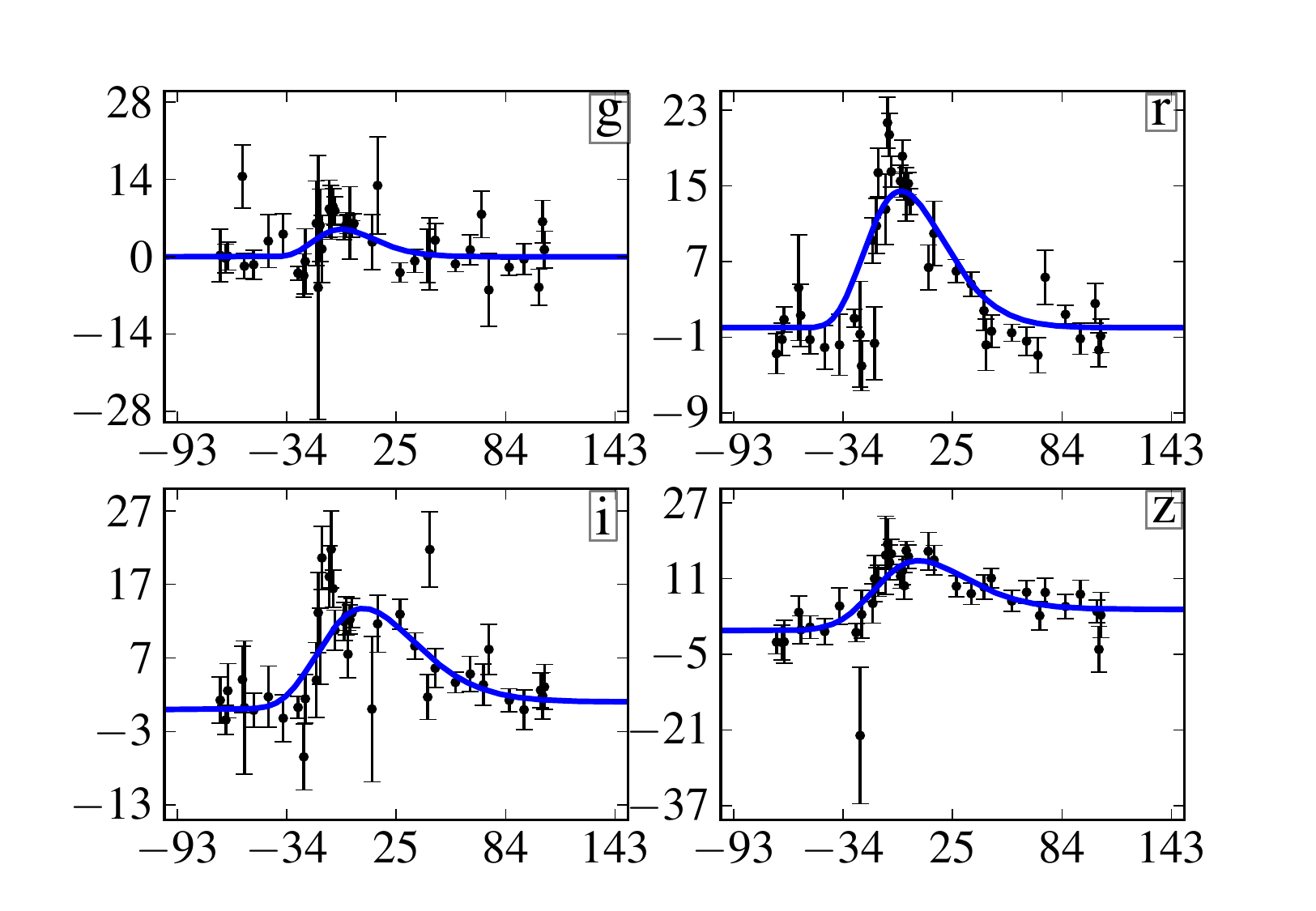}
	\end{center}
\caption{(\textit{Above}) A typical well-sampled SNIa lightcurve, in this case at redshift z = 0.694. (\textit{Below}) The lightcurve of a typical well-sampled non-Ia SN at z = 0.663. Overplotted are the best-fitting curves using Eq.~\ref{gamma}.}
\label{fig:Ia_ex.pdf}
\end{figure}

Figure~\ref{fig:Ia_ex.pdf} shows the multi-band lightcurve data for a randomly selected Ia and non-Ia supernova. To these measurements, a parametric curve has been fitted as discussed in Section~\ref{sec:gamma}.

\subsubsection{Training Samples}\label{subsubsec:training}
The aim of the SNPCC was for the participants to classify each of the simulated SNe into Ia or non-Ia (and non-Ia sub-classes if they desired) with the aim of minimising false Ia detections and maximising correct Ia detections. To aid this, a spectroscopic training sample of $\sim 1000$ SNe with known type was provided which is a simulation of expected spectroscopic observations on a 4 meter class telescope with a limiting magnitude of $r \sim 21.5$, and an 8 meter class telescope with limiting $i$ band magnitude of $23.5$. Because spectroscopy is harder than photometry the distribution of SNe in this spectroscopic sample is much brighter on average than the full photometric sample, and hence is 
not representative of the full sample. This is a crucial point to appreciate and as a result in this paper we refer to this sample as the \textit{non-representative training sample}. 

We will often compare with the results from a representative sample, generated by spectroscopically following up a sample of objects that is representative of the full photometric SN population. To produce an unbiased training sample, at the conclusion of the SNPCC when the types of each SNPCC object were revealed, we randomly selected $\sim 1000$ SNe from the entire SNPCC dataset, and considered the effect of using this as our training sample. This is referred to in the text as the \textit{representative training sample}. We refer to the SNe that require classification as as the \textit{unclassified set}.

\subsection{Post-processed Data}\label{sec:processed-data}
\subsubsection{Fitting a parameterised curve}\label{sec:gamma}
\begin{figure}
	\begin{center}
		\includegraphics[width=3.5in]{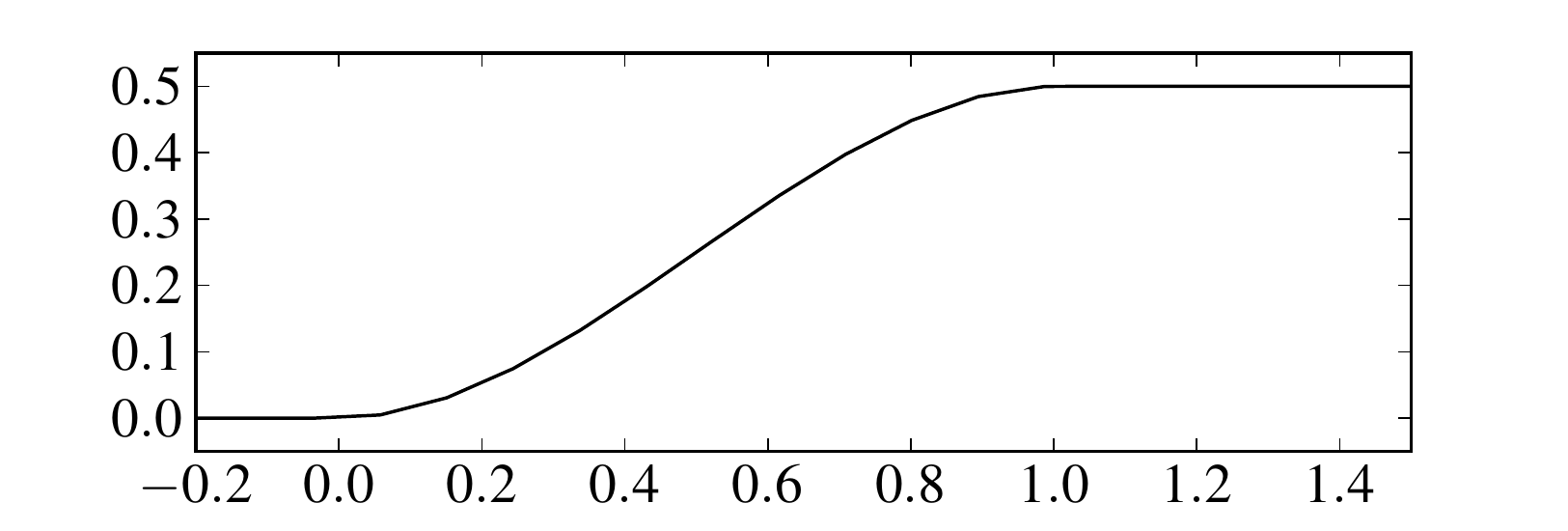}
	\end{center}
	\caption{The tail function $\Psi$, which is used in fitting Eq.~\ref{gamma}. Parameters $(\psi, \phi, \tau)$ are kept fixed at $(0.5,  0, 1)$ here.}
	\label{fig:nettail}
\end{figure}

In the provided photometric data the number, sampling times, frequency and accuracy of the sampled magnitudes varies greatly for each supernova, as illustrated in Figure~\ref{fig:Ia_ex.pdf}. In order to standardize the raw data we fit, by weighted least squares, a parameterised function to the lightcurves in each of the four colour bands. Our parameters are $(A, \phi, \psi, k, \sigma)$ and the flux in each band is taken to be \footnote{This function has a single maximum and therefore cannot fit examples which have a double peak. However, for the data we use in this paper this turns out not to be an important limitation.}:
\begin{equation}
F(t)=A\left(\frac{t-\phi}{\sigma}\right)^k\exp\left(-\frac{t-\phi}{\sigma}\right)k^{-k} e^{k}+ \Psi(t).
\label{gamma}
\end{equation}
The five parameters to be fit in each band have the following interpretations: $A$ + $\psi$ is the peak flux, $\phi$ is the starting time of the explosion, $k$ determines relative rise and decay times and $\sigma$ is a temporal stretch term. $\tau$, the time of peak flux, is determined by these parameters via $\tau = k\cdot\sigma + \phi$. The function $\Psi$ is a ``tail" function such that $F(t) \rightarrow \psi$ as $t \rightarrow \infty$. The exact form (illustrated in Figure~\ref{fig:nettail}) of $\Psi$ is:

\[
\Psi(t) = \left\{ 
\begin{array}{l l}
  0 & \quad -\infty<t<\phi\\
  \mbox{cubic spline} & \quad \;\;\;\;\phi<t<\tau\\
  \psi & \quad \;\;\;\;\tau <t <\infty 
  \end{array} \right. 
\]

\begin{figure}
	\begin{center}
		\includegraphics[width=3.5in]{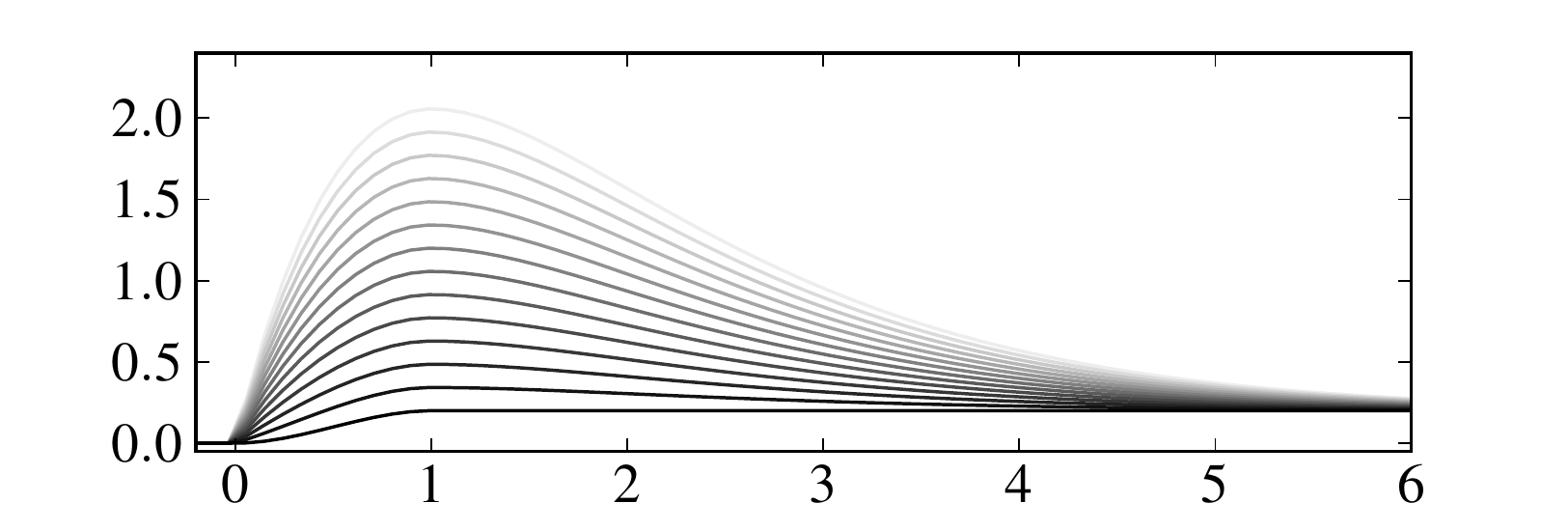}
	\end{center}
	\caption{The effect of varying $A$ on the function $F(t)$ from low (dark) to high (light). We keep the parameters $(k,\sigma,\phi,\psi)$ fixed at $(1,1,0,0)$.}
	\label{fig:A_effect}

\begin{center}
		\includegraphics[width=3.5in]{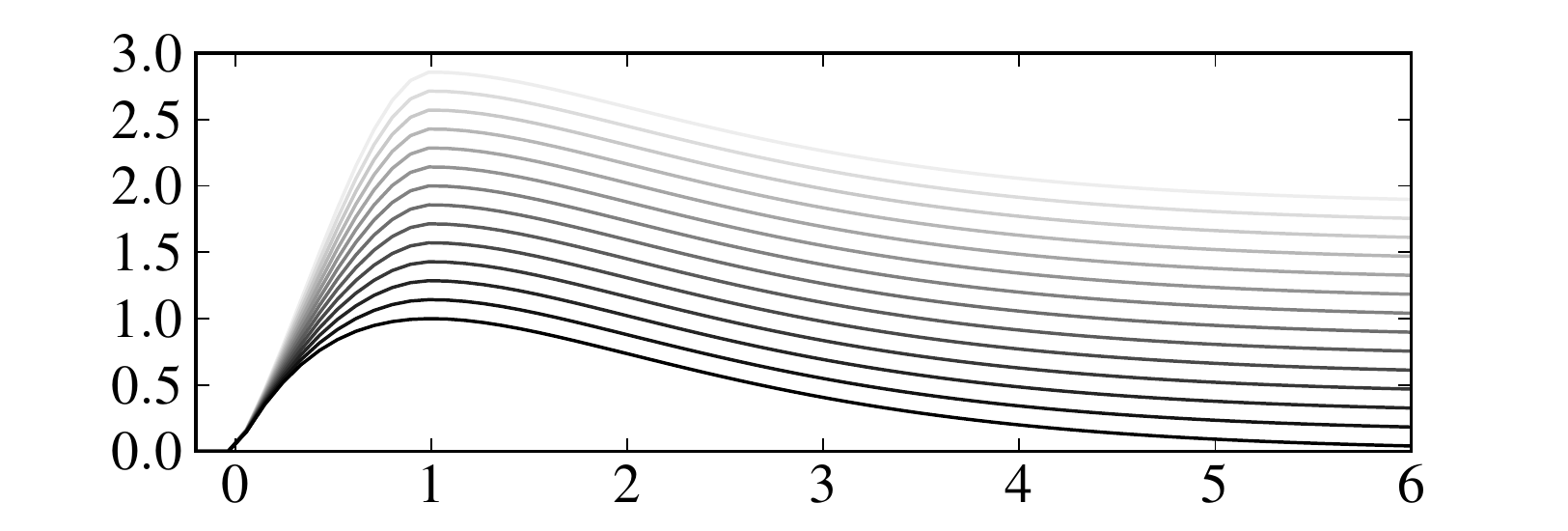}
	\end{center}
	\caption{The effect of varying $\psi$ on the function $F(t)$ from low (dark) to high (light). We keep the parameters $(k,\sigma,\phi,A)$ fixed at $(1,1,0,1)$.}
	\label{fig:tail_effect}
\end{figure}	
	
\begin{figure}	
	\begin{center}
		\includegraphics[width=3.5in]{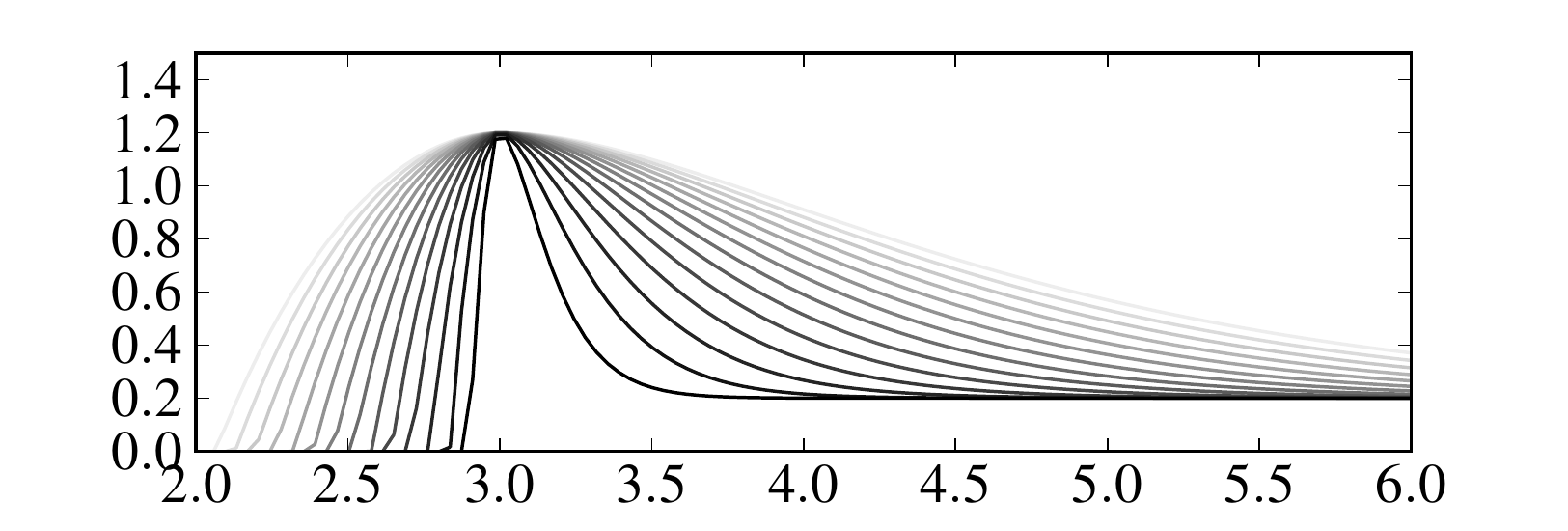}
	\end{center}
	\caption{The effect of varying $\sigma$ on the function $F(t)$  from 0.1 (dark) to 1.0 (light). Increasing $\sigma$ linearly stretches the curve away from the $t = \phi$. We keep the parameters $(A,\phi,k,\tau)$ fixed at $(1,0,1,3)$.}
	\label{fig:sig_effect}

	\begin{center}
		\includegraphics[width=3.5in]{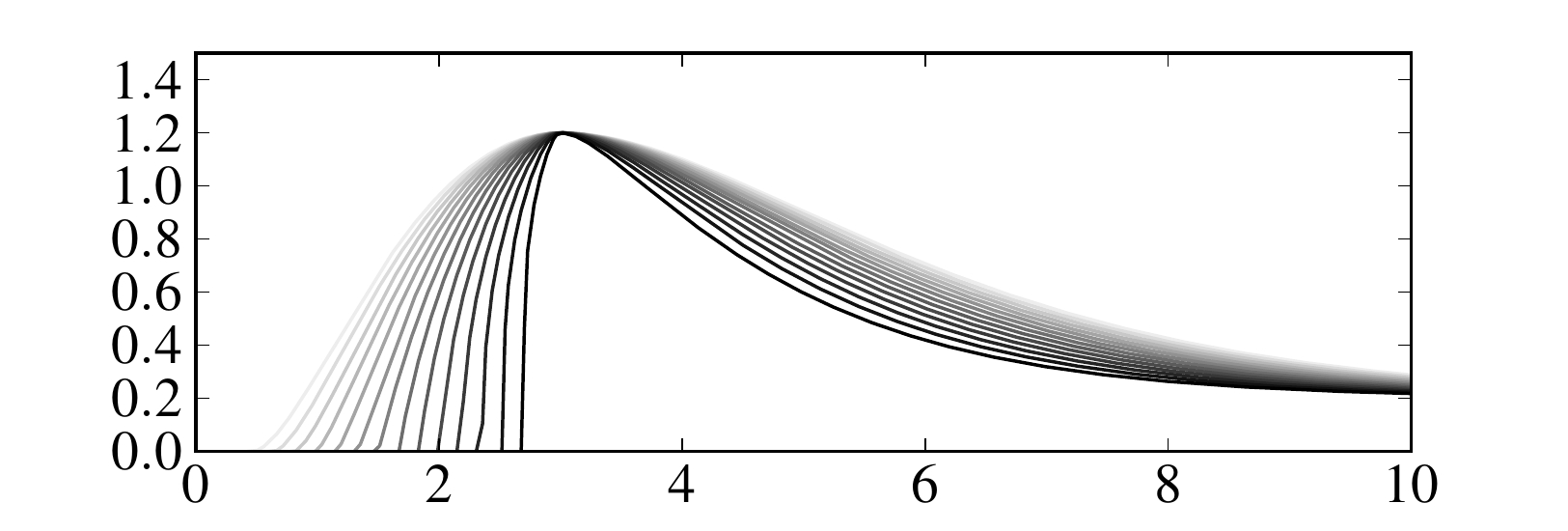}
	\end{center}
	\caption{The effect of varying $k$ on the function $F(t)$ from 0.2 (dark) to 1.8 (light). Increasing $k$ decreases the ratio of rise to decay time (Rapid rise relative to decay means low $k$).  We keep the parameters $(A,\sigma,\phi,\tau)$ fixed at $(1,1.5, 0,3)$.}
	\label{fig:k_effect}
\end{figure} 

where the cubic spline is uniquely determined to have zero derivative at $t = \phi$ and $t = \tau$. The effect of each parameter is illustrated in Figures~\ref{fig:A_effect} to \ref{fig:k_effect}. We have also posted two files at \cite{COSMOAIMS}, each containing 200 randomly selected and fitted SNe to illustrate the range of fits possible. With five free parameters, $A$, $\psi$, $\phi$, $k$ and $\sigma$ in each colour band and a host redshift (in \texttt{+HOSTZ} challenge), we have 21 parameters specifying each SN. We do not require that there be any correlation between the derived parameters in any band, e.g. between explosion time, time at peak or stretch. This is a natural extension to study in future work. 

\subsubsection{Sparse datasets}\label{sec:sparse}
About 5\% of all the SNe had fewer than 8 observations in one or more of the four bands. To avoid overfitting, we did not fit these SNe with Eq. \ref{gamma}. Instead, these sparsely sampled SNe were each fit to a 5 dimensional point - the maximum flux in each of the four colour bands plus the host redshift. The KDE and boosting methods (Section~\ref{sec:methods}) were applied to these SNe in the same way as was done in the 21 dimensional case (Sections \ref{sec:application:21-D}, \ref{sec:application:boosting}). Unless otherwise stated, discussions and illustrations will all reference the 95\% of SNe which had 8 or more observations in all bands and hence were fit with 21 parameters.

\subsubsection{SALT fits}\label{sec:SALT}
\begin{figure}
	\begin{center}
		\includegraphics[width=3.5in]{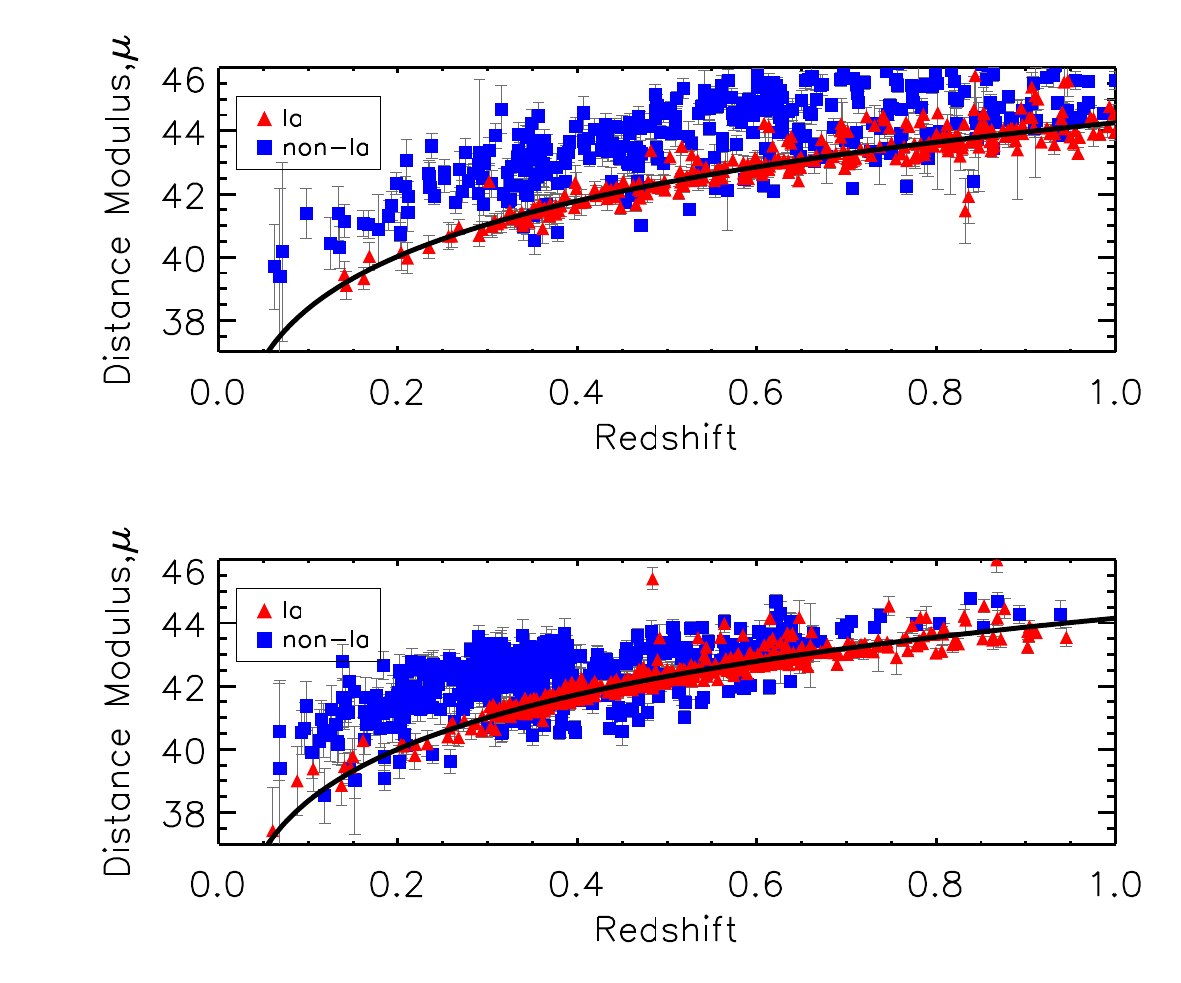}
	\end{center}
	\caption{
	Hubble diagrams for the 2 training samples considered in this paper. SNeIa are shown as red triangles, while non-Ia SNe are plotted as blue squares. Also shown is the best-fit cosmology to each SNIa sample. (\textit{Above}) The representative training sample, with $\Omega_m=0.23$. (\textit{Below}) The non-representative training sample, as provided for the SNPCC, with $\Omega_m = 0.3$.}
	\label{fig:training_hubble}
\end{figure}

In Section~\ref{sec:hkde}
, we consider classification methods that require information on the distances to SNe to constrain their type. Distance moduli for all SNPCC SNe were derived using the publicly available lightcurve fitter SALT2 \citep{2007AA46611G}.
Fits were carried out using the \textit{g}, \textit{r} and \textit{i} passbands (i.e. \textit{z} colour band data was not included). All available SNe were considered, which is significantly more liberal than the usual data-quality cuts applied during past SN cosmology 
analyses \citep{sdss}.
In this way, we maximized the number of SNe available for this work. 
We applied SALT2 to 1256 SNe available in the non-representative training sample. Immediately, we found that 165 SNe failed to pass through SALT2 with the reported error of the lightcurve either having too low a S/N or missing g-band data. We did not investigate these errors further and simply exclude these SNe. Furthermore, when the S/N is low, SALT2 fits some
SNe but returns a default upper limit magnitude of 99 and is unable to produce meaningful parameters from the lightcurve fit. This affected 62 SNe in the training sample, which were also removed from the sample. For the 1029 SNe that were successfully fitted, SALT2 returned a best fit value for four parameters $M, X_0, X_1$ and $c$ for each event (which relate the peak magnitude and stretch/colour corrections to the lightcurve). The best-fit Ia model lightcurve was also returned in the observer frame, which we used
to calculate the $\chi^2$ value for each SN in each passband (\textit{g},\textit{r},\textit{i}) which are used in Section~\ref{sec:hkde} 
 to classify SNe. Distance moduli are calculated with
\begin{equation}
\mu = (m_B - M)+\alpha x_1 - \beta  c.
\label{eq1}
\end{equation}
\noindent where we used values of $\alpha=0.1$, $\beta=2.77$ and $M=30.1$ to calculate the distance moduli, as discussed in \citet{2010ApJ722566L}. These values are consistent with those found in other analyses 
and were not expected to significantly affect our results. Figure~\ref{fig:training_hubble} shows the Hubble diagrams for the two training samples considered in this analysis. Also shown is the best-fit cosmology to each Ia dataset assuming a flat $\Lambda$CDM model. The non-representative training sample has a best-fit value of $\Omega_m = 0.30$ compared to a value of $\Omega_m = 0.23$ for the representative training sample. In the non-representative training sample, non-Ia SNe are predominately found at lower redshifts than the representative training sample due to the effective magnitude cuts coming from the spectroscopic requirement of the non-representative sample.

\section{New Classification Methods} \label{sec:methods}
We now describe in very general terms the classification algorithms we have used to facilitate application to other areas of cosmology and astrophysics. In order to classify a given object $Y$ as either Ia or non-Ia, one would like the posterior probabilities $\textrm{P}\left(Y=\textrm{Ia}|x\right)$ and $\textrm{P}(Y= \textrm{non-Ia}|x) =1-\textrm{P}(Y=\textrm{Ia} | x)$. Here $x$ are the parameters or features that characterize the supernova. Knowing these posterior probabilities is equivalent to knowing the \textit{odds}:
\begin{equation*}
\textit{odds}(x) = \frac{\textrm{P}(Y=\textrm{Ia} | x)}{\textrm{P}(Y=\textrm{non-Ia} | x)}.
\end{equation*}

Now one classifies $Y$ as a Ia for example if $odds(x)>1$, i.e if  $\textrm{P}(Y=\textrm{Ia} | x) > 0.5$. The two methods we discuss in this Section approximate the \textit{odds} in different ways:

1) Kernel Density Estimation estimates $\textrm{P}(x | Y=\textrm{Ia})$ and $\textrm{P}(x | Y =\textrm{non-Ia})$, the density of the features in classes Ia and non-Ia respectively, and then uses Bayes formula to give
\begin{equation*}
odds(x) =  \frac{\textrm{P}(x | Y=\textrm{Ia}) \cdot \textrm{P}(Y=\textrm{Ia})}{\textrm{P}(x | Y=\textrm{non-Ia})\cdot \textrm{P}(Y=\textrm{non-Ia})}.
\end{equation*}

2) Boosting directly estimates  \textit{odds}(\textit{x}) through regression methods, as a sum of small trees built by a type of functional gradient descent.

These methods are discussed in detail below.

\subsection{Kernel Density Estimation (KDE)}\label{sec:KDE}
Kernel Density Estimation (KDE) is a non-parametric method for estimating the probability density function (pdf) of a sequence of observables. Within this paper, the probability densities of the post-processed data described in Sections~\ref{sec:gamma} - \ref{sec:SALT} are used for classification. Pdfs are useful as we may base a classification rule upon the relative probabilities that a candidate supernova is either type Ia or not type Ia. Such a classification rule will require both the Ia and the non-Ia probability densities for the observed SN data. KDE enables us to derive these pdfs in a fairly model-independent manner, as we now discuss.

\begin{figure}
	\begin{center}
		\includegraphics[width=3.5in]{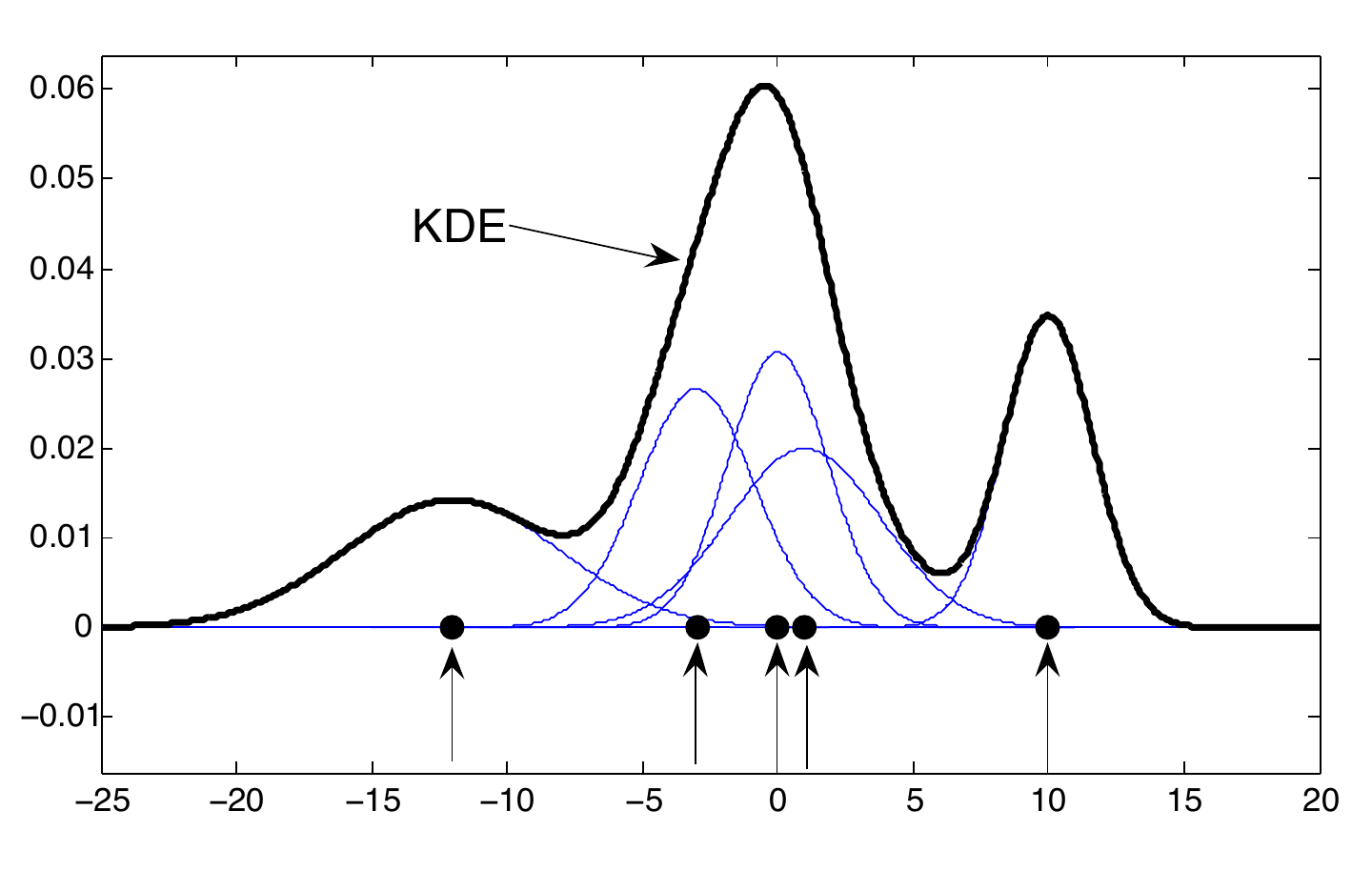}
	\end{center}
\caption{Schematic figure illustrating the idea of a KDE in one dimension. The training data points are shown as dark points with arrows. The Gaussian kernels are shown together with the sum of the kernels. Note that the KDE is not normalised in this figure and is thus close to what we actually use in this paper.}
\label{fig:KDEillus}
\end{figure}

\begin{figure}
	\begin{center}
		\includegraphics[width=3.5in]{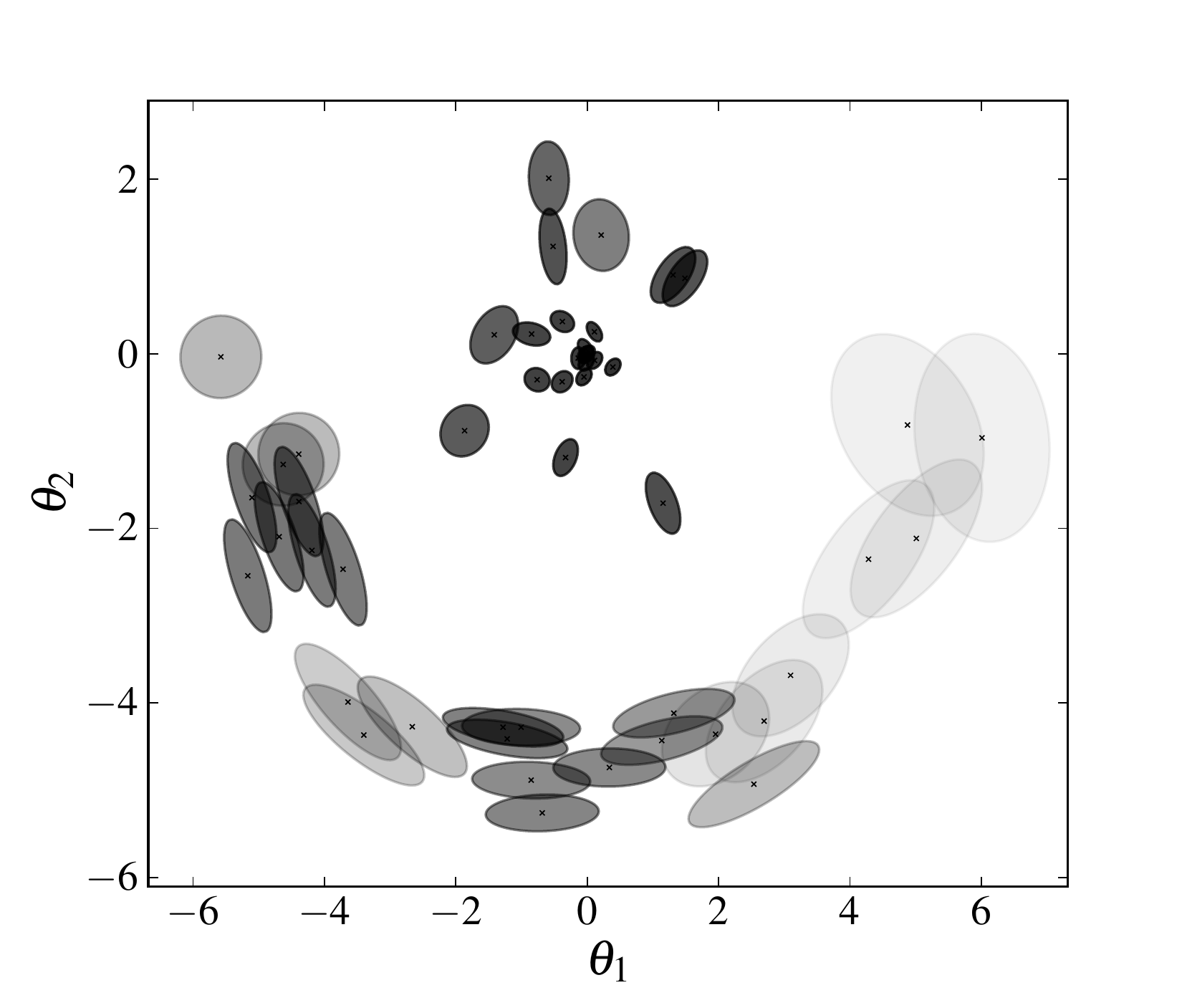}
	\end{center}
\caption{A realisation of fifty points from an unusal distribution. Around each observed point a kernel is constructed. The axes of each kernel are the eigenvalues of the point's (2 $\times$ 2) $\Sigma_i$  matrix (Eq.~\ref{siggysiggy}). Each $\Sigma_i$ is the covariance matrix of the nearest $\ell$ points multiplied by the global bandwidth, $h$. Here $h=~0.6$ and $\ell = 10$.} 
\label{fig:vbw}
\end{figure}

Suppose we have a set of $d$ observables and that we would like to estimate the value of the pdf at a point $\vec{x}$ in this $d$-dimensional space. Given a training set with $n$ observations, i.e. $n$ points $\vec{X}_i$ in this $d$-dimensional space, the Kernel Density Estimate (KDE) is given by
\begin{equation}
\hat{f}_h\left(\vec{x}\right)=\frac{1}{n}\sum_{i=1}^n{\frac{1}{h^d}K_i\left(\frac{\vec{x}-\vec{X}_{i}}{h}\right)},
\label{kernel}
\end{equation}
where $\hat{f}_h\left(\vec{x}\right)$ is the KDE, $\vec{X}_{i}$ is the $i$-th training observation, $K_i$ is the kernel function for the $i$-th training observation and $h$ is the global kernel bandwidth. $h$ is a tuning parameter: the kernels become more ``peaked" about the training observations as $h$ becomes smaller. The optimal bandwidth may be obtained by cross-validation (see Appendix~\ref{app:cross-val}). The choice of kernel is arbitrary, except that any proposed kernel should satisfy the following two conditions:
\begin{itemize}
	\item $\int K(\vec{x})d\vec{x} = 1$
	\item $K(-\vec{x}) = K(\vec{x}) \mbox{\ \ \ }$
\end{itemize}
The first condition ensures that the KDE integrates to unity and that all observations carry equal weight, whilst the second condition ensures that the KDE is unbiased and is centered about one of the $n$ $d$-dimensional training data points. The basic idea of the KDE method is illustrated in Figure~\ref{fig:KDEillus} in a simple 1D example. A commonly used kernel (and the kernel that we will use throughout this paper) is a multivariate Gaussian, normalised to unit volume:
\begin{eqnarray}\label{siggysiggy}
&&K\left(\frac{\vec{x}-\vec{X}_i}{h}, \Sigma_i\right) = \\ \nonumber
&&\frac{1}{\sqrt{(2\pi)^d |\Sigma_i|}} \exp\left\{-\frac{1}{2} \left(\frac{\vec{x}-\vec{X}_i}{h}\right)^T\Sigma_i^{-1} \left(\frac{\vec{x}-\vec{X}_i}{h}\right)
\right\} .
\label{Normal}
\end{eqnarray}
Here $\vec{x}$ and $\vec{X}_i$ are $d$ dimensional vectors and $\Sigma_i$ is a $d \times d$ covariance matrix that changes the orientation and shape of the kernel around each training observation $i$; for example the covariance matrix $\Sigma_i$ can be estimated from the nearest $\ell$ neighbours of a training data point, which is what we do, as described in Section~\ref{res:21-D KDE} and as illustrated in Figure~\ref{fig:vbw}. This provides the possibility of adapting the kernel to local variation. In contrast the bandwidth parameter $h$ affects the global behaviour of the kernels.
While it is more common to choose the covariances to be equal, for the SNPCC and the current application this would have been a bad choice (as described in Section~\ref{res:21-D KDE}). 

\subsubsection{Integration over data errors}\label{sec:integrating_errors}

In order to classify a supernova with lightcurve measurements $\vec{x}$, we must evaluate the KDE at $\vec{x}$. However in our case we are not sure where $\vec{x}$ lies in parameter space as the lightcurve measurements have errors and are not perfectly sampled.

Using a Gaussian kernel, we write the KDE as
\begin{equation}
\hat{f}(x) = \frac{1}{n} \sum_i \frac{1}{h^d} K\left(\frac{x-X_i}{h},\Sigma_i\right)
\end{equation}
For simplicity we suppress vector notation but all quantities (other than $h$) are $d$ dimensional vectors or matrices, and the index $i$ runs over the points in the training set.

Now assume that the location of a point in the $d$ dimensional space is not known exactly and is instead given by a Gaussian pdf. We take the mean to be $x$ and the covariance matrix to be $Y$. The KDE value is then given by integrating the KDE over the unknown pdf of the point being classified:
\begin{eqnarray}
&&\int dz K\left(z-x,Y\right) \hat{f}(z) = \\ \nonumber
&&\frac{1}{n} \sum_i \frac{1}{h^d} \int dz K\left(z-x,Y\right) K\left(\frac{z-X_i}{h},\Sigma_i\right) .
\end{eqnarray}
We notice that this reverts back to the original value if $K$ is a delta-function located at $x$. Further, the function being integrated is a product of two Gaussians, which is itself another Gaussian. The KDE value then simplifies to
\begin{equation}
\label{eqn:gKDE}
\hat{f}(x) = \frac{1}{n} \sum_i \frac{1}{h^d} K\left(\frac{x-X_i}{h};\Sigma_i+h^{-2d}Y\right) ,
\end{equation}
i.e. the KDE kernels simply have an increased variance, given by the sum of their covariance matrix and the covariance matrix of the point being evaluated, scaled by $h^{-2d}$. 
The importance of including this increased variance for uncertain observations should not be ignored, especially when the variances of the points being classified are large (as is the case in this paper). Correctly implementing Eq.~(\ref{eqn:gKDE}) can significantly improve classification performance. In Section~\ref{res:21-D KDE} we compare analyses on the SN data including and ignoring the covariance $Y$.

\subsection{Boosting}\label{sec:boosting}
Boosting (\cite{Freund95}) is a learning algorithm for classification.  Until recently the most popular boosting algorithm was AdaBoost (\cite{Freund1997119}). AdaBoost works by combining weak-classifiers into a committee, whose combined decision is significantly better than that of individual weak-classifiers. The precise workings behind AdaBoost's success remained hazy until it was shown (\cite{FHT2000}) that boosting produces the powerful committee by sequentially adding together weak-classifiers calculated by steepest descent. The further ideas of slow learning~\citep{F2001} and bagging (\cite{Friedman2002367}) were later introduced into boosting, culminating eventually in the Gradient Boosting Machine (GBM) algorithm. The algorithm, implemented as a package in the statistical programming language R\footnote{R and its associated packages can be downloaded from http://www.r-project.org.}, is described in Section \ref{sec:gbm}. A brief discussion of trees and loss functions is presented in Sections \ref{sec:trees} and \ref{sec:classification} in preparation for the presentation of the GBM algorithm.

\subsubsection{Tree functions}\label{sec:trees}
\begin{figure}
	\begin{center}
		\includegraphics[width=3.5in]{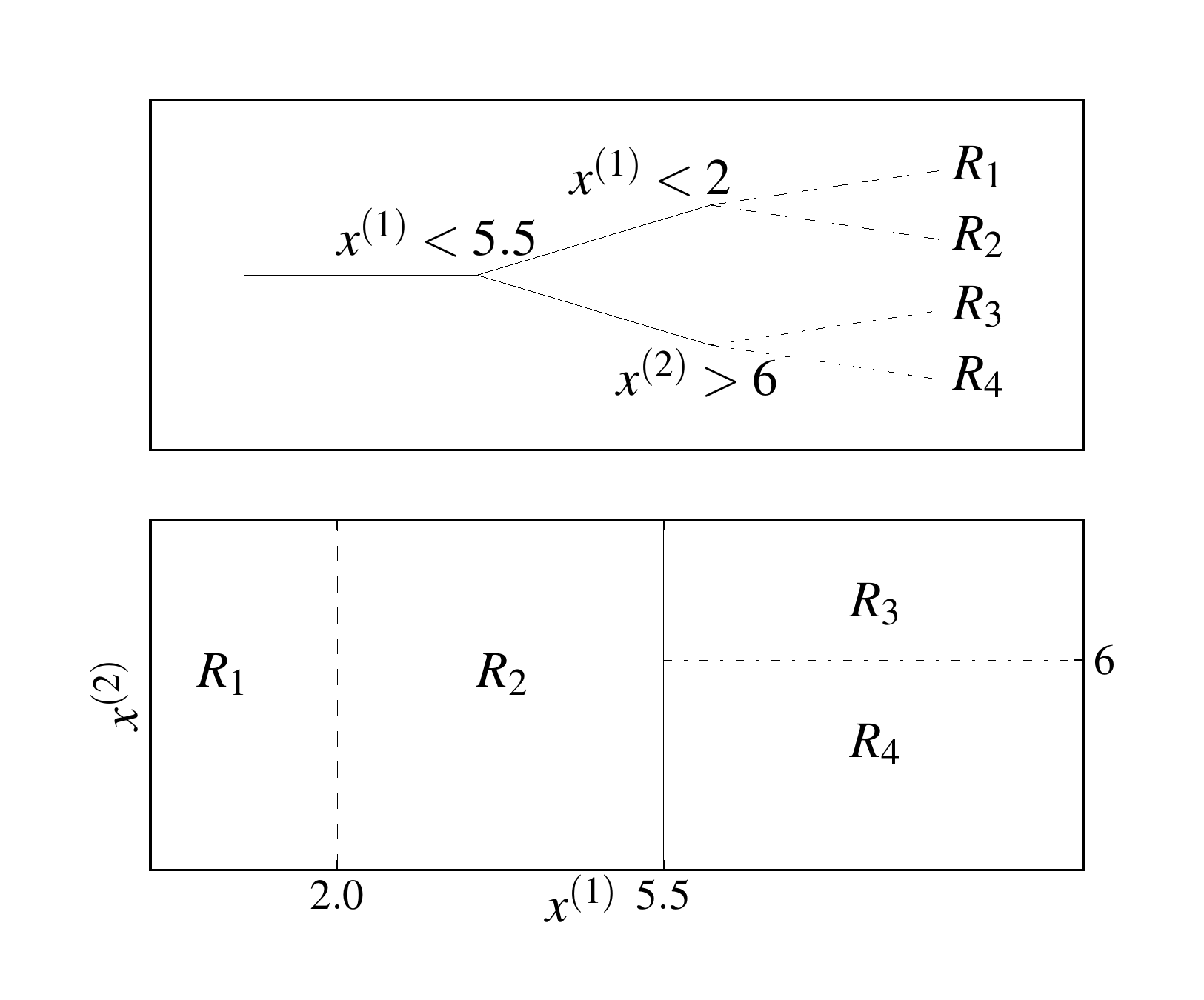}
	\end{center}
\caption{(\textit{Above}) A tree of depth 2 for classifying an object into one of $2^2$ regions. (\textit{Below}) The tree domain containing $2^2$ distinct regions as defined by the tree.}
\label{fig:basic_tree}
\end{figure}

The most widely used weak-classifiers (a.k.a. basis functions) in boosting are trees. Trees are discontinuous functions which take discrete values in different regions of a domain. That is to say, a tree $T$ has the form:

\[T(\vec{x}) = \left\{ 
\begin{array}{l l}
  z_1 & \quad \mbox{if $\vec{x} \in R_1$}\\
  \vdots\\  
  z_K & \quad \mbox{if $\vec{x} \in R_n$}\\ \end{array} \right. \]
  
where the $K$ distinct regions $R_1 \cdots R_K$ together partition $\vec{x}$-space. The region boundaries can be described through the branchings of a tree, as illustrated in Figure~\ref{fig:basic_tree}. For boosting, it is common to only use trees of a very simple form, that is only trees with branchings of the form  $x^{(i)} < v$, where $x^{(i)}$ is one of the dimensions of $\vec{x}$-space and $v$ is a real number. In the case of the SNPCC, $\vec{x}$ are the parameters fitted to the lightcurves in Section~\ref{sec:gamma}.

\subsubsection{Loss function for classification}\label{sec:classification}

Suppose we have observed $n$ training points, each consisting of data and type: ($\vec{X}_i,\tau_i)$, where the data $\vec{X}_i$ is a $d$-dimensional vector, and the type $\tau_i$ is $\pm 1$, corresponding to the two classes. Suppose that we are required to find a function $F:R^{d}\rightarrow R$ which minimises the following \textit{loss function}:
\begin{equation}
\label{function:loss}
L\left( F \right) = \sum_{i=1}^{n} \log\left(1 + \exp\left[-2F(\vec{X}_i)\tau_i\right]\right).
\end{equation} 

The specific form chosen for the loss function~\ref{function:loss} can be explained by considering its partial derivatives with respect to $F(\vec{X}_i)$. Doing so~\citep{hastie_09_elements-of.statistical-learning}, it can be shown that the form of $F$ which minimises (\ref{function:loss}) is given by:

\begin{equation}
\label{function:logoddsapprox}
F(\vec{X}_i) = \frac{1}{2}\log\ \frac{\mbox{$\#$ observations:\hspace{4mm}}  \vec{X}_i, \tau = 1}{\mbox{$\#$ observations:\hspace{3.5mm}} \vec{X}_i, \tau = -1} 
\end{equation} 
 
This is an approximation to half the \textit{log odds} (the log of the \textit{odds}):
\begin{equation}
\label{function:logodds}
\textit{log odds} \equiv \log\frac{P(\tau_i = 1\vert \vec{x} = \vec{X}_i)}{P(\tau_i = -1 \vert \vec{x} = \vec{X}_i)}.
\end{equation}

\noindent This is the key result: a function which minimises the loss function~(\ref{function:loss}) is a good approximation to half the \textit{log odds}. 
 A good approximation to the \textit{log odds} is exactly what is needed for classification problems. The boosting algorithm aims to approximately minimise this loss function and in so doing arrive at an approximation of the \textit{log odds} which can then be used for classification.

If you have observations at every possible data point, you can directly approximate the \textit{log odds} through (\ref{function:logoddsapprox}). In reality, you will not have observations at all possible data points, and so cannot do this. This trivially corresponds to not having observed all possible lightcurves, and so needing to make inferences from simililar lightcurves. Boosting does this inference through constrained minimisation of the loss function, as described in the following section.

\subsubsection{The Gradient Boosting Machine}\label{sec:gbm}
The Gradient Boosting Machine~\citep{F2001} works by sequentially adding new trees to a function $F$, each addition reducing $L(F)$~(\ref{function:loss}) and so improving the approximation of $F$ to half the \textit{log odds}. 

The trees, which have depth $D$, are appended to $F$ at each of the $M$ iterations of the GBM algorithm. Choosing larger $M$ and $D$ values results in a final $L(F)$ nearer to the global minimum value (\ref{function:logoddsapprox}). However, our end objective is not to reach the global minimum but to construct a good approximation to the \textit{log odds}, and trees of lower depth are generally better suited to this end.

Algorithm 1 (below) outlines an implementation of the GBM. A few subtleties have been omitted from it here, and we refer you to Appendix~\ref{app:gbm} for a fuller description. We recommend watching our demonstrative animation of the algorithm while reading Algorithm 1. The animation can be be found at~\cite{COSMOAIMS}.

\medskip
\medskip

\noindent \textbf{Algorithm 1 - Gradient Boosting Machine}

\medskip

\noindent $\cdot$Input: $\vec{X}_i, \tau_i$ for observations $i = 1$ to $n$.

\noindent $\cdot$Initalise: $ F_0(\vec{x}) \leftarrow \displaystyle \frac{1}{2}\log{\displaystyle\frac{1 + \bar{\tau}}{1 - \bar{\tau}}}
 \hspace{8mm} $where $ \bar{\tau} = \displaystyle\frac{1}{n} \displaystyle\sum_{i=1}^n \tau_i$. \\
\noindent $\cdot$Initalise: $ z_i \leftarrow 0$ for observations $i = 1$ to $n$. The $z_i$'s will measure how much of a ``misfit" each observation is.
\vspace*{1mm}
\newline
\noindent $\cdot$Choose tree depth $D$ and number of trees $M$.
\vspace*{1mm}
\newline
\noindent $\cdot$for $m = 1$ to $M$: \\
\hspace*{3mm}1) for $i = 1$ to $n$, update $z_i$: \\
\vspace*{0.5mm}
\hspace*{8mm} 
\hspace*{8mm}$z_i \leftarrow -\displaystyle\frac{\partial L\hspace{7mm}}{\partial F_{m-1}\left(\vec{X}_i\right)} = \displaystyle\frac{2\tau_i}{1 + \exp{\left[2F(\vec{X}_i)\tau_i\right]}}$ \newline
\vspace*{1.5mm}

\noindent \hspace*{3mm}2) Fit by least squares $T_m$, the new tree: $z_i \sim T_m(\vec{X}_i)$.
\vspace*{0.5mm}
\newline
\hspace*{6mm}(where $T_m$ has regions $R_{m,1} \cdots R_{m,2^D}$ fitted to minimise \newline
\hspace*{7mm}the ingroup variance: see Appendix~\ref{app:boosting} for details.)
\vspace*{1.5mm}
\newline 
\hspace*{3mm}3) Choose constants $\gamma_{m,1} \cdots \gamma_{m,2^D}$ for $R_{m,1} \cdots R_{m,2^D}$.
\hspace*{6mm}(chosen to minimise $L\left(F_{m-1} + T_m\right)$)
\vspace*{0.5mm}
\newline
\hspace*{3mm}4) $F_{m} \leftarrow F_{m-1} + T_m$
\vspace*{0.5mm}
\newline
\noindent $\cdot$Finally, $F \leftarrow F_{M}.$
\vspace*{1.0mm}
\newline
\noindent $F$ is our final approximation to half the \textit{log odds}, and it can now be used to classify with a simple rule of the form:
\vspace*{1.0mm}
\newline
\noindent\hspace*{1mm}IF $F(\vec{x}_i) > v \Rightarrow  \tau_i = 1;\hspace{1mm}\textrm{ELSE}\hspace{2mm} \tau_i = -1$,
\vspace*{1.0mm}
\newline
where the optimal $v$ depends on the Figure of Merit. 

Notice that the variable $z_i$, updated in step 1, is positive if $\tau_i=1$ and negative if $\tau_i = -1$. For this reason, when $T_m$ is fit to the $z_i$'s at step 2, observations of the same type are more likely to fall into the same region of $T_m$. Moreover, observations with large $z_i$'s carry more weight while fitting $T_m$, and hence are even more likely to be placed with objects of the same type. This acts to place special attention on unusual objects, or objects whose type is not clear. 

While values are fitted for each tree region in step 2 (as described in Appendix~\ref{app:boosting}), these values will not necessarily result in a reduced $L\left(F_{m-1} + T_m\right)$. Hence at step 3 of the algorithm, $\gamma_{m,k}$ values are explicitly chosen to minimise $L\left(F_{m-1} + T_m\right)$. In effect, only the tree \textit{shape} is taken from step 2.
  
\section{Results}\label{sec:results}

The entries in the SNPCC were evaluated using the Figure of Merit (FoM): 
\begin{eqnarray*}
f(N_{Ia}^{\tickYes}, N_{non-Ia}^{\textrm{\tickNo}}) &=& \textrm{efficiency} \times \textrm{pseudo-purity}\\
 &=&\left( \frac{N_{Ia}^{\tickYes}}{N_{Ia}^{TOT}}\right) \times  \left(\frac{ N_{Ia}^{\tickYes} }{N_{Ia}^{\tickYes} + 3\cdot  N_{non-Ia}^{\textrm{\tickNo}}}\right),
\label{eq:FoM}
\end{eqnarray*}

where $N_{Ia}^{\tickYes}$ is the number of correctly classified SNeIa, $N_{non-Ia}^{\textrm{\tickNo}}$  is the number of non-Ia SNe classified as SNeIa, and $N_{Ia}^{TOT}$ is the total number of SNeIa. Had the coefficient of $N_{non-Ia}^{\textrm{\tickNo}}$ in the denominator of the pseudo-purity term been 1 and not 3 the term would have been true purity, i.e. the proportion of SNeIa in the final Ia-classified group. How relevant this FoM is to cosmology is not absolutely clear, but it is a robust measure of how well a classification algorithm penalises both missed detections and false discoveries. For applications such as BEAMS~\citep{beams} a FoM which takes type probabilities as inputs would be more useful. 

In this section we discuss the implementation and performance of each of our methods. Unless stated otherwise, the scores given in this section refer to the SNPCC, while all figures are using the post-SNPCC data described in Section~\ref{sec:SNPCC}. Of particular interest to us is the comparison of results obtained when the training is done with representative and non-representative samples. We also briefly mention applications that these methods have previously found in cosmology and related fields.
 
\subsection{21D KDE}\label{res:21-D KDE}

\subsubsection{Application}\label{sec:application:21-D}
Kernel Density Estimators have been used before in astronomy for estimating the probability density function from a discrete or noisy data set \citep{Fadda1998,Bissantz2007,Ascasibar2010}, identifying groups \citep{Balogh2004} and clusters \citep{Valtchanov2004} in galaxy surveys, and determining the timings of millisecond pulsars \citep{Carstairs1991} and gamma-ray bursts \citep{deJager1986}, to name a few examples.

In Section~\ref{sec:gamma} we described how we fit the SN lightcurves in each of the 4 bands using the parameterised function~(\ref{gamma}), resulting in twenty lightcurve parameters. With the addition of host redshift in the case of the \texttt{+HOSTZ} challenge, each SN is described by a 21 dimensional (21D) point. We use KDE to approximate the 21D Ia and non-Ia probability density functions (pdfs) based on the training data.
 
We allowed the 21D training points to have different covariance matrices, as described in Section~\ref{sec:KDE}. As previously mentioned a single global covariance is most common for KDE, but in cases where a pdf has large regions of high and low probability, this can be problematic. In low probability regions the kernel density will be too ``spikey" while in high probability regions it will be too smooth. To understand this, consider what would happen if, in Figure~\ref{fig:vbw}, the ellipses were constrained to all be of the same size. Chosen too small and the low probability region would have ``bumps", too large and the high probability region would lose features. The 21D points for the SNPCC are not uniformly distributed, as illustrated by the cumulative plots of Appendix~\ref{sec:pdists}, and so are susceptible to this problem. Using cross-validation we chose $\ell = 10$ and $h = 0.6$ (using the notation from Section~\ref{sec:KDE}).

\begin{figure}
	\begin{center}
		\includegraphics[width=3.5in]{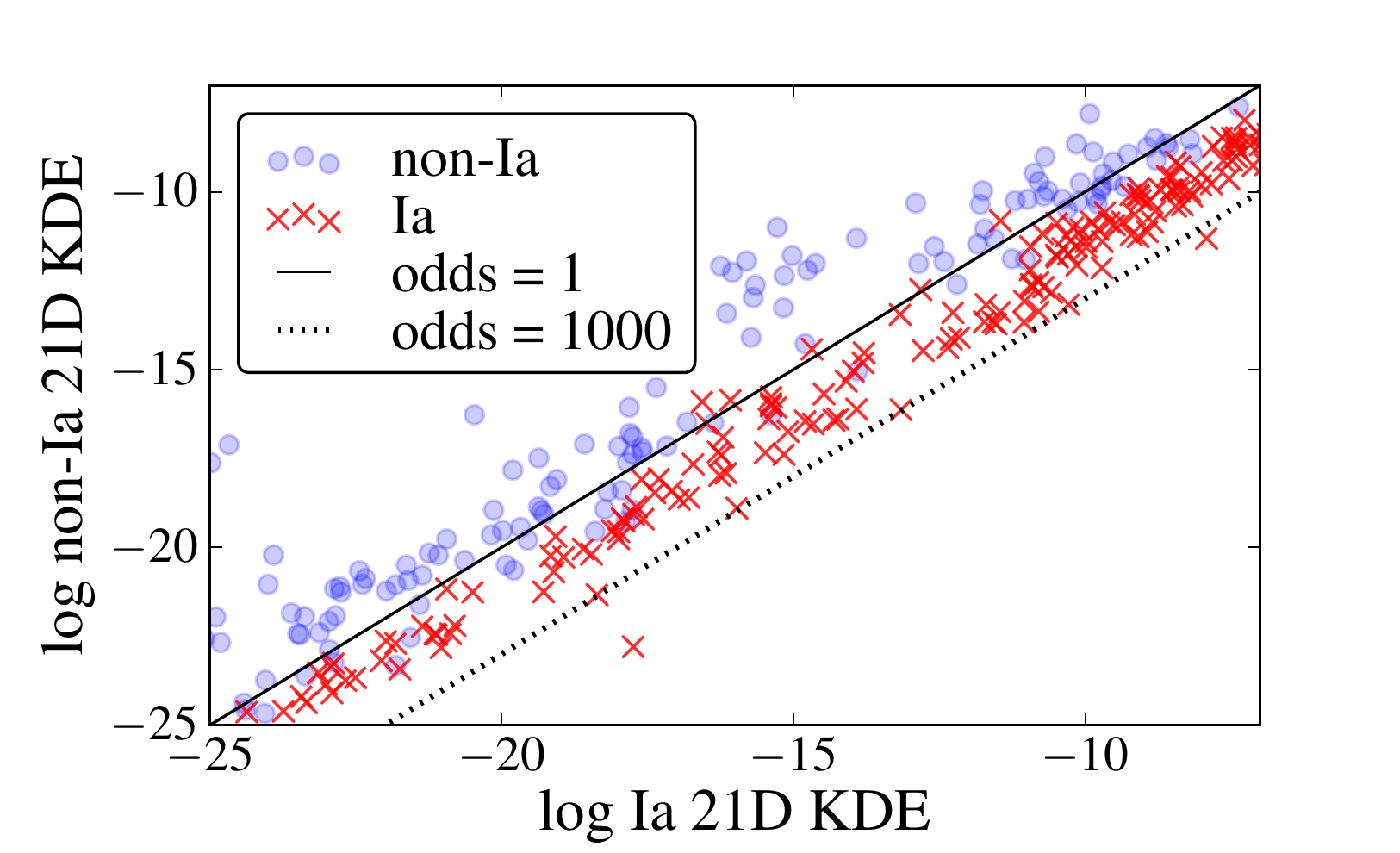}
	\end{center}
	\caption{Ia (red crosses) and non-Ia (blue circles) in the non-representative training sample. The KDE values at calculated using tenfold cross-validation.}
\label{fig:KDEvs}
\end{figure}		

Having constructed two KDEs from a training sample, each unclassified SN may be classified as follows: 
\begin{enumerate}
	\item Fit Eq.~\ref{gamma} to each of the four lightcurves thus obtaining a 21D point for the candidate.
	\item Evaluate the Ia and non-Ia kernel probabilities derived from the training sample at the 21D point, and then evaluate the \textit{odds}.
	\item If the \textit{odds} (or \textit{log odds}) is above some threshold, classify as Ia. 
\end{enumerate}
In cases where one or both of the KDEs are a poor representation of the underlying pdf, it may be preferable to modify step (iii). For example if one of the KDEs is particularly inaccurate, one may prefer to classify by using only the other KDE. For the SNPCC leaving step (iii) unchanged was advisable, as can be deduced from Figure~\ref{fig:KDEvs}. The lines in Figure~\ref{fig:KDEvs} are lines of constant \textit{odds}. If KDEs are accurate approximations to pdfs, a line of constant \textit{odds} is optimal for discriminating between Ia's (below the line) and non-Ia's (above the line), irrespective of the FoM used. Furthermore, if the KDEs are accurate approximations to pdfs, there should be an equal number of Ia's and non-Ia's on the line  \textit{odds}~=~1 and 1000 times more Ia's as non-Ia's on the line \textit{odds}~=~1000. This is roughly observed in Figure~\ref{fig:KDEvs} and so we can proceed to choose the \textit{odds} line which maximizes the SNPCC FoM.

\begin{figure}
	\begin{center}
		\includegraphics[width=3.35in]{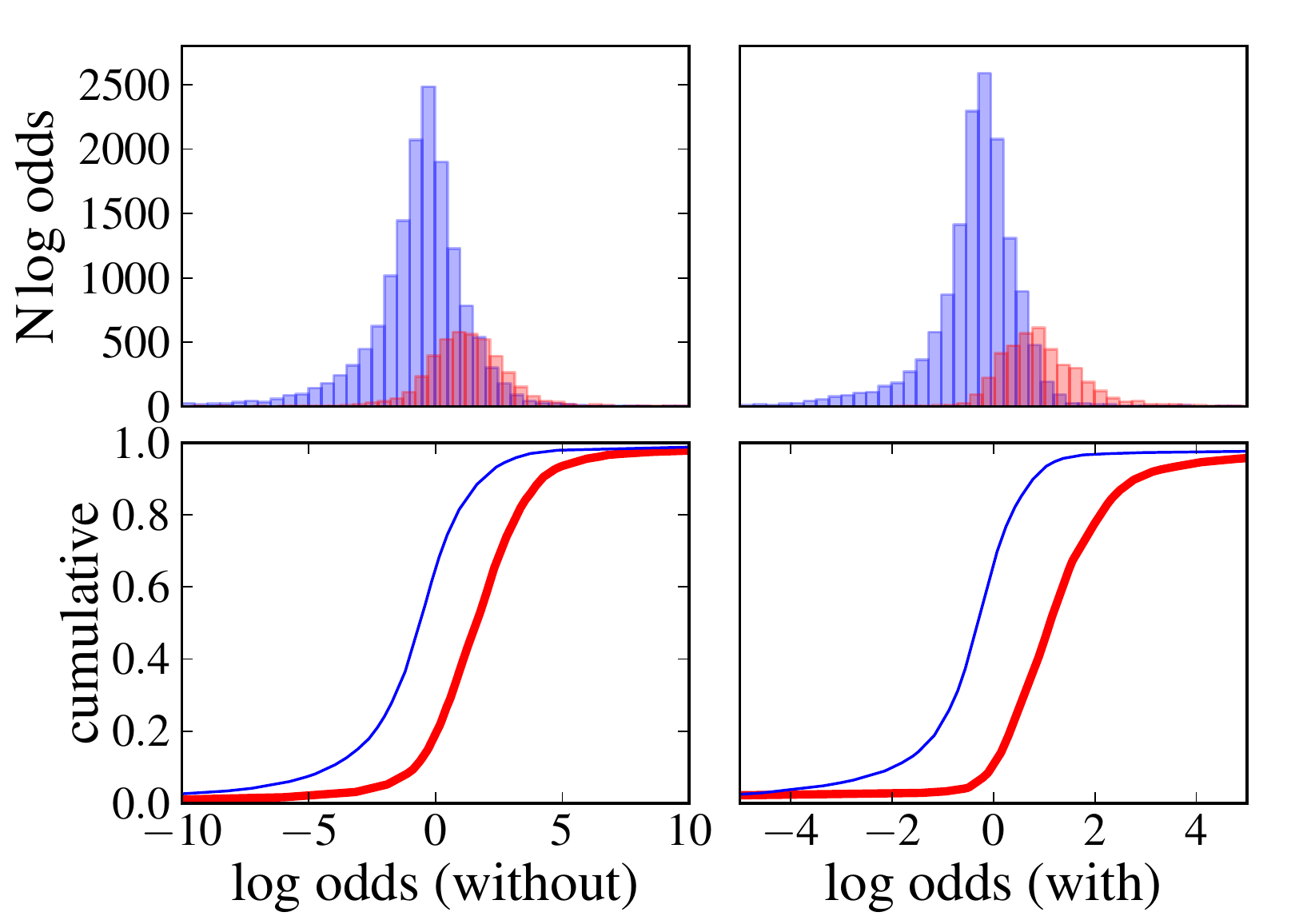}
	\end{center}
\caption{(\textit{Above}) Histograms and (\textit{Below}) cumulative plots of the 21D (representatively constructed) KDE \textit{log odds}. (\textit{Left}) The parameter covariance matrix is not included in KDE evaluation as proposed in Section~\ref{sec:integrating_errors}. (\textit{Right}) The parameter covariance matrix is included in KDE evaluation. } 
\label{fig:with_without_cm}
\end{figure}

For the entry in the SNPCC, we failed to include the parameter covariance matrices when calculating KDE values (in effect, we set $Y$ to be a matrix of zeros in Eq.~\ref{eqn:gKDE}). Our final score suffered as a result - the benefit of correctly implementing the calculation~(\ref{eqn:gKDE}) is illustrated in Figure~\ref{fig:with_without_cm}, where we see from both the histograms and the cumulative plots an increased separation between Ia's and non-Ia's when Eq.~\ref{eqn:gKDE}) is correctly implemented. We find a $15\%$ increase in score when correctly implemented on the post-SNPCC data.  
The KDE method still obtained the second and third highest scores in the \texttt{-HOSTZ} and \texttt{+HOSTZ} competitions respectively, with scores of $0.37$ and $0.39$. Of interest is that the 20D KDE (\texttt{-HOSTZ}) is almost as good at classifying as the 21D KDE (\texttt{+HOSTZ}). The winning competition scores \citep{sntc_results} were  $0.51$ (\texttt{-HOSTZ}) and $0.53$ (\texttt{+HOSTZ}).

\subsubsection{Non-representative vs representative}

\begin{figure}
	\begin{center}
		\includegraphics[width=3.5in]{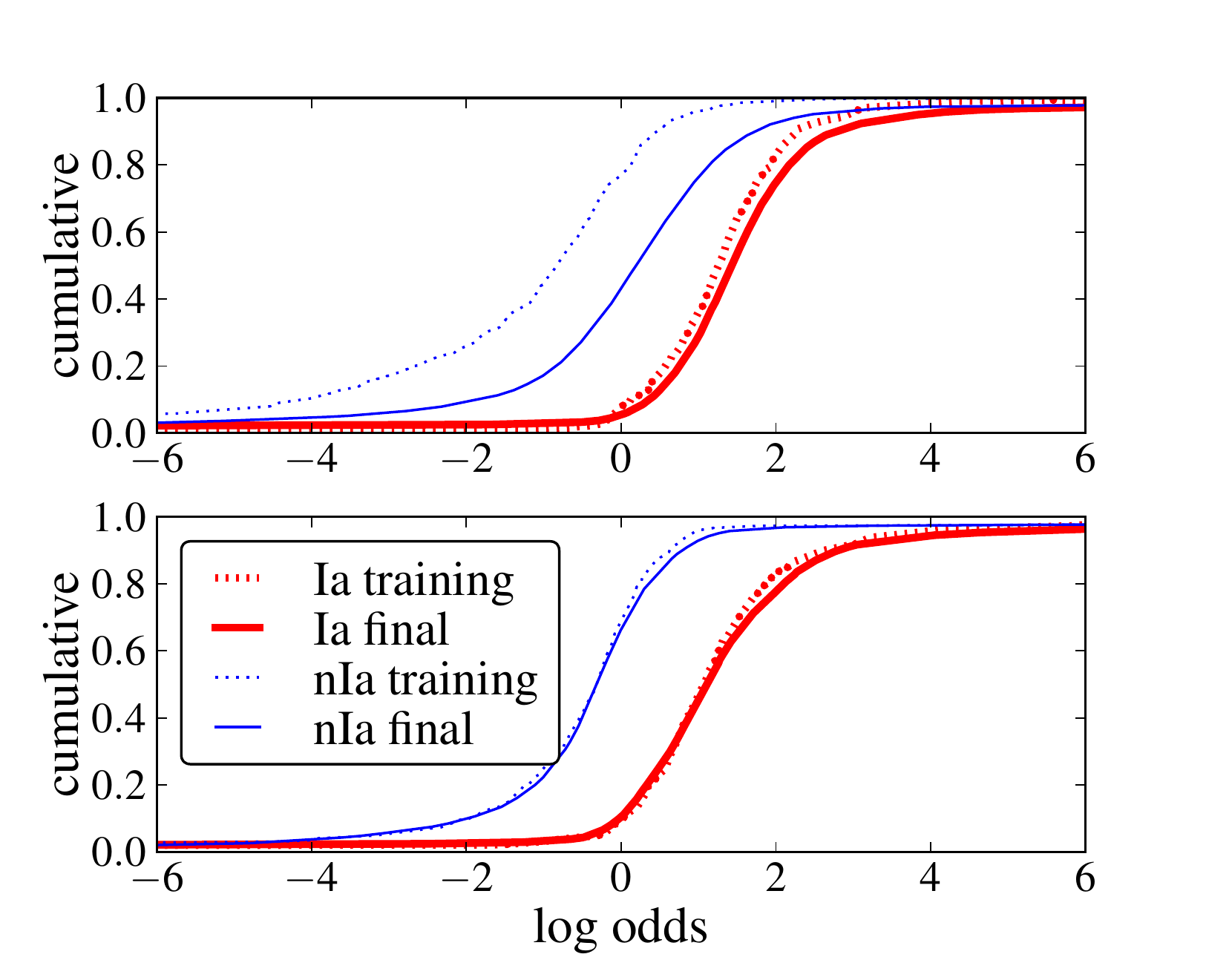}
	\end{center}
	\caption{The cumulative frequency of \textit{log odds} for non-Ia (blue) and Ia (red) SNe, for the training (dashed) and test (solid) samples. The training \textit{log odds} were calculated using tenfold cross-validation. (\textit{Above}) Using non-representative training and (\textit{Below}) using representative training. }
\label{fig:KDE_4_cums_tr_nr}
\end{figure}

As with all of our methods, we constructed classifiers using both the non-representative sample provided and a representative sample of equal size, as described in Section~\ref{subsubsec:training}. In each case, the remaining unclassified SNe were used as a test of the performance of the classifier. 

Figure~\ref{fig:KDE_4_cums_tr_nr} carries useful information about the performance of the non-representatively trained KDEs and representatively trained KDEs. For example, the efficiency of classifying Ia's with a \textit{log odds} threshold of 2 is simply the cumulative value of the unclassified Ia's (solid red) at \textit{log odds}~=~2. For both representatively and non-representatively trained KDEs this is about 0.75, meaning that about 75\% of SNeIa are correctly classified when a threshold of \textit{log odds}~=~2 is used. 

To obtain high purity, the \textit{log odds} threshold must be chosen such that the non-Ia cumulative frequency is low compared to the Ia cumulative frequency. To obtain high efficiency, the \textit{log odds} threshold must be chosen such that the Ia cumulative frequency is high. Putting these together, to obtain both high purity and high efficiency, a \textit{log odds} threshold must be found at which the non-Ia cumulative frequency is low and the Ia cumulative frequency is high.

The dashed lines in Figure~\ref{fig:KDE_4_cums_tr_nr} are the cumulatives of the training data using tenfold cross-validation. In the case of representative training, we see that these are accurate predictors of the true cumulatives. But in the case of non-representative training, the non-Ia cumulatives of training and unclassified SNe are vastly different. If in the case of non-representative training one assumed that the training sample were in fact representative, one would predict a non-Ia misclassification rate of under of 10\% using a \textit{log odds} cutoff of 1. In reality it is 30\%. Such dangerous predictions are impossible to make if a representative sample is used in KDE construction, as illustrated by the hugging of the solid lines to the dotted lines.

\subsection{Boosting}\label{sec:boosting}

\subsubsection{Application}\label{sec:application:boosting}

Boosting has been used in particle physics, for example by the MiniBooNE neutrino oscillation experiment
\citep{Roe2005}
and is implemented in the photometric redshift package AborZ \citep{Gerdes2010}. In the SNPCC we applied boosting to the twenty fitted lightcurve parameters for the \texttt{-HOSTZ} competition, and the twenty-one parameters for the \texttt{+HOSTZ} competition. Using tenfold cross-validation we chose to use 4000 trees to maximize the FoM (\ref{eq:FoM}). We chose the learning rate to be 0.05 and the bagging fraction to be 0.5 (these parameters are described in Appendix~\ref{app:gbm}).

During the training phase of the SNPCC we expected, based on the idea that the training sample was representative, that 
boosting would significantly outperform the 21D KDE. In reality boosting performed more poorly than the 21D KDE, 
obtaining scores of $0.20$ (\texttt{-HOSTZ}) and $0.25$ (\texttt{+HOSTZ}) \citep{sntc_results} strongly suggesting that the 21D KDE method is more robust to biases in the training set than boosting. 

In the case of the post-SNPCC data, the score obtained with non-representative training is even lower $(0.15)$ (\texttt{+HOSTZ}) due to bugs in the original SNPCC data such as too dim non-Ias which made classification easier, as described in \citet{sntc_results}. As a result comparison of scores in this paper with those in the competition cannot be made directly. 


\begin{figure}
	\begin{center}
		\includegraphics[width=3.5in]{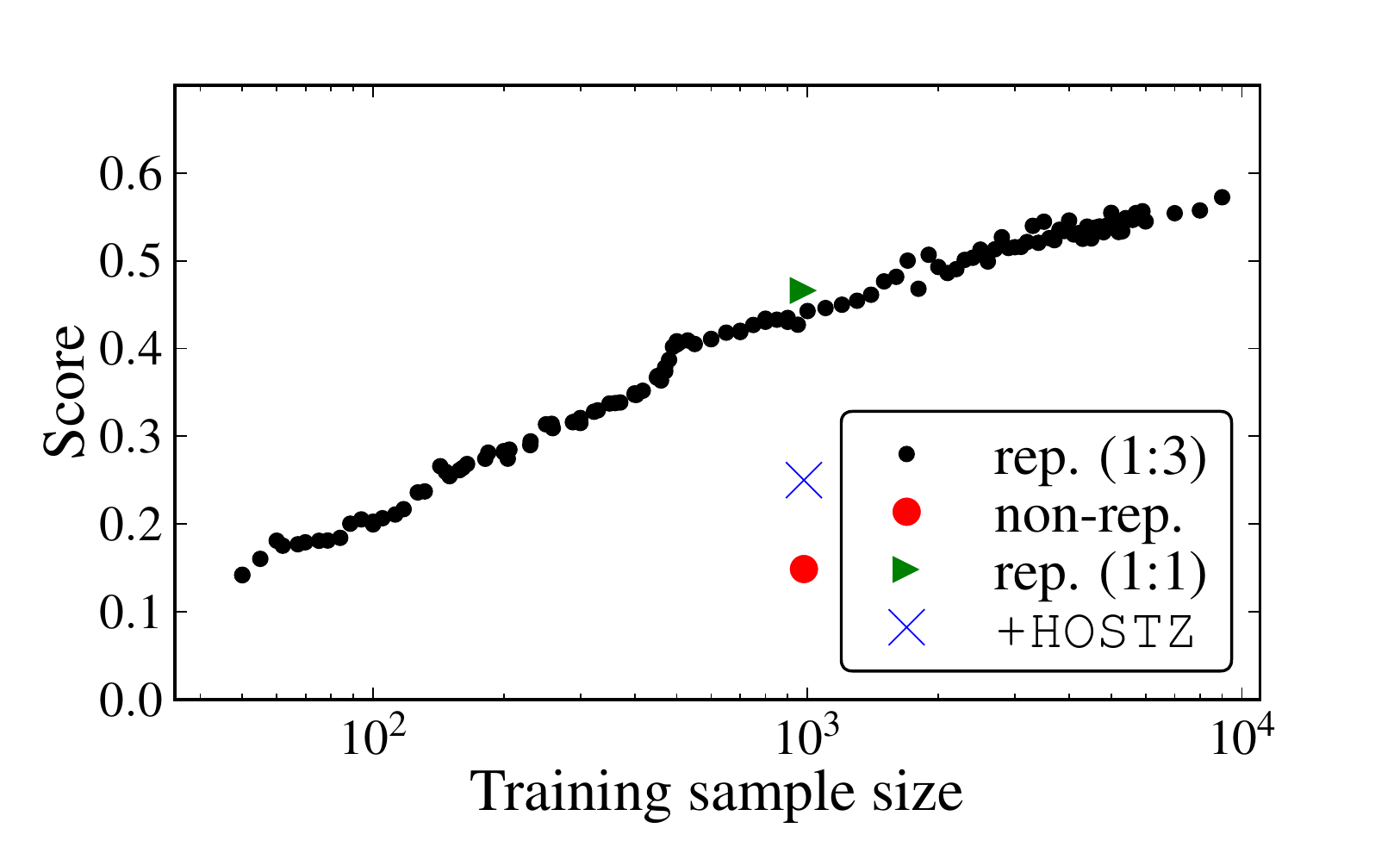}
	\end{center}
\caption{(Small black circles) The score obtained by boosting when trained with random repreresentative samples of varying size (100 to 6000 SNe). (Large red circle) Training on the given non-representative sample. (Blue cross) The score obtained in the \texttt{+HOSTZ} competition. (Green triangle) The performance when trained with a ``random" sample with non-random Ia:non-Ia ratio of 1:1 as opposed to true ratio Ia:non-Ia $\sim$ 1:3.}
\label{score_FoM_boosting_random}
\end{figure}

\subsubsection{Non-representative vs representative}\label{sec:boostrnr}

\begin{figure}
	\begin{center}
		\includegraphics[width=3.35in]{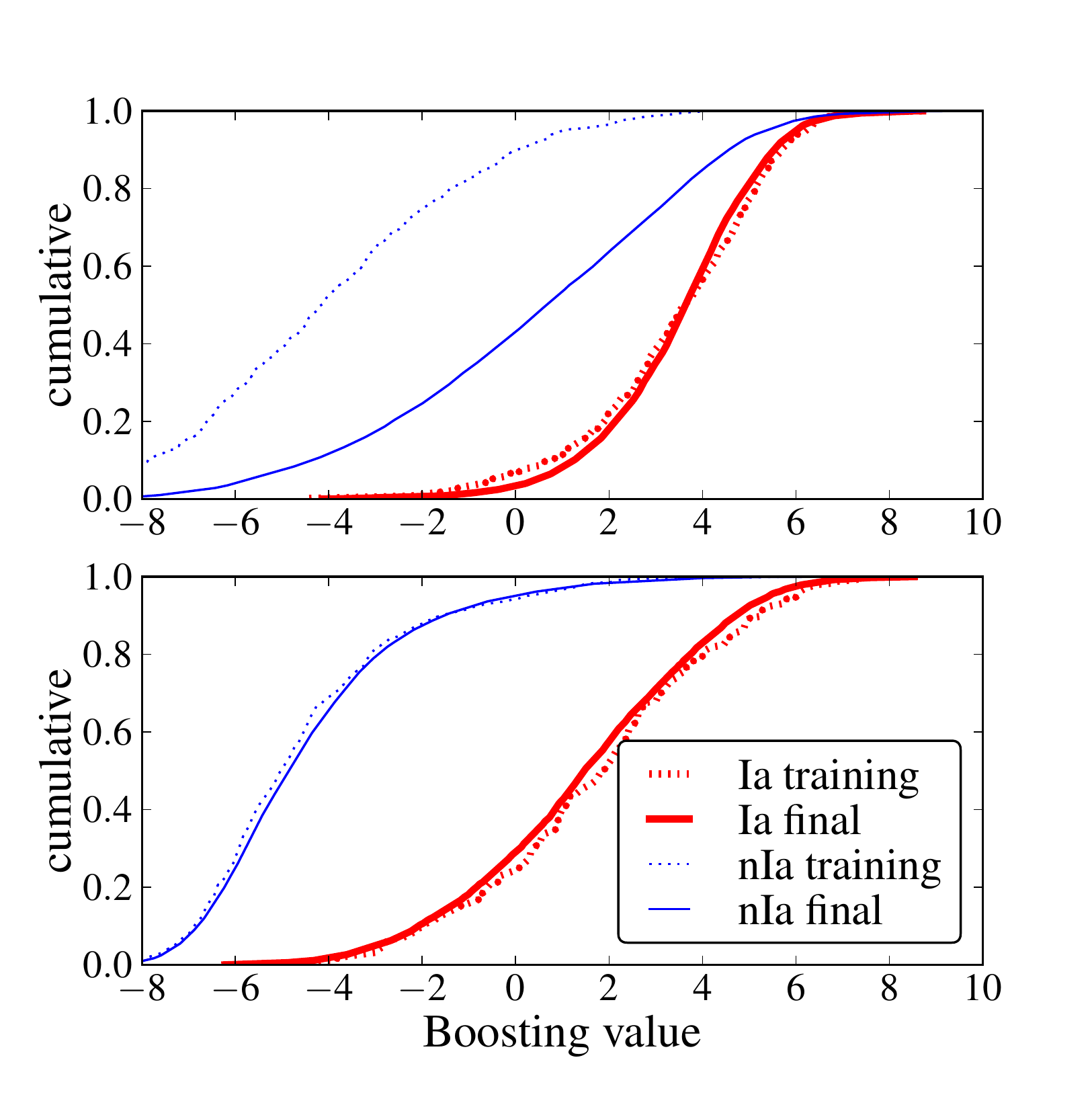}
	\end{center}
\caption{Boosting values obtained using (\textit{Above}) the non-representative training sample and (\textit{Below}) the representative training sample. The boosting values are approximations to $\frac{1}{2}~log~odds.$ } 
\label{boost1_cum}
\end{figure}

Our failure to correctly predict our 
score in the SNPCC was a result of the biases in the training sample. Boosting appears to be even more sensitive to training sample bias than the 21D KDE method. This is illustrated by the large deviation in Figure~\ref{boost1_cum} of the unclassified non-Ia curve from the training non-Ia curve with non-representative training. 

While boosting is more sensitive to bias in the training sample than the 21D KDE, it is a superior classifier when a representative training sample is used. This is illustrated in Figure~\ref{boost1_cum} by the large vertical separation between non-Ia and Ia cumulative curves when a representative sample is used. The vertical separation between the Ia and non-Ia curves is larger in the case of the boosting than the 21D KDE, resulting in a lower contamination rate and higher efficiency when boosting is used.

We see from Figure~\ref{score_FoM_boosting_random} that training with 1000 representative SNe results in a score 3 times greater than training with 1000 non-representative SNe. We also see from Figure~\ref{score_FoM_boosting_random} that training with a non-representative sample of size 1000 can be matched by training with only 50 representative SNe. The score obtained when 500 representative Ia and 500 representative non-Ia SNe are used for training, as opposed to the truly representative case where the Ia:non-Ia training ratio is 1:3, is only slightly higher; the advantage of extra Ia's at the cost of non-Ia's is marginal. 

We did not include the parameter covariance matrices in any way in boosting. It is not clear how this inclusion would best be done, but the noticeable improvement to the 21D KDE score when the covariance is included suggests that it is worthwhile considering this question for future implementations. Two possibilities are a) `supersampling' - converting each training point into 100 training points drawn from a distribution with covariance given by the parameter covariance matrix, and b) including the covariance matrix determinant as a 22nd boosting parameter.

We find that with boosting if a non-representative training sample is used the cumulative frequency lines of the unclassified SNe do not follow those of the training sample. On the other hand if a representative sample is used, tenfold cross-validation provides accurate predictions for the unclassified SNe boosting values, as illustrated by the close hugging of training and unclassified cumulative lines in Figure~\ref{boost1_cum}.

We see that boosting the 21D lightcurve parameters with a representative sample results in a robust photometric classifier. To illustrate this point we have created on online archive of 200 randomly selected unclassified SNe, and labeled them according to boosting's output~\cite{COSMOAIMS}. In some cases it is difficult to identify obvious Ia or non-Ia features, yet the algorithm classifies correctly. 

\subsection{Parameter importance}

One advantage of the boosting algorithm is its ability to quantify the importance of parameters in classification (see Appendix~\ref{app:parameter_i} for details). In this section we look at these quantities in an effort to discover which fitted parameters are most useful for classification. We also ask which are the parameters that distinguish the non-representative training sample from the representative training sample, i.e. what makes the non-representative Ia's and non-Ia's a biased sample. We answer this question by performing boosting on a sample of representative and non-representative Ia's, as if the SNPCC had been a competition to determine if a SN attains a spectrum or not. 

\begin{figure}
	\begin{center}
		\includegraphics[width=3.5in]{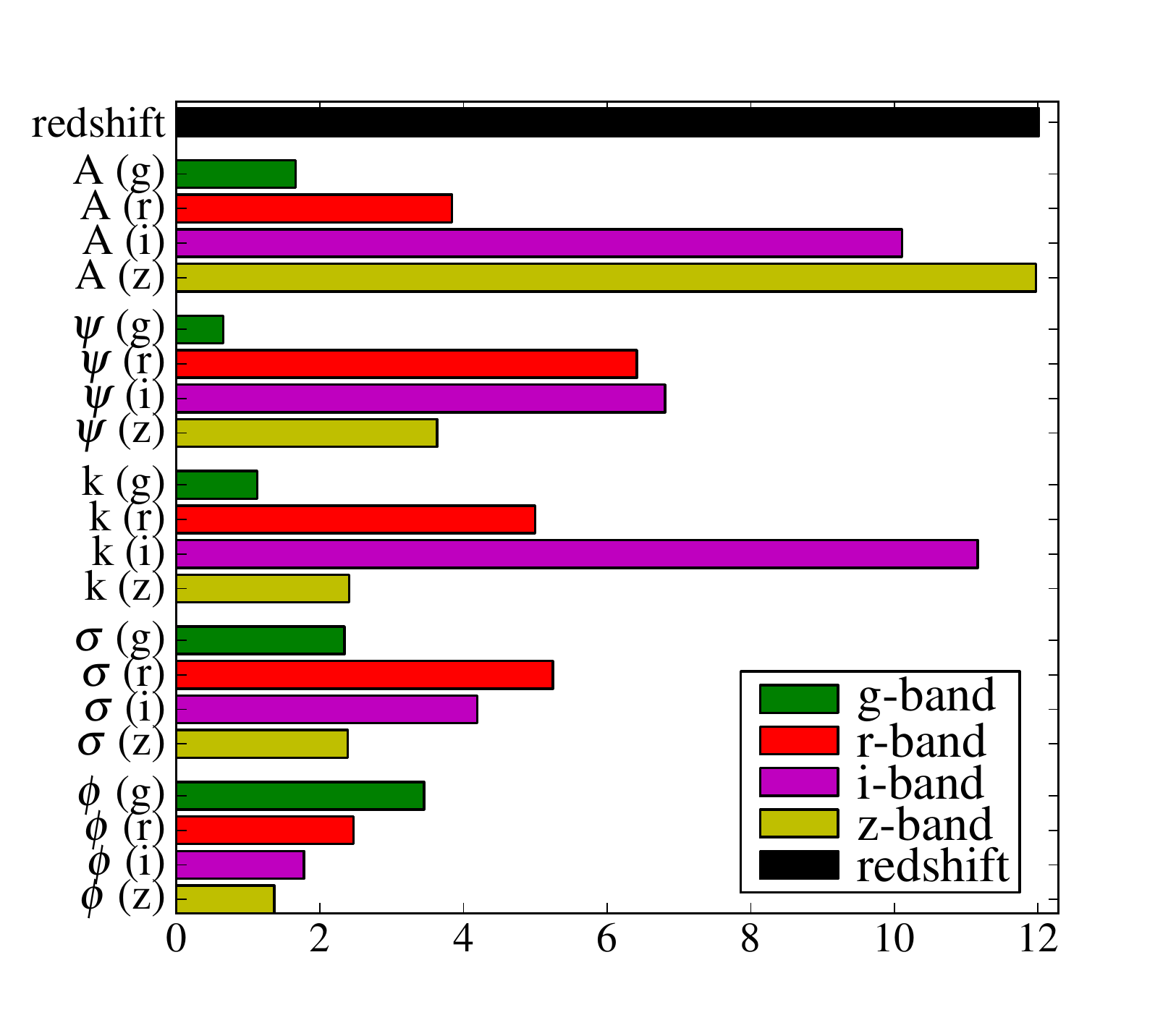}
	\end{center}
\caption{The importance of each of the 21 parameters in classifying SNe as Ia (or not) using boosting on the representative training sample.}
\label{fig:boost_Ia_nIa}
\end{figure}

Figure~\ref{fig:boost_Ia_nIa} illustrates which parameters are most useful in distinguishing Ia from non-Ia in the representative training sample. One interesting feature illustrated in Figure~\ref{fig:boost_Ia_nIa} is that every parameter appears to carry information. 



The third most important parameter (after redshift and and $A$ in $z$-band) is the parameter $k$ in the $i$-band. To interpret this piece of information, we first see in Figure~\ref{fig:k_cumf_tr_nr} that non-Ia SNe have on average lower $k$ values than Ia's. From this we then infer from Figure~\ref{fig:k_effect} that Ia's have a higher rise-time to decay-time ratio than non-Ia SNe. 

The equivalent figure for the non-representative training (Figure~\ref{boosting_imp_tau_nr}  in Appendix~\ref{app:figs}) paints a similar picture with one noticeable difference: The information for distinguishing between Ia and non-Ia SNe in the non-representative training sample is carried almost exclusively in the \textit{r}-band.

\begin{figure}
	\begin{center}
		\includegraphics[width=3.5in]{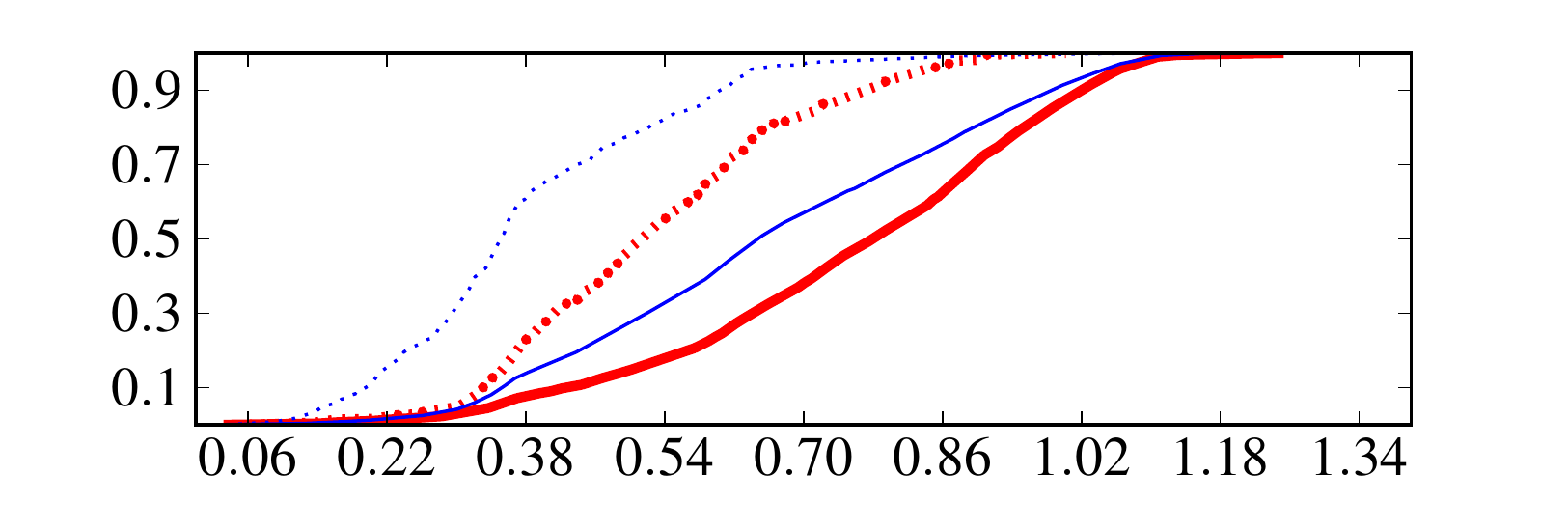}
	\end{center}
\caption{Cumulative plot of redshift, non-representative training (dashed) vs unclassified (solid) and Ia (red, thick) vs non-Ia (blue, thin).}
\label{fig:redshift_cumf_tr_nr}
\end{figure}

\begin{figure}
	\begin{center}
		\includegraphics[width=3.5in]{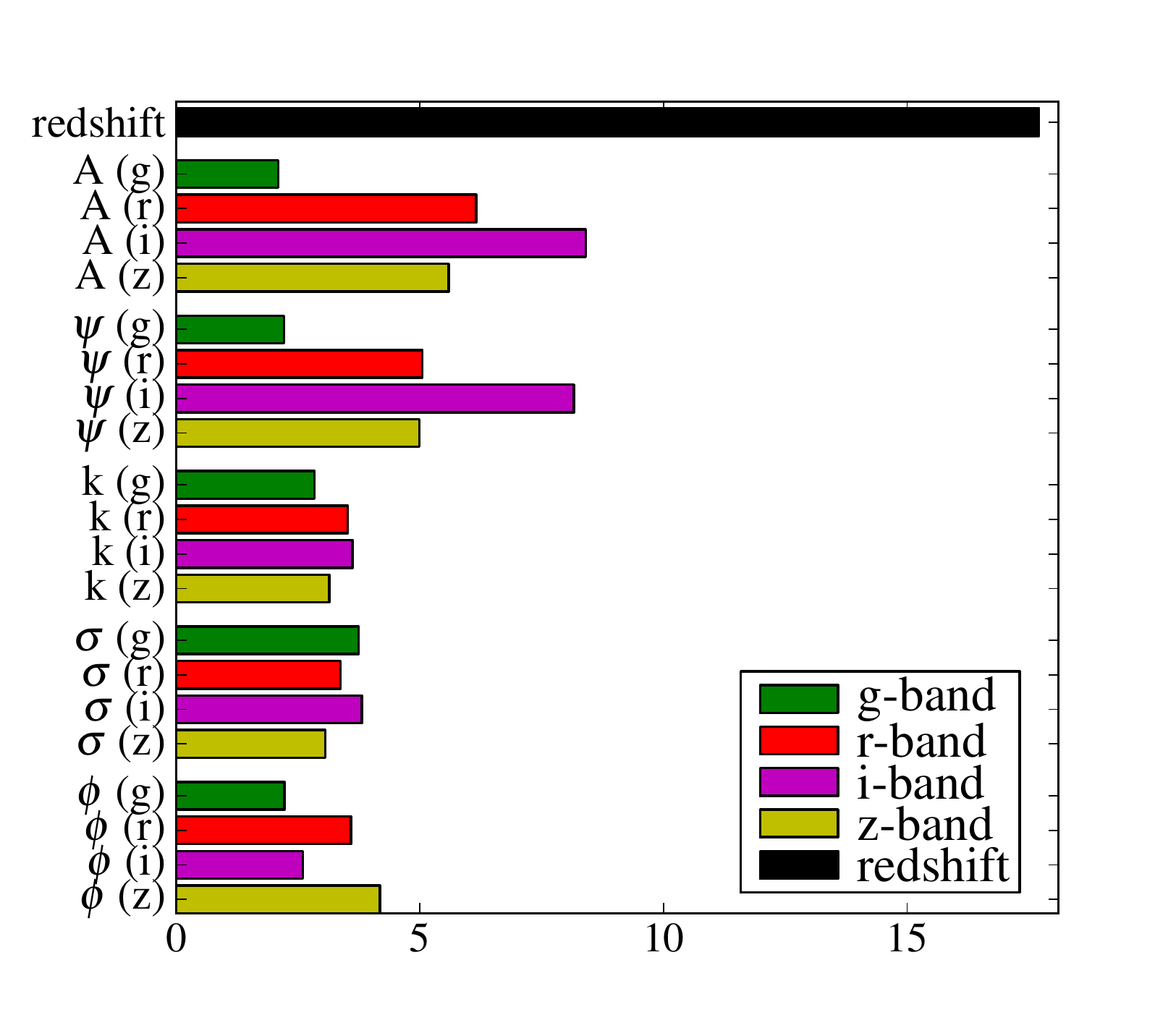}
	\end{center}
\caption{The importance of parameters in distinguishing representative from non-represesntative SNeIa using boosting.}
\label{fig:boost_te_tr_Ia}
\end{figure}

We now turn to the comparison of representative and non-representative SNe. Figure~\ref{fig:boost_te_tr_Ia} suggests that the most biased parameter in the non-representative training sample is redshift. This is not surprising given that we know that the non-representative SNeIa are at lower redshift than the true Ia population (Figure~\ref{fig:redshift_cumf_tr_nr}). Indeed, we see from Figure~\ref{fig:redshift_cumf_tr_nr} 70\% of Ia SNe in the non-representative training set are at a redshift of less than 0.6, while only 20\% of Ias in the unclassified set are within this redshift. 

In the case of non-Ia's SNe (Figure~\ref{fig:boost_te_tr_nIa} in Appendix~\ref{app:figs}) boosting allocates the majority of the bias in the non-representative sample to the $A$'s. This is also unsurprising given that we are more likely to obtain a spectrum from bright objects than dim objects. It is not clear to us why boosting designates non-Ia bias to the $A$'s and Ia bias to redshift. 

\subsection{Hubble KDE}\label{sec:hkde}

\begin{figure}
	\begin{center}
	\includegraphics[width=3.5in]{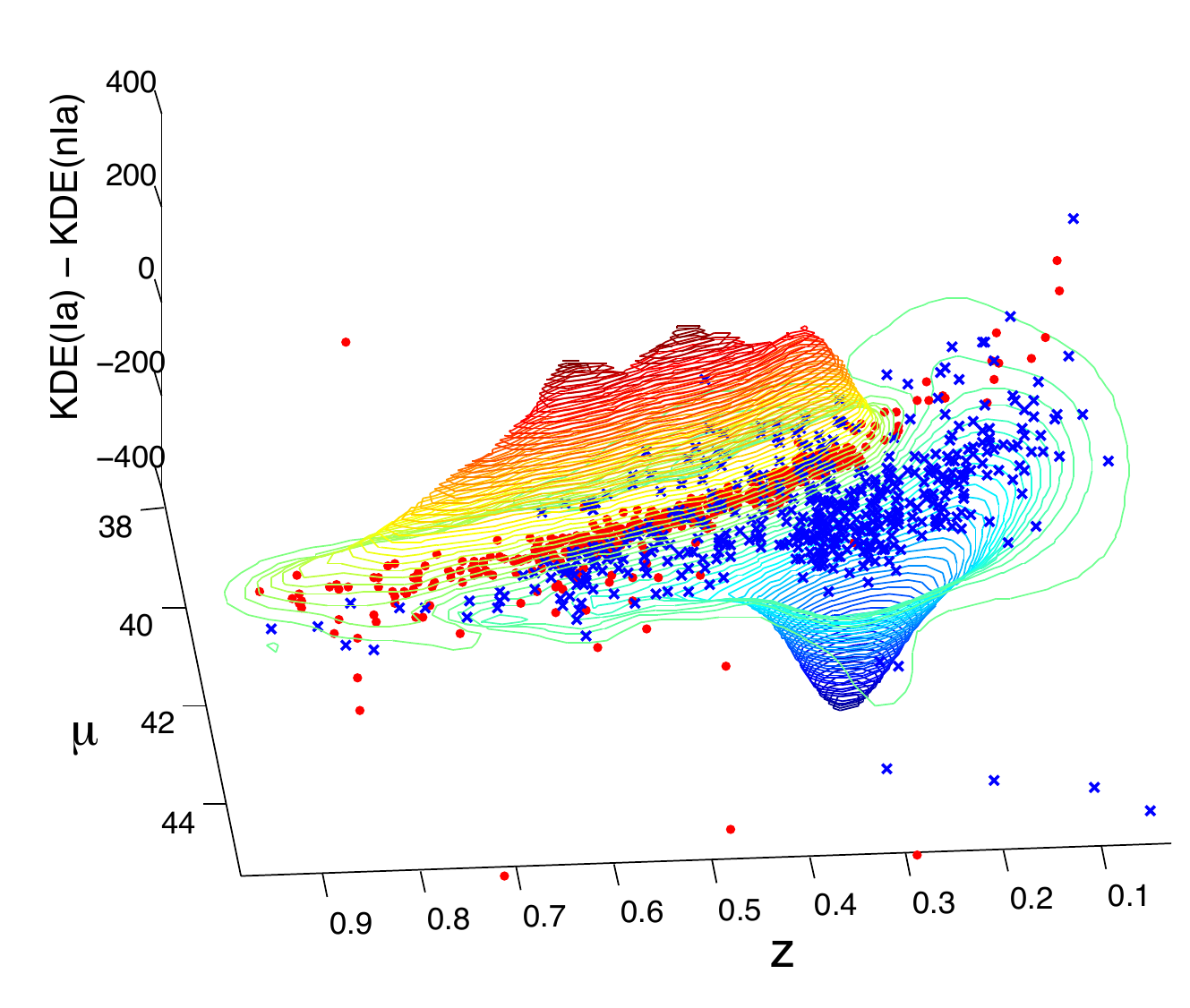}
	\end{center}
\caption{3D contours of the difference between the Ia and non-Ia Hubble diagram KDEs as a function of redshift and distance modulus ($\mu$) together with the actual non-representative training data used to produce the KDEs. The data used to construct the KDEs are also shown: Ia data as red circles and non-Ia data as blue crosses. There is a clear offset in the two KDEs reflecting the fact that in this training data the non-Ia's are fainter, hence predominantly at lower redshift and with a much larger scatter than the Ia's.  }
\label{fig:HubKDE1}
\end{figure}

%

\subsubsection{Applications}
An alternative method for using the idea of KDEs is to use the SALT2 lightcurve fitter (with $\alpha = 0.1$ and $\beta = 2.77$ as in \citet{Hicken}) to estimate distance moduli, $\mu_i$ and errors $\sigma_i$ for all the objects in both the training and test data, assuming that all the data are SNeIa. We can then construct two 2D KDEs for the training data: one consisting of all the known SNeIa and one from all the non-Ia data. Each kernel is normalised to have a total volume of unity and we use a slight modification of the standard KDE formalism because we do not normalise the KDE. Instead the heights of the summed KDEs are proportional to the number of SNeIa and non-Ia respectively. In this way we include prior information related to the supernova rates. A redshift range where there are many more SNIa than non-Ia will automatically tend to lead to a larger Ia KDE as a result. Of course, this does increase sensitivity of the method to biases/non-representativity in the the training sample rates.   

The 2D Gaussian kernel chosen for the Hubble KDE algorithm had a fixed bandwidth (standard deviation) in the redshift direction of $0.05$ (chosen simply to avoid being too peaked but small enough that the distance modulus does not change significantly across it) while the bandwidth in the $\mu$ direction was determined by the error $\sigma_i$, on the distance modulus coming from SALT2, and also includes a 0.12 mag intrinsic dispersion error as usual. This means that points with large errors contribute very broad, low amplitude humps to the final KDE, while points with small errors are much more peaked, reflecting our confidence in that point. For illustrative purposes we plot the difference KDE$_{Ia}$ -KDE$_{non-Ia}$ of the two KDEs in Figure~\ref{fig:HubKDE1}. Positive values correspond to places where the Ia KDE dominates, negative values to where the non-Ia KDE dominates. In addition we plot the training data used to construct the KDEs.

Classification using these KDEs is then simple. For any candidate object, we run it through SALT2 to give an estimated $\mu$ and $\sigma$. We can only use this approach on the data with a redshift estimate, $z$, unlike the 21D KDE and boosting algorithms which do not require a redshift. We then simply find the values of the two KDEs at that $(\mu, z)$ to yield probabilities of the object being a Ia or non-Ia. As in the other KDE method, one should fold in the error $\sigma$ on the candidates which, assuming Gaussianity, is simple, as described in Section~\ref{sec:integrating_errors}. The result of this analysis is that each candidate has a pair of probabilities: $(P_{Ia}, P_{non-Ia})$ that can be used to classify the candidate.   

\subsubsection{Non-representative training sample}

We applied this methodology to the whole sample of unknown SNe supplied. In total, we started with 17065 SNe and lost 4578 SNe as junk because of SALT2 failures previously mentioned (of which 2619 were complete failures and 1959 failed to return meaningful parameters from the Ia lightcurve fit), leaving 12487 SNe for further analysis. 

Essentially this Hubble KDE approach simply checks whether or not an object lies close to the true cosmology curve on the Hubble diagram (defined by the Ia KDE) at that redshift. However, there are many non-Ia's which lie close to the true cosmology curve. As a result one either has to be very strict with cuts (and therefore lose many true Ia's) or one has to accept a large number of false positives: non-Ia's that are classified as Ia's. 

Because there are so many non-Ia's this, and similar Hubble-diagram based methods (such as the Portsmouth entry to the SNPCC), are less competitive as classifiers. In addition they also require a redshift estimate for the SNe and are hence doubly inferior compared with the 21D KDE and boosting. 

\subsection{Combining 21D and Hubble KDEs}\label{sec:combining}

In Section~\ref{res:21-D KDE} we described the 21D KDE approach, and in Section~\ref{sec:hkde} we described the Hubble KDE approach. In this section, we describe how we combined these approaches. As outlined in Appendix~\ref{app:bayes}, there are several ways of combining \textit{odds} from different algorithms to construct a better combined classifier. For our combination entries in the SNPCC, we constrained our classifier to be of the form:
\begin{equation}\label{eq:block}
\mathrm{(Hubble~odds)}^{\alpha} \cdot \mathrm{(21D~odds)}^{\beta} > \eta.
\end{equation}

\begin{figure}
	\begin{center}
		\includegraphics[width=3.35in]{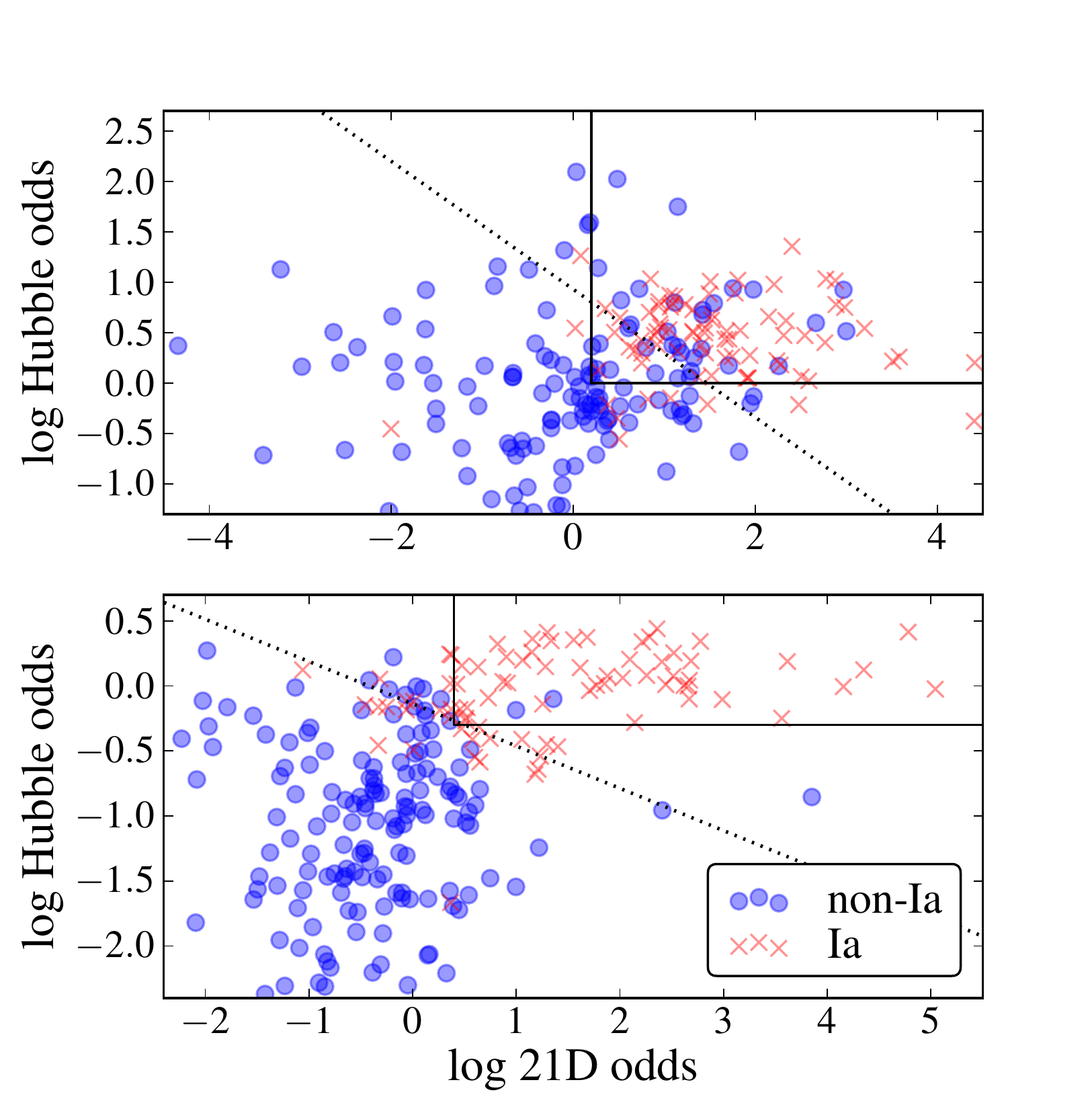}
	\end{center}
\caption{SNe of type Ia (cross) and non-Ia (circle), located according to their 21D \textit{odds} (x-axis) and Hubble \textit{odds} (y-axis). (\textit{Above}) \textit{Odds} were calculated from KDEs constructed using the non-representative training sample. (\textit{Below}) A corresponding plot where KDEs were constructed with the representative training sample. We see that the separation obtained is smaller when non-representative training is used, and indeed the score obtained in the non-representative case is significantly lower. Note that the SNe in this figure are a random sample of the $\sim$ 12500 with a meaningful SALT2 fit.}
\label{Bayes_Bayes_line}
\end{figure}

This corresponds to a straight line in Figure~\ref{Bayes_Bayes_line}. The scores for the combination entry was $0.28$. Surprisingly, this was less than the score obtained using the 21D KDE alone, and so we believe that the line chosen for the SNPCC was poor. A straight line does seem to be a good choice for the distribution of values in Figure~\ref{Bayes_Bayes_line}, but perhaps a better choice would be of the form: 
\begin{equation}\label{eq:line}
\mathrm{Hubble~odds} > \gamma_1 \hspace{2mm}  \mathrm{and} \hspace{2mm} \mathrm{21D~odds} >\gamma_2.
\end{equation}

A pure 21D \textit{odds} classifier would rely on a vertical decision line, and a pure Hubble \textit{odds} classifier would rely on a horizontal line, but it is clear from Figure~\ref{Bayes_Bayes_line} that a classifier of the form~\ref{eq:block} (dashed) or~\ref{eq:line} (solid) should work better. Figure~\ref{Bayes_Bayes_line} shows the separation of Ia's and non-Ia's that come from using the Hubble KDE \textit{odds} and 21D KDE \textit{odds} with the integration of errors presented in Section~\ref{sec:integrating_errors}. The optimal lines of forms~(\ref{eq:block}) and~(\ref{eq:line}) result in scores of 0.24 and 0.22 respectively in the case of non-representative training and 0.45 and 0.42 respectively in the case of representative training. These scores are calculated using a purely 21D \textit{odds} classification for the $\sim 8000$ SNe without SALT2 fits, and a 21D-Hubble combination for the remaining $\sim 12500$ SNe with SALT2 fits. As with boosting, the 21D KDE classifier is significantly worse using the post-SNPCC data as previously discussed in section~\ref{sec:boostrnr}, and so comparison between these post-SNPCC scores and other SNPCC scores should not be made until further analyses have been done.

To be in the top-right corner of Figure~\ref{Bayes_Bayes_line}, and therefore be classified as Ia, requires that a candidate must simultaneously lie close to the true cosmology distance modulus and have multiband lightcurves that have the right shape; a very natural approach to SNIa classification. It would be interesting to combine the Hubble \textit{odds} with 21D boosting instead of 21D KDE, as boosting the twenty parameters produces better results, as seen in Section~\ref{sec:boosting}.

An obvious extension, if one wanted to combine the outputs from more than two classifiers, would be to use them as inputs to a new boosting analysis. The \textit{odds} from the 21D KDE, the Hubble KDE, the 21D boosting, and indeed any classifier of sufficient ability can be used as weak classifiers in boosting. A reason to exercise caution in using boosting or a neural network as a final classifier in this way is the possibility of overtraining, but this can be prevented by using tenfold cross-validation.

\section{Discussion and Conclusions}

In this paper we have discussed the problem of classifying supernovae (SNe) into sub-classes (Type Ia or non-Ia) based on photometric lightcurve data alone, that is, multi-band fluxes as a function of time. This will be necessary for future surveys which will detect vastly more candidates than will be possible to follow up spectroscopically. 

We have investigated two novel classes of classification algorithms, Kernel Density Estimation (KDE) and boosting, and applied them to simulated SNe lightcurve data, finding that the methods performed impressively as long as they were trained on a representative sample. Using the KDE approach, we considered both a 21 dimensional case based on lightcurve parameters from all bands and a 2 dimensional version based on fits to the Hubble diagram, using redshift information and an estimate of the distance modulus obtained using standard lightcurve fitting software.

A key issue for the classification methods we used was the issue of the training data sets. We compared the results based on training on two very different data sets: the first, a non-representative set, mimicking the kind of spectroscopic sample available as part of the follow-up program of a typical current-generation SN survey. The second was a representative sample of the same size where training objects were selected at random from the full sample. 

In general we found that the training on the representative sample produced exceptionally good results and that cross-validation on the training sample was able to accurately predict the purity and efficiency of the method on the full sample.  On the other hand, training on the non-representative sample lead to relatively poor performance on the full data set. The importance of having an unbiased, representative sample is illustrated by the fact that for boosting, 
representative samples larger than about 50 objects outperformed the full non-representative sample of 1000 objects, as shown in Figure~\ref{score_FoM_boosting_random}. 

Our primary result and recommendation therefore is that boosting and KDE are powerful methods for SN classification, with remarkably little astrophysical input. However, they require training samples that are as unbiased and representative as possible. Further, we found that a small unbiased  training sample outperforms a much larger, but biased, training sample. 

Our other main result is that neither boosting nor the 21D KDE method suffered particularly when the SN redshift information was unavailable. This is particularly gratifying given that accurate SN/host galaxy redshifts will not be available for most candidates in the future and that methods based on the Hubble diagram critically require redshift information to perform successfully.  

While the algorithms we have presented were successful, there are modifications to our boosting implementation that should be experimented with, for example different choices of lightcurve parameterisation. Further it would be very useful to investigate methods to reduce sensitivity to biases in the training data.
 
Finally it is perhaps useful to comment on how our methods compare to the winner of the SNPCC (the methods we described in this paper finished second and third in the competition) which used a template-based method and performed very well.  Our first comment is that comparison is hard because there was an overlap between the templates used by Sako and those used to generate the SNPCC, as described in \cite{sntc_results}, so it is not clear how the method would perform on completely independent data. Secondly, it is not clear how the various methods would perform with different Figures of Merit, an important issue which we do not discuss here. It is clear that finding the best approach to supernova classification, and the best way to combine results from different classifiers, will be an active area of research in the coming decade.  

\begin{figure}
	\begin{center}
	    \includegraphics[width=3.4in]{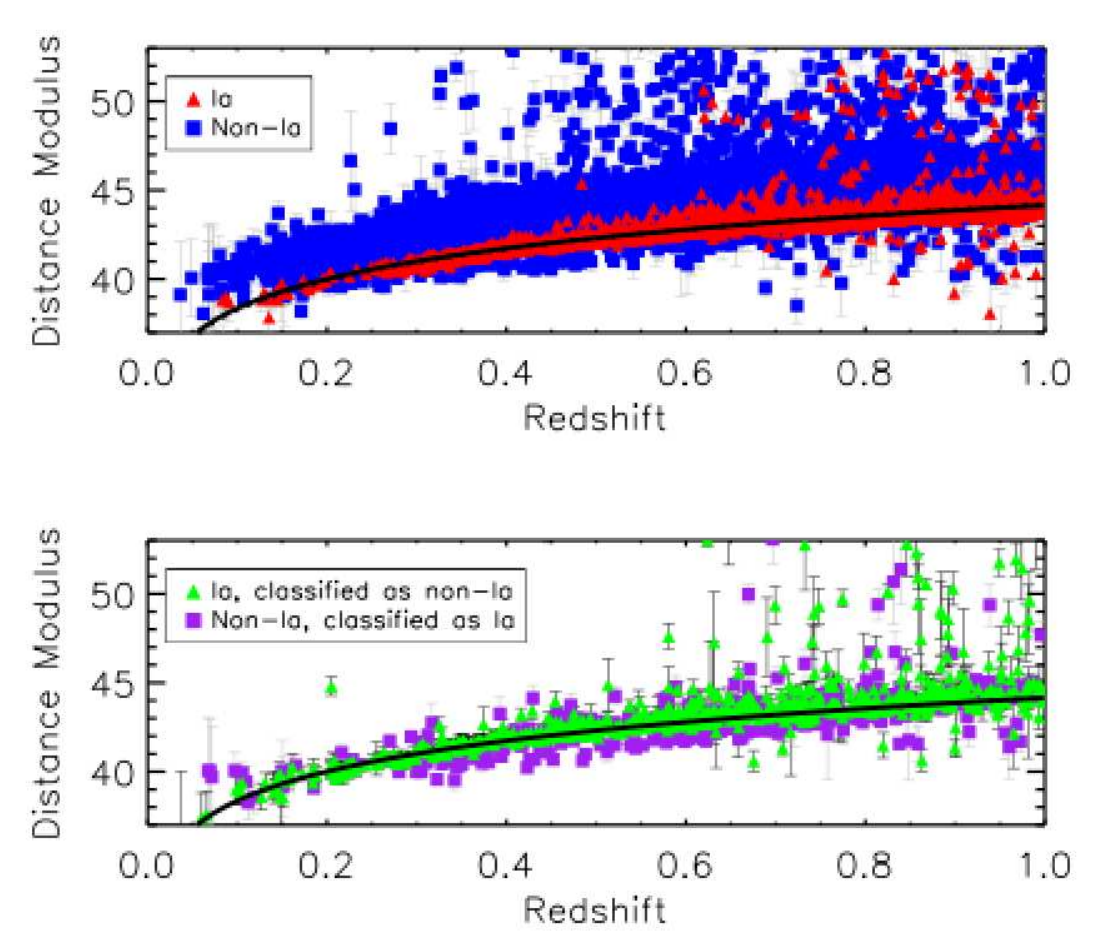}
	\end{center}
\caption{Hubble diagrams for the boosting results using the representative training sample. (\textit{Above}) Objects that were correctly identified by the boosting method. SNeIa are plotted as red triangles, with non-Ia SNe shown as blue squares. (\textit{Below}) SNe that were incorrectly typed by boosting. SNeIa that were considered to be non-Ia SNe by boosting are shown as green triangles, with incorrectly typed non-Ia SNe shown as purple squares. Overplotted on each graph is the best fitting cosmological model inferred from the representative training sample.}
\end{figure}
%
\begin{figure}
	\begin{center}
		\includegraphics[width=3.5in]{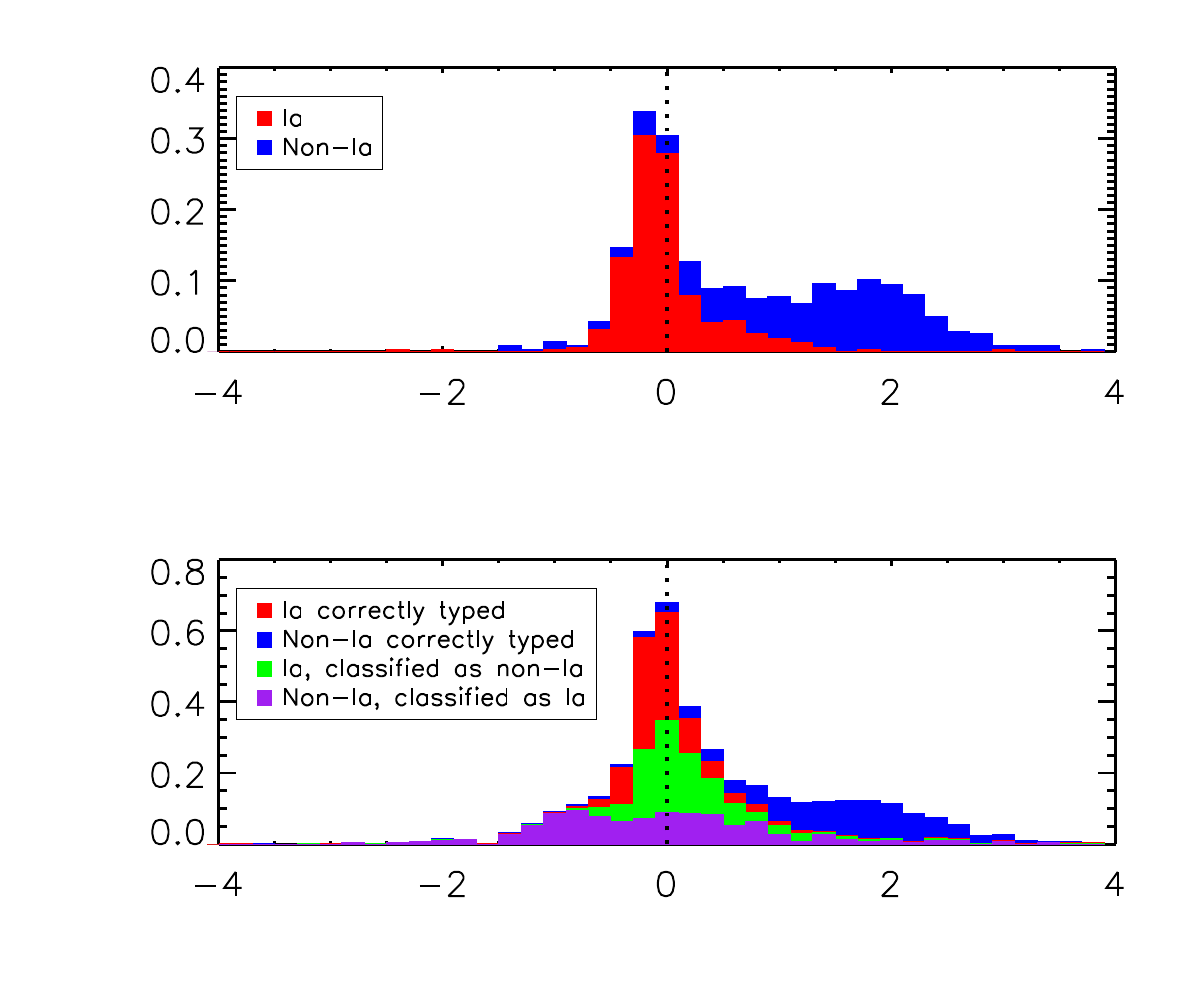}
	\end{center}
\caption{Cumulative histograms of the residuals from the best-fit Hubble diagram, determined using the SNeIa in the representative training sample. (\textit{Above}) Residuals for the representative training sample. SNeIa are plotted in blue, with non-Ia SNe shown in red. (\textit{Below}) Residuals for the boosting results. SNeIa that were correctly typed are shown in blue, with correctly typed non-Ia SNe shown in blue. SNeIa that were considered to be non-Ia SNe by boosting are shown in green, with incorrectly typed non-Ia SNe shown in purple.}
\label{geen4}
\end{figure}


\section{Acknowledgements}
We thank Trevor Hastie for suggesting the use of boosting and comments on the draft and Matt Hilton and Prina Patel for initial work on the project. We also thank Rick Kessler for comments on the draft. This work 
was begun during the JEDI4 workshop in Cape Town and is supported by the NRF and 
a Royal Society-NRF bilateral exchange grant. JN has a SKA bursary and MS is funded by a SKA fellowship. BB acknowledges funding from the NRF and Royal Society. MK acknowledges financial support by the Swiss NSF. DP was supported by the Science and Technology Facilities Council [grant number ST/F002858/1]. RH acknowledges funding from the Rhodes Trust.
\appendix

\section{cross-validation}\label{app:cross-val}
Cross-validation is a statistical technique that enables one to tune model parameters so as to optimize model prediction. Within the context of the 21D KDE, both the kernel bandwidth $h$, the number of nearest neighbours $k$ and the odds threshold may be optimized for some figure of Merit (FoM) by tenfold cross-validation. This entails partitioning the training set into 10 roughly equal parts. One may then use nine-tenths of the data to estimate the Ia and non-Ia probability densities and then use these probability densities to classify the remaining one-tenth of the training set. This may be repeated ten times, predicting the class for each of the ten partitions of the data using the KDEs estimated from the remaining nine partitions. Since we know the supernova types of the training set, we can then find a combination of the aforementioned three parameters that maximizes the FoM. Cross-validation can be used in a similar way for boosting. Figure~\ref{fig:num_trees} uses cross-validation to determine that 4000 trees will be near optimal.

\section{Probabilistic interpretation and combination of probabilities}\label{app:bayes}
By evaluating each KDE, we may obtain the probability of observing a lightcurve (with the lightcurve
data denoted as $x$) conditioned on the supernova being a Ia or
not, i.e. we get $p_1 = p(x|\textrm{Ia})$ and
$p_2=p(x|$non-Ia$)$. The ratio of $p_1$ to $p_2$ is known as the \textit{Bayes factor}, $B_{12}$. What interests us, however, is the relative probability of the observation $x$ being from a Ia supernova versus
another type. That relative probability, called the \textit{Odds ratio},  $odds(x)$ is

\begin{eqnarray}
P(\mathrm{Ia}|x) &=& p_1 \frac{P(\mathrm{Ia})}{P(x)} \\
P(\textrm{non-Ia}|x) &=& p_2 \frac{P(\textrm{non-Ia})}{P(x)} \\
odds(x) &=& \frac{P(\mathrm{Ia}|x)}{P(\textrm{non-Ia}|x)} = \frac{p_1 P(\mathrm{Ia})}{p_2 P(\textrm{non-Ia})}\\ \nonumber
&=&B_{12}\frac{P(\mathrm{Ia})}{P(\textrm{non-Ia})}.
\end{eqnarray}

\noindent The probabilities $P(\mathrm{Ia})$ and $P(\textrm{non-Ia})$ are the prior probabilities to observe a Ia supernova or one of another type respectively.

To convert the relative probability back into absolute probabilities we
need use the fact that there are only two possibilities (Ia or not), so
that $P_2 = (1-P_1)$. In this case we have that 

\begin{equation}
P_1 = odds(x)/(1+odds(x)).
\label{B_to_P}
\end{equation}

If we have two independent observations $x$ and $y$ then we can
update the relative probability $odds(x)$ from observation $x$:
\begin{equation}
odds(x,y) = odds(x) \frac{p(y|\mathrm{Ia})}{p(y|\textrm{non-Ia})} .
\label{combining_Bs}
\end{equation}
We can use this to combine for example the probability from the
21 dimensional KDEs with information from the Hubble diagram, but we have
to be careful if the 21D KDEs already contain some of the Hubble
information implicitly, e.g. through the evolution of the overall
amplitudes of the lightcurves as a function of redshift.

It is possible that the KDEs should not be interpreted as probabilities. This may be due to oversmoothing of too wide kernels, or shot noise from too narrow kernels. With a sufficiently large training set one can test how accurately the KDEs represent probabilities - the proportion of SNeIa in a (calculated) \textit{odds} bin should equal that predicted by Eq.~\ref{B_to_P}. If it is not, one can consider making a mapping from the calculated \textit{odds} to the true \textit{odds}.
    
If in combining probabilities one does not want to assume independence, or does not trust the probabilities and doesn't want to make a mapping to true probabilities, there are several alternatives to Eq.~\ref{combining_Bs}. Some of these include capping unreliable \textit{odds} at 1, using linear combinations of \textit{odds} instead of products, using p-values instead of probabilities and down-weighting particularly small/large Bayes' factors. Often an optimal method can be decided on by considering a scatter plot (like Figure~\ref{Bayes_Bayes_line}) of the training set. In Section~\ref{sec:combining} we considered two new ways of combining \textit{odds}, equations~\ref{eq:block} and~\ref{eq:line}.

\section{Best trees}\label{app:boosting}
\begin{figure}
	\begin{center}
		\includegraphics[width=3.5in]{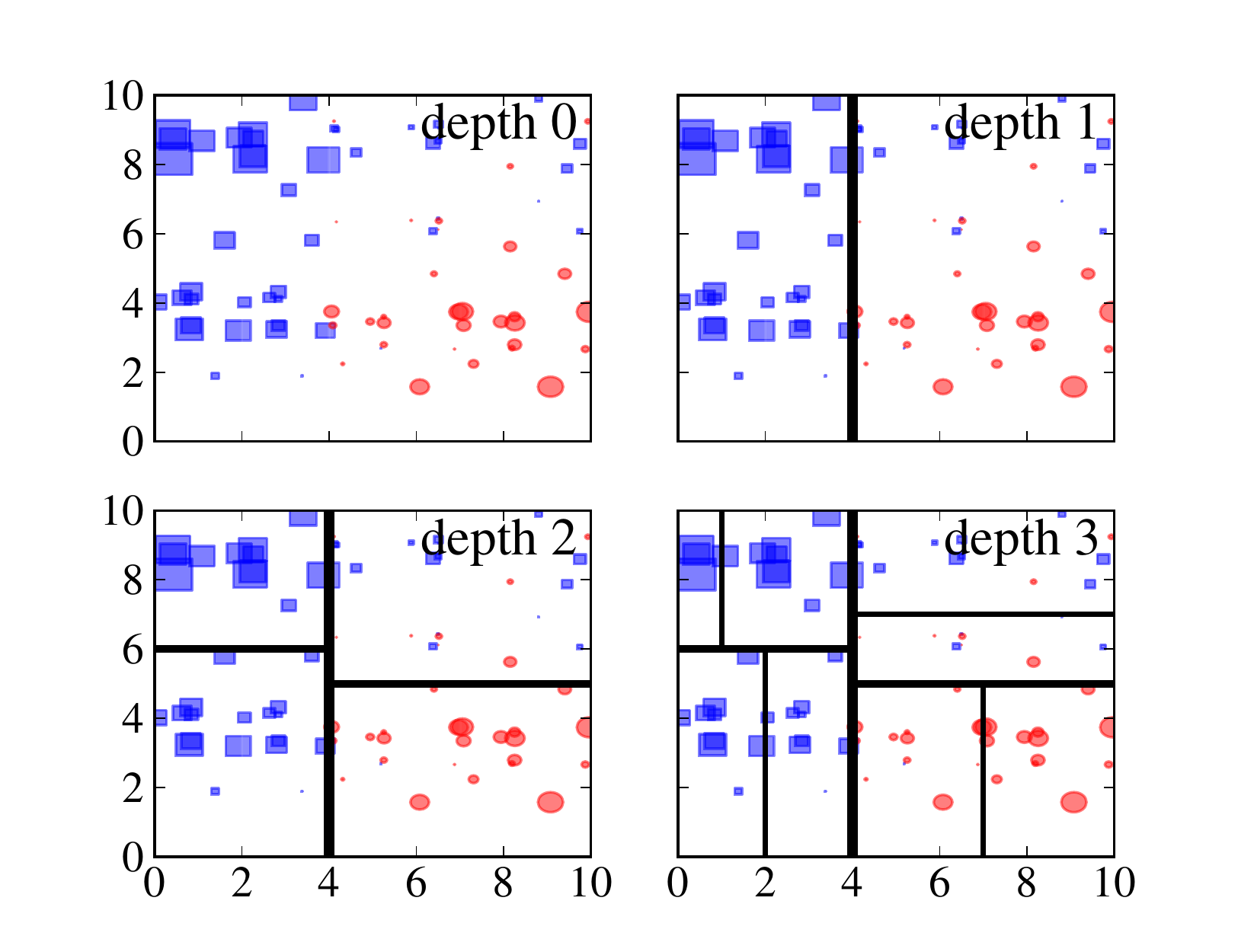}
	\end{center}
\caption{(\textit{Above left}) At several $\vec{X}_i \in R^2$ we have a value $z_i \in R$, represented by a rectangle if negative and a circle if positive, with the size of the shape being proportional to the magnitude of $z_i$. We want to split the set of observations by $X^{(1)}$ or $X^{(2)}$ to minimise the average in-group variance. (\textit{Above right}) After considering all vertical and horizontal lines, we settle on this vertical line as our first ``branching" as it minimizes in-group variance. (\textit{Below left}) Sub-branches are chosen to minimize in-in-group variance. (\textit{Below right}) A tree of depth 3.}

\label{fig:treeg}
\end{figure}

Suppose we have some data $\vec{X}_i \in R^2$, $z_i \in R$, and we would like to fit $z_i \sim \vec{X}$ using a tree. To be precise, we would like to find a tree which minimises $\sum_{i = 1}^n (T(\vec{X}_i) - z_i)^2$, where $T(\vec{X}_i) = v_k$ when $\vec{X}_i$ falls into node $k$ of the tree. We therefore need to find two things, the tree shape and the ``leaf" values (the $v_k$s). Figure~\ref{fig:treeg} illustrates the idea of ``greedy" tree construction. Note that this may not be \textit{the} best depth 3 tree. The greedy approach ignores several potential trees. However it is quick and easy, and for boosting where thousands of trees are made it is not necessary to have exactly minimising trees at each step. See also our animation of tree construction on the arxiv at \cite{COSMOAIMS}.

\section{Calculating parameter importance}\label{app:parameter_i}
To measure how much information each parameter carries in the boosting classifier, we can do the following. For each branching within each tree constructed from the training data, calculate how much total ingroup variance was reduced by this branching. Then for each parameter, for all the branchings which it defines add up the ingroup variance reductions. This value is a good indicator of a parameter's importance in classificaction.

\section{GBM in full}\label{app:gbm}
In this appendix we complete the GBM algorithm presented in Section~\ref{sec:gbm}. There was no mention in Section~\ref{sec:gbm} of the learning rate $\nu$, or the bagging fraction $\phi$. The learning rate $\nu \in \left[ 0,1 \right]$ should appear in step 4 of the main loop. Originally given as,
\begin{equation*}
F_{k} \leftarrow F_{k-1} + T_k
\end{equation*}
\noindent step 4 should appear as, 
\begin{equation*}
F_{k} \leftarrow F_{k-1} + \nu T_k
\end{equation*}

The learning rate should be set quite low, we used 0.05. It acts to reduce the sensitivity of $F$ to the initial tree choice. 

The use of bagging has been shown to improve the efficiency of the GBM algorithm and the accuracy of the final classifier \citep{Friedman2002367}. The idea of bagging  is that instead of all the training data being used for every tree construction, a fraction ($\phi$) is randomly chosen to fit the tree at each step. For the SNPCC we used $\phi = 0.5$. To include bagging, the inner \texttt{for} loop should be modified to read,
\begin{equation*}
\mathrm{for } \;i\; \mathrm{ in}\; \{ \mathrm{sample~of~size~} \; \phi\cdot N \; \mathrm{from~integers~1~to~N} \}
\end{equation*}

The last modification that needs to be made to complete the GBM algorithm is at step 3 of the main loop. Full line searches for optimal $\gamma_{k, j}$'s are not used, instead to speed up the algorithm only the initial step of Newton's method is used: 
\begin{equation}
\gamma_{k,j} = \frac{\displaystyle\sum_{\vec{X}_i\in R_{k,j}}z_i }{\displaystyle\sum_{\vec{X}_i\in R_{k,j}}\vert z_i\vert\left( 2 - \vert z_i\vert \right)}
\end{equation}  

\begin{figure}
	\begin{center}
		\includegraphics[width=3.5in]{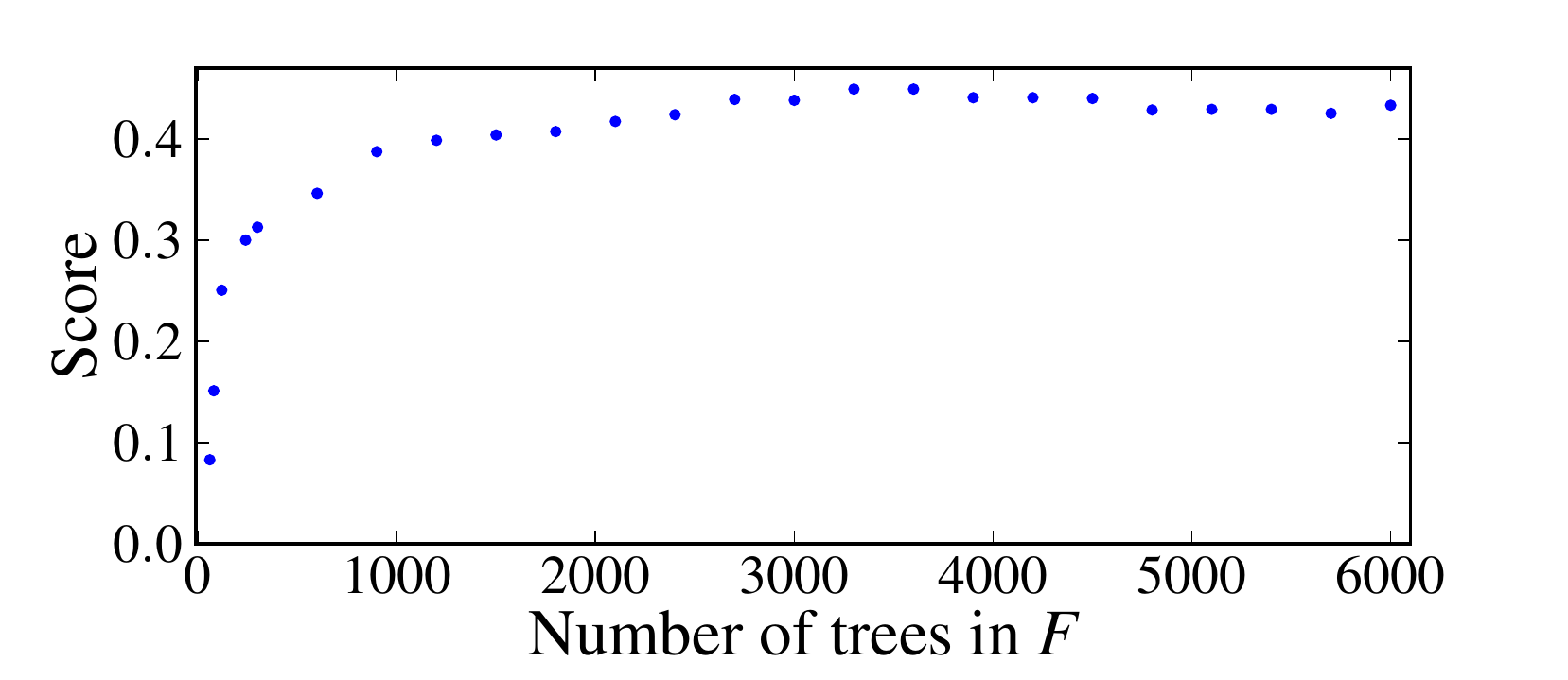}
	\end{center}
\caption{The score, predicted using tenfold validation, on the representative training sample (1.0). Seems to be no overfitting as more trees are added. We go with number of trees = 4000.} 
\label{fig:num_trees}
\end{figure}

\section{Parameter distributions}\label{sec:pdists}

\begin{figure}
	\begin{center}
		\includegraphics[width=3.5in]{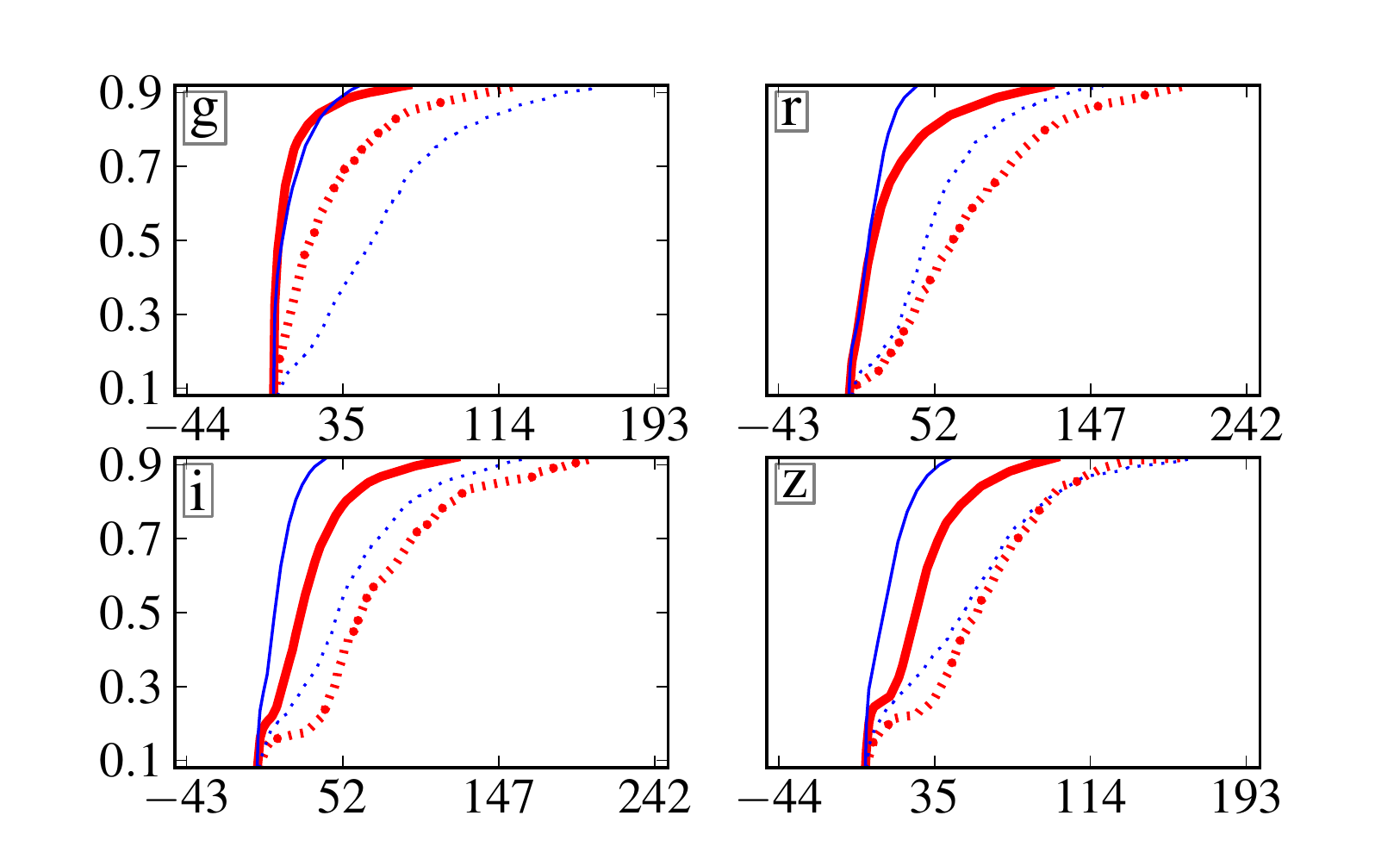}
	\end{center}
\caption{Cumulative plots of parameter $A$ in bands \textit{g,r,i,z}. Non-representative training (dashed) vs unclassified (solid) and Ia (red, thick) vs non-Ia (blue, thin). In all bands, the magnitude $A$ of SNe is far larger in the non-representative training set than in the unclassified set.}
\label{fig:A_cumf_tr_nr}
\end{figure}

\begin{figure}
	\begin{center}
		\includegraphics[width=3.5in]{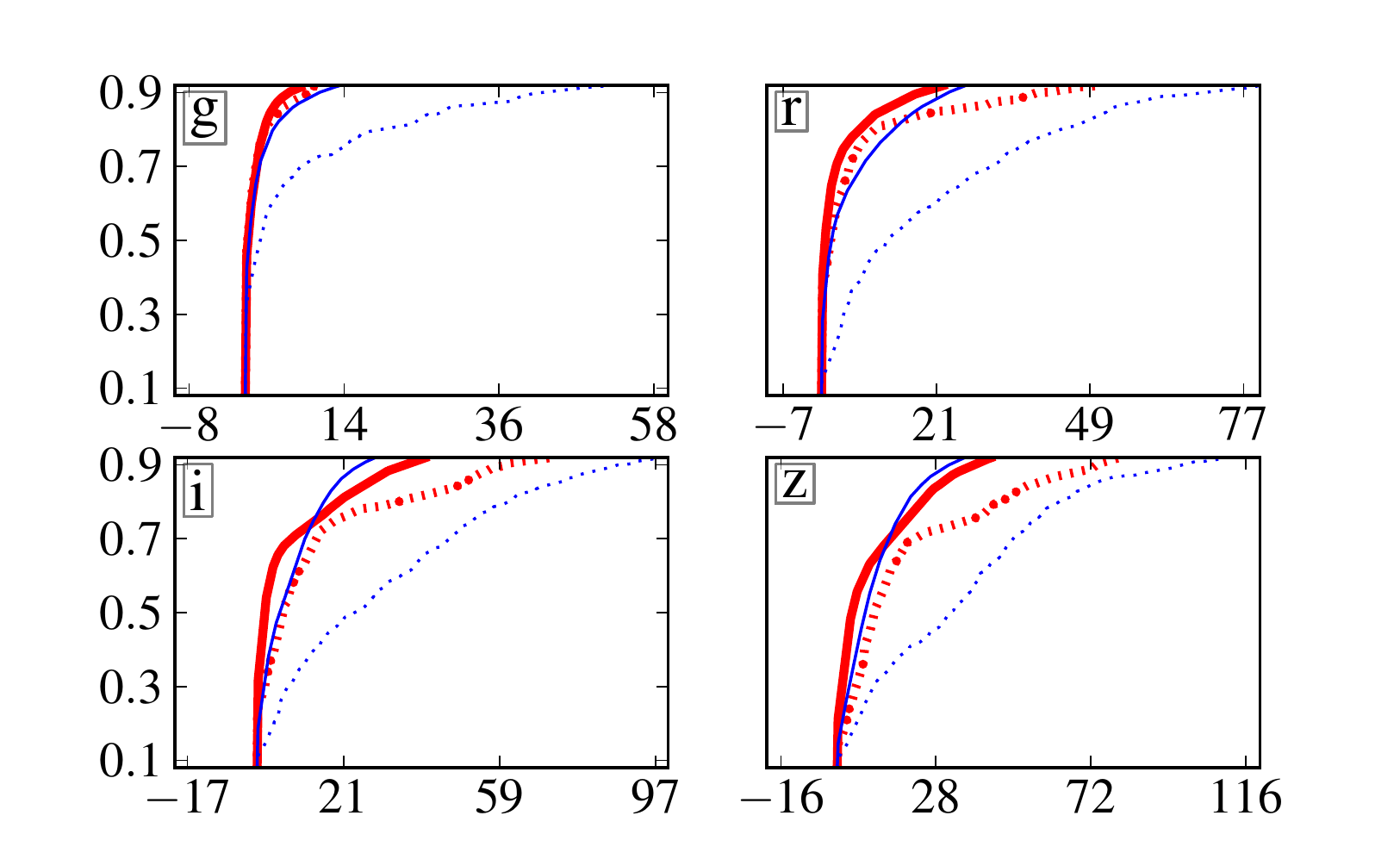}
	\end{center}
\caption{Cumulative plot of parameter tail in bands \textit{g,r,i,z}. Non-representative training (dashed) vs unclassified (solid) and Ia (red, thick) vs non-Ia (blue, thin).}
\label{fig:tail_cumf_tr_nr}
\end{figure}

\begin{figure}
	\begin{center}
		\includegraphics[width=3.5in]{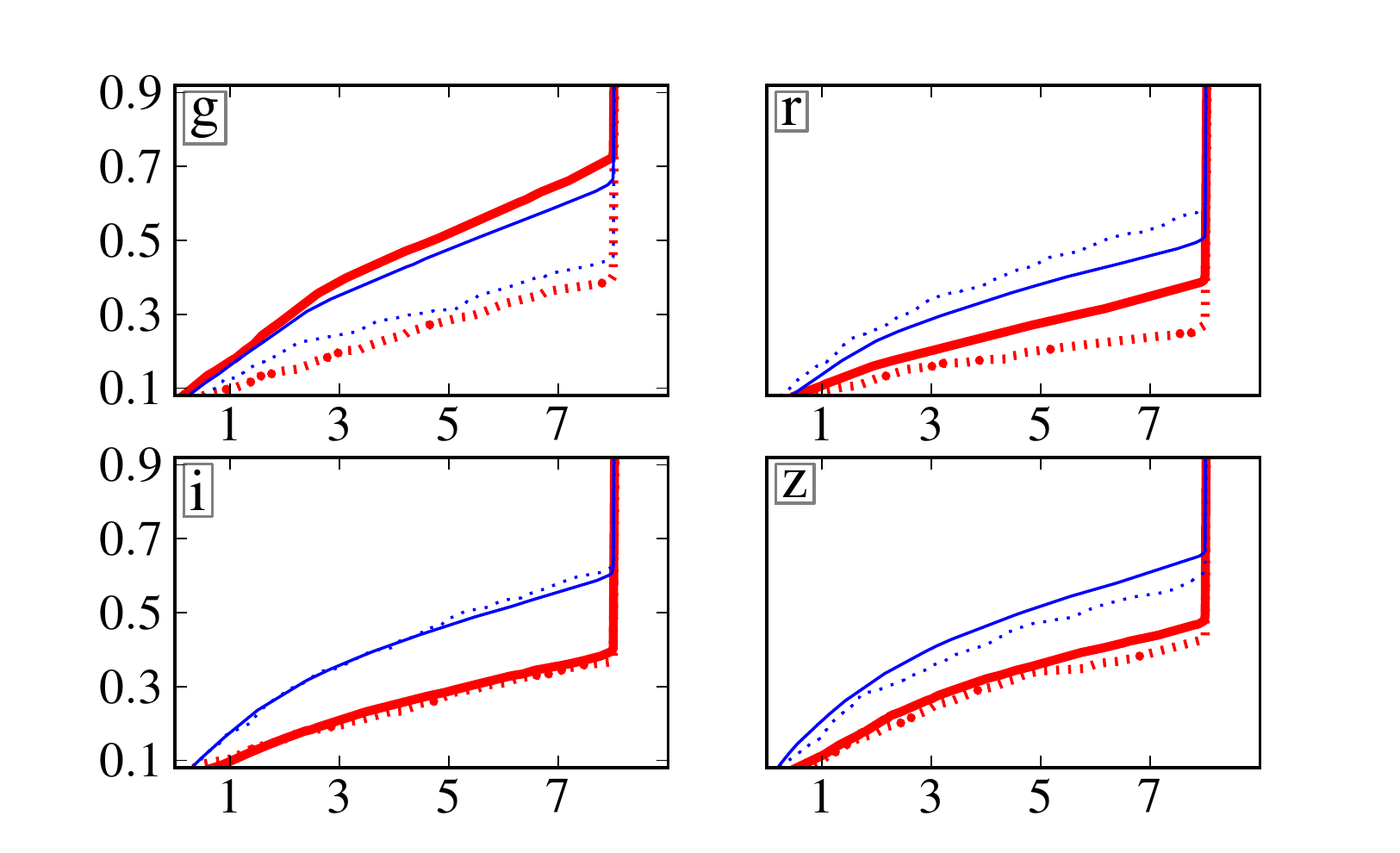}
	\end{center}
\caption{Cumulative plot of parameter $k$ in bands \textit{g,r,i,z}. Non-representative training (dashed) vs unclassified (solid) and Ia (red, thick) vs non-Ia (blue, thin).}
\label{fig:k_cumf_tr_nr}
\end{figure}

\begin{figure}
	\begin{center}
		\includegraphics[width=3.5in]{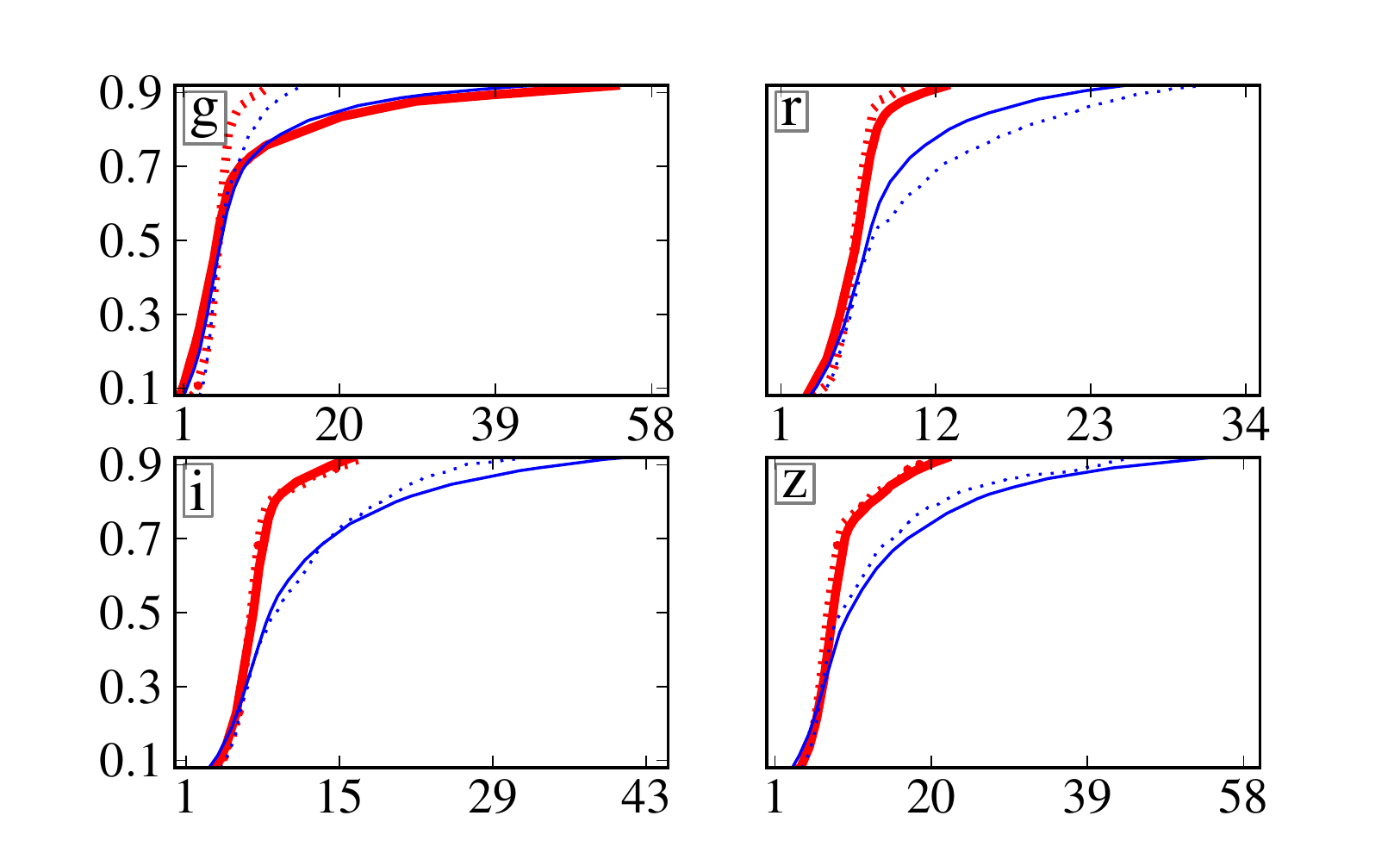}
	\end{center}
\caption{Cumulative plot of parameter $\sigma$ in bands \textit{g,r,i,z}. Non-representative training (dashed) vs unclassified (solid) and Ia (red, thick) vs non-Ia (blue, thin).}
\label{fig:sigma_cumf_tr_nr}
\end{figure}

\begin{figure}
	\begin{center}
		\includegraphics[width=3.5in]{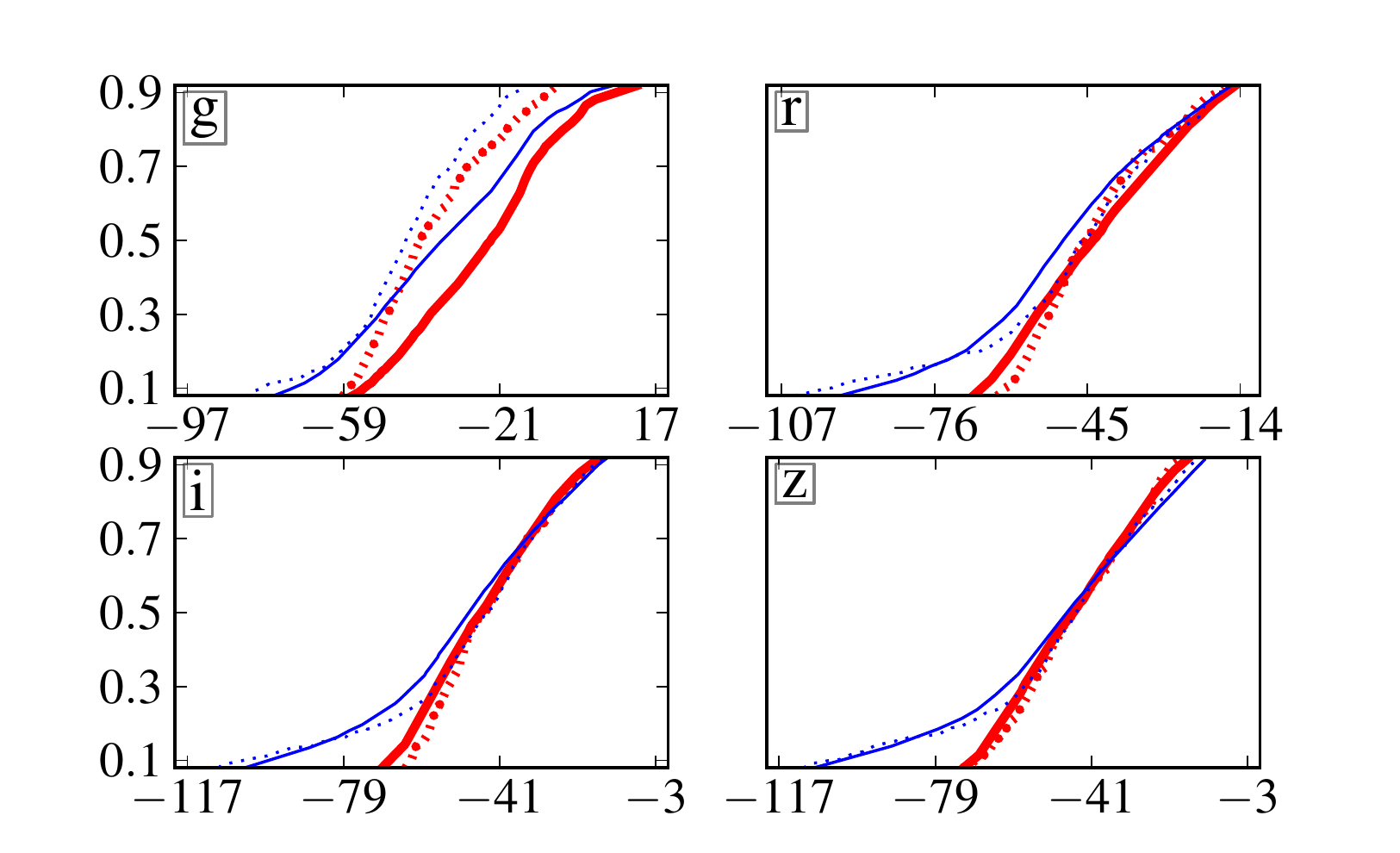}
	\end{center}
\caption{Cumulative plot of parameter $\phi$ in bands \textit{g,r,i,z}. Non-representative training (dashed) vs unclassified (solid) and Ia (red, thick) vs non-Ia (blue, thin).}
\label{fig:phi_cumf_tr_nr}
\end{figure}

In this appendix we look at how the five lightcurve fitting parameters and redshift differ between Ia's and non-Ia's, and between training and test SNe. Ia SNe cumulative frequency lines are red and thick, while non-Ia SNe are blue and thin. The cumulative frequency lines for training SNe are dotted, while the cumulative frequency lines for the unspecified SNe are solid. This appendix comprises Figures~\ref{fig:redshift_cumf_tr_nr} to~\ref{fig:phi_cumf_tr_nr}.

\section{Additional Figures}\label{app:figs}
\begin{figure}
	\begin{center}
		\includegraphics[width=3.5in]{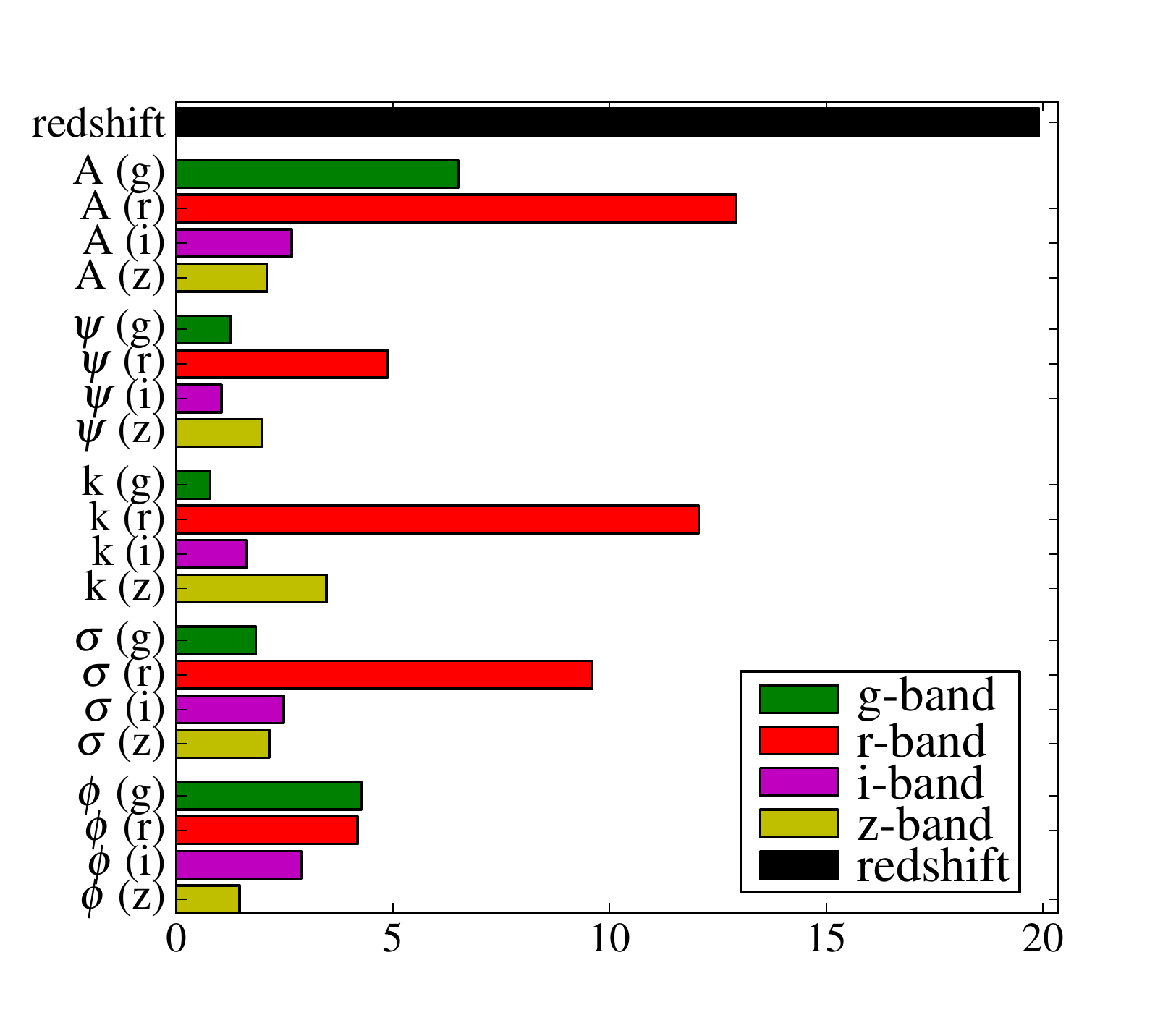}
	\end{center}
\caption{The importance of parameters in distingushing Ia from non-Ia in the non-representative training sample.}
\label{boosting_imp_tau_nr}
\end{figure}

\begin{figure}
	\begin{center}
		\includegraphics[width=3.5in]{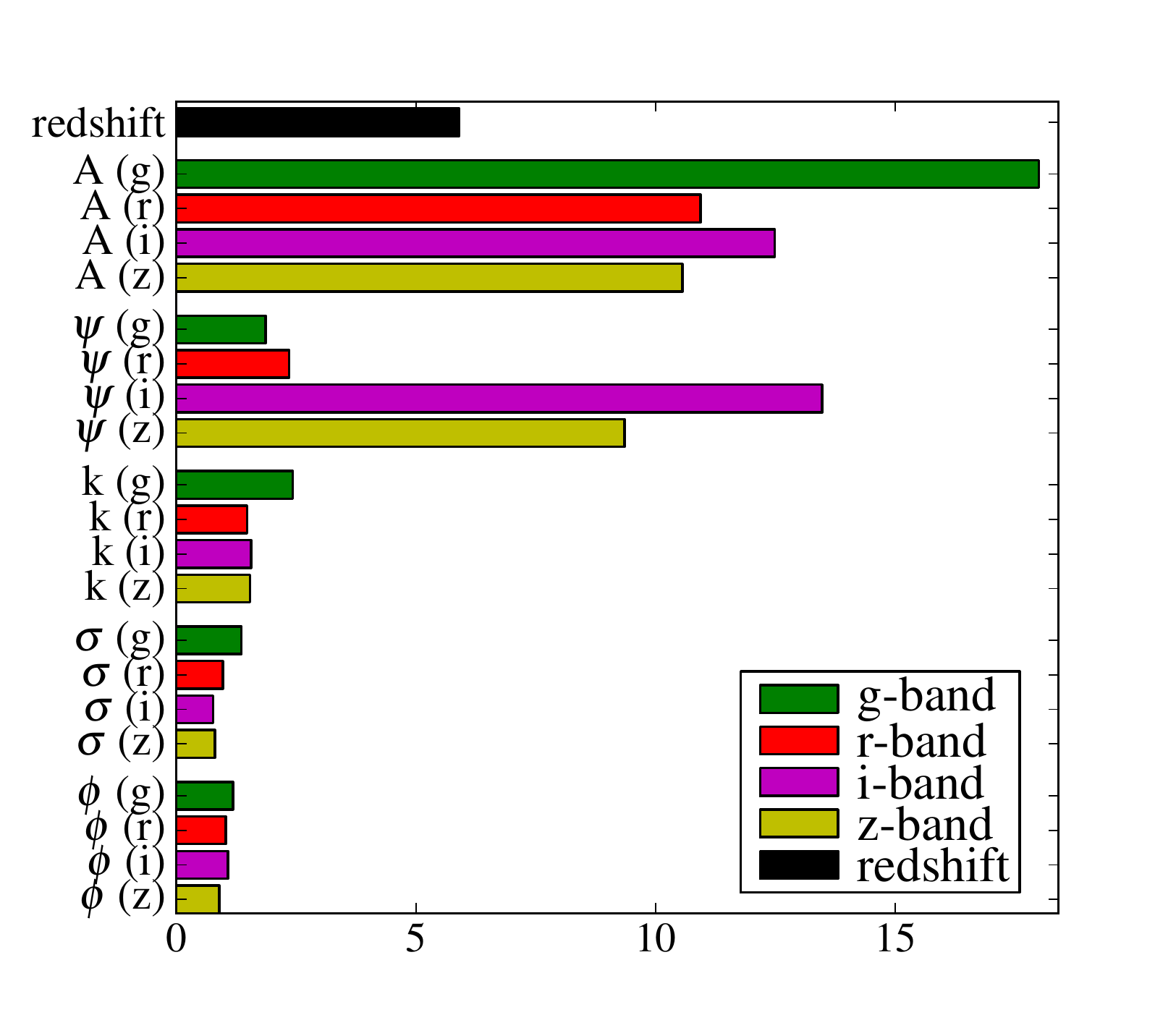}
	\end{center}
\caption{The importance of parameters in distinguishing non-representative training (with spectrum) from unclassified (without spectrum) non-Ia SNe using boosting.}
\label{fig:boost_te_tr_nIa}
\end{figure}

This appendix contains two more boosting parameter importance Figures:~\ref{boosting_imp_tau_nr} and~\ref{fig:boost_te_tr_nIa}

\section{Random SNe}\label{app:SNe}
This appendix contains a random selection of unclassified Ia and non-Ia SNe and their boosting values from representative training. Also, had a threshold of zero been used on the boosting value, would the classification have been correct (\tickYes) or incorrect (\tickNo). An extension of this appendix (200 SNe) can be found online at~\citet{COSMOAIMS}.

\begin{figure}
	\begin{center}
		\includegraphics[width=3.5in]{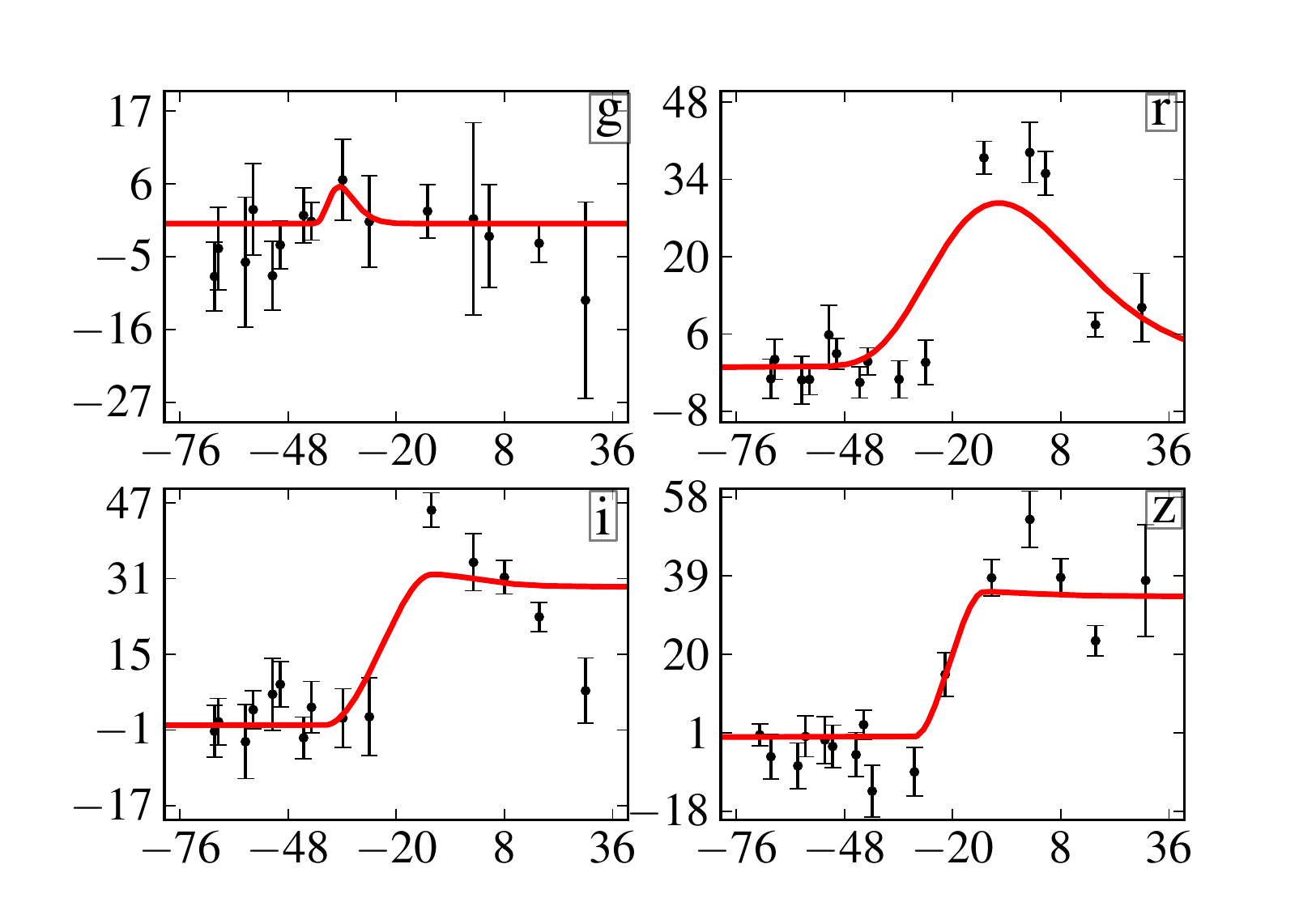}
	\end{center}
\caption{Ia SN at z = 0.64. Boosting value of 1.78. \tickYes}
\label{fig:Ia_4309.pdf}
\end{figure}

\begin{figure}
	\begin{center}
		\includegraphics[width=3.5in]{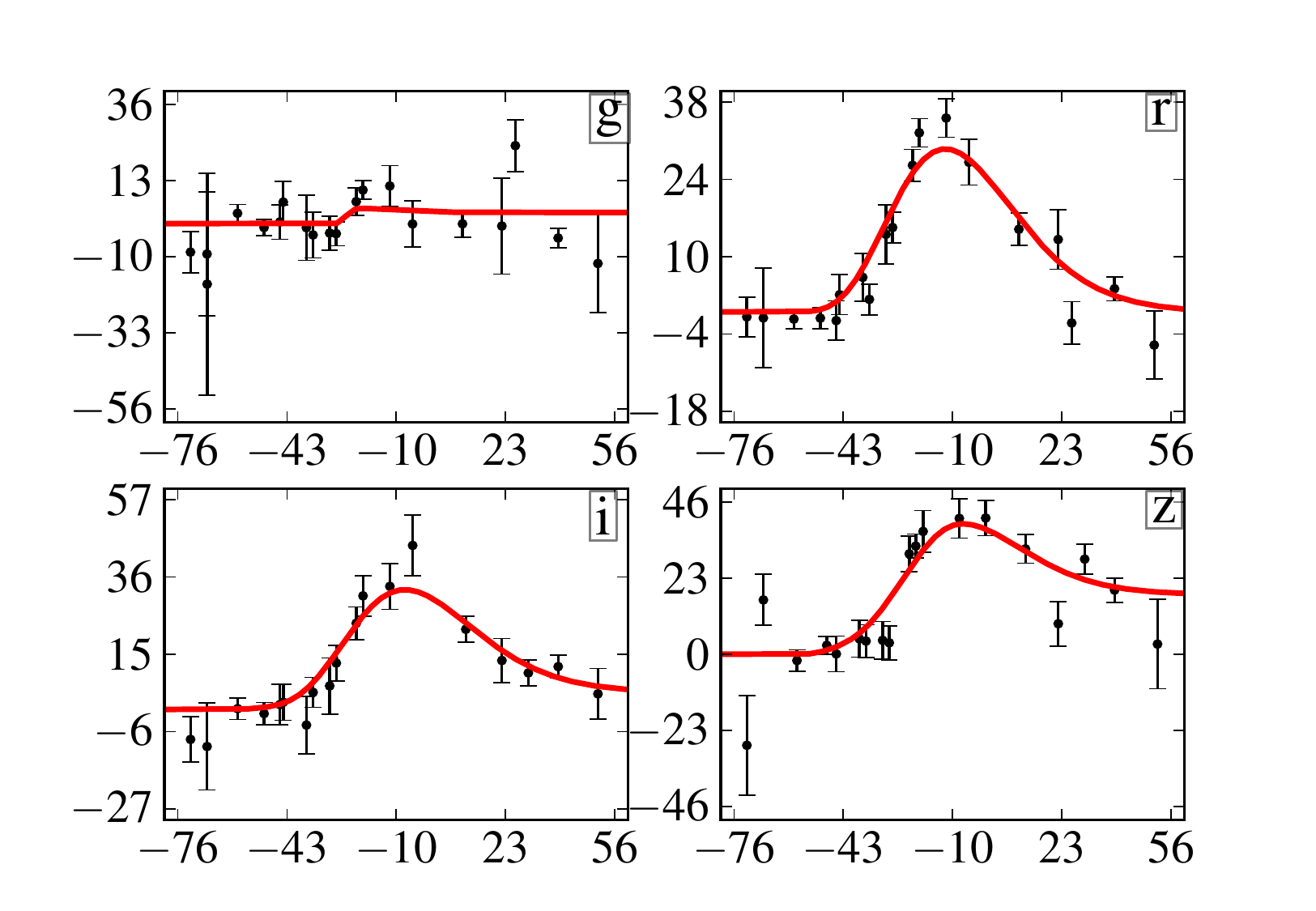}
	\end{center}
\caption{Ia SN at z = 1.08. Boosting value of 4.29. \tickYes}
\label{fig:Ia_12259.pdf}
\end{figure}

\begin{figure}
	\begin{center}
		\includegraphics[width=3.5in]{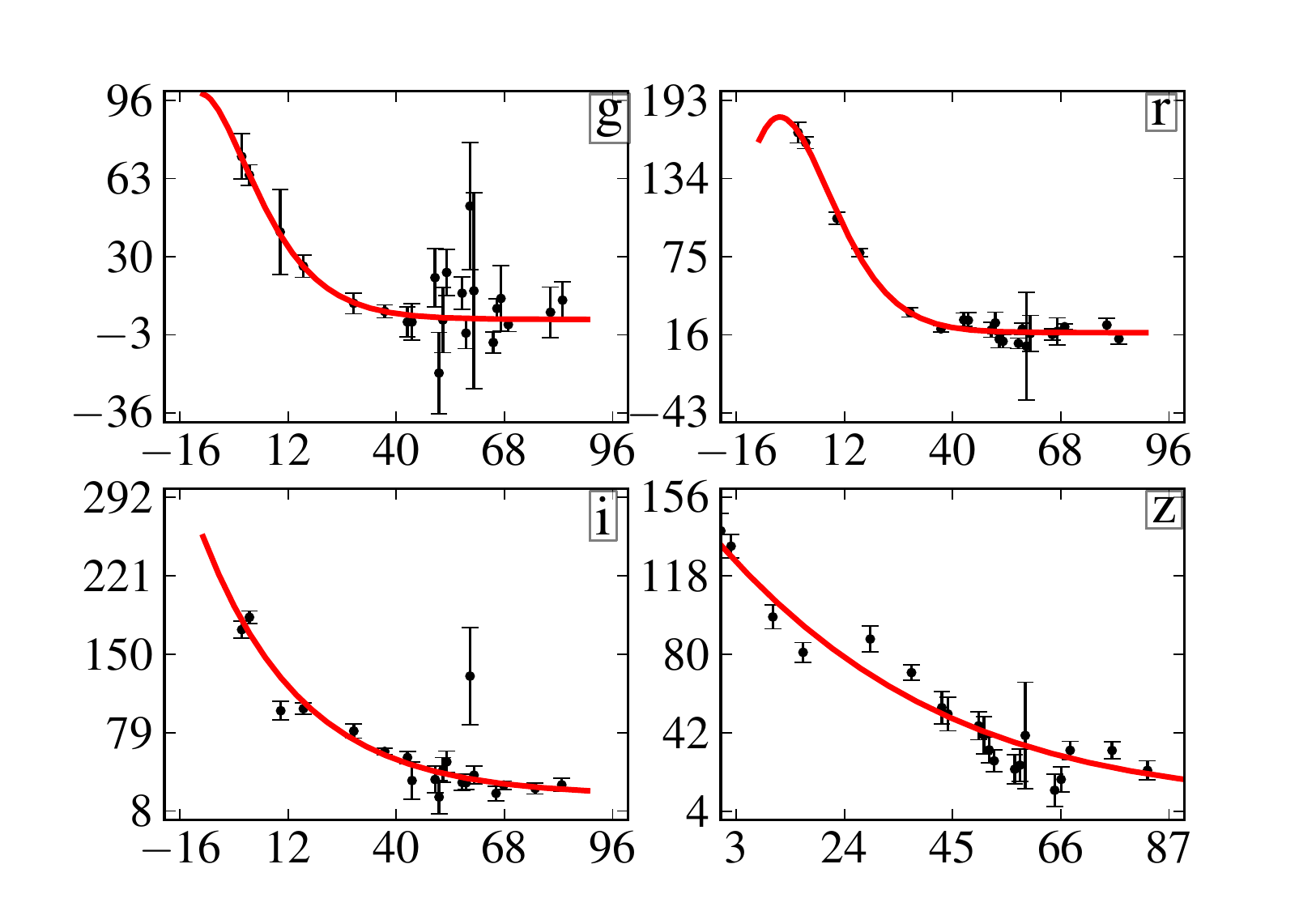}
	\end{center}
\caption{Ia SN at z = 0.439. Boosting value of 1.15. \tickYes}
\label{fig:Ia_3257.pdf}
\end{figure}

\begin{figure}
	\begin{center}
		\includegraphics[width=3.5in]{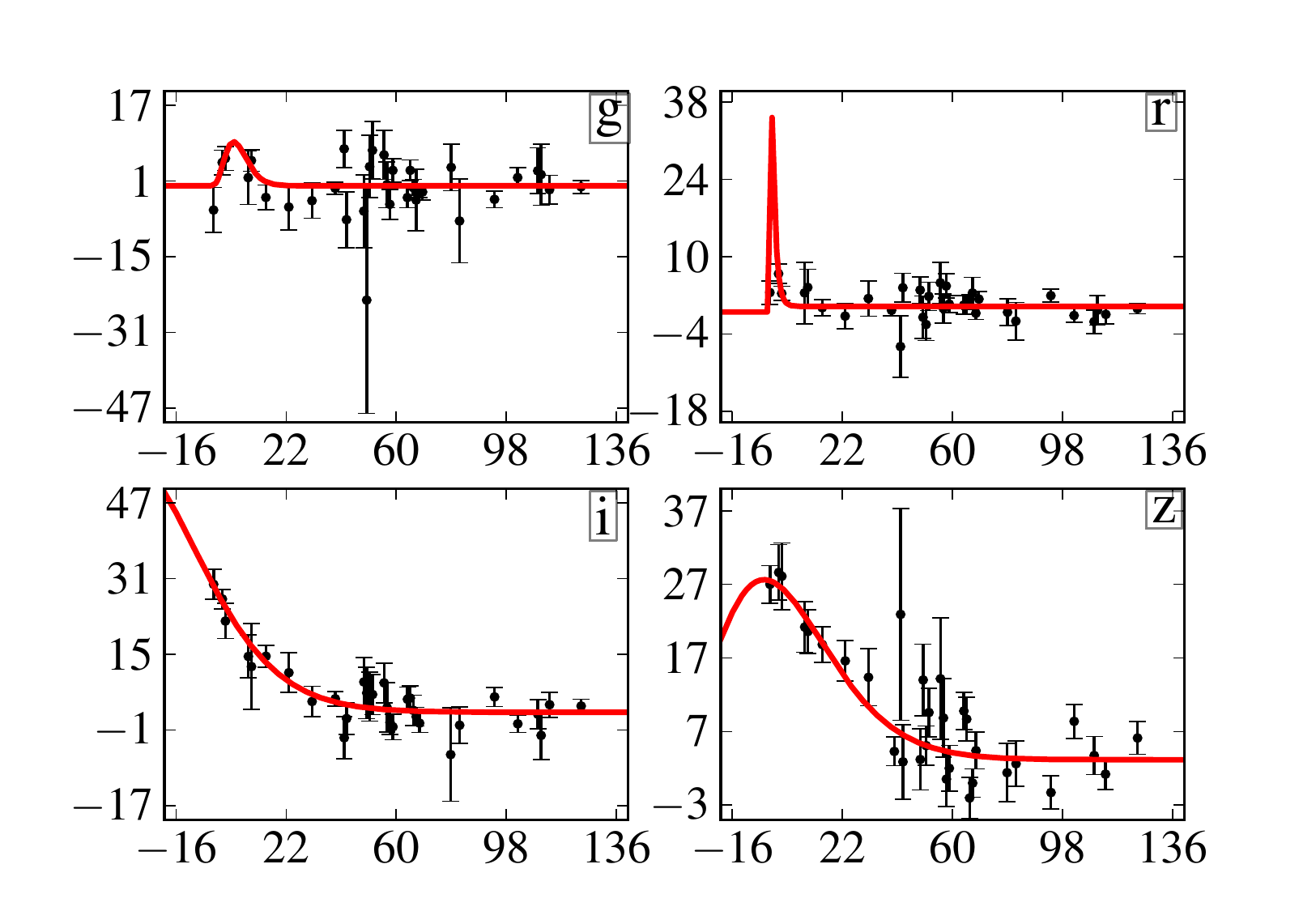}
	\end{center}
\caption{Ia SN at z = 1.01. Boosting value of 5.94. \tickYes}
\label{fig:Ia_20312.pdf}
\end{figure}

\begin{figure}
	\begin{center}
		\includegraphics[width=3.5in]{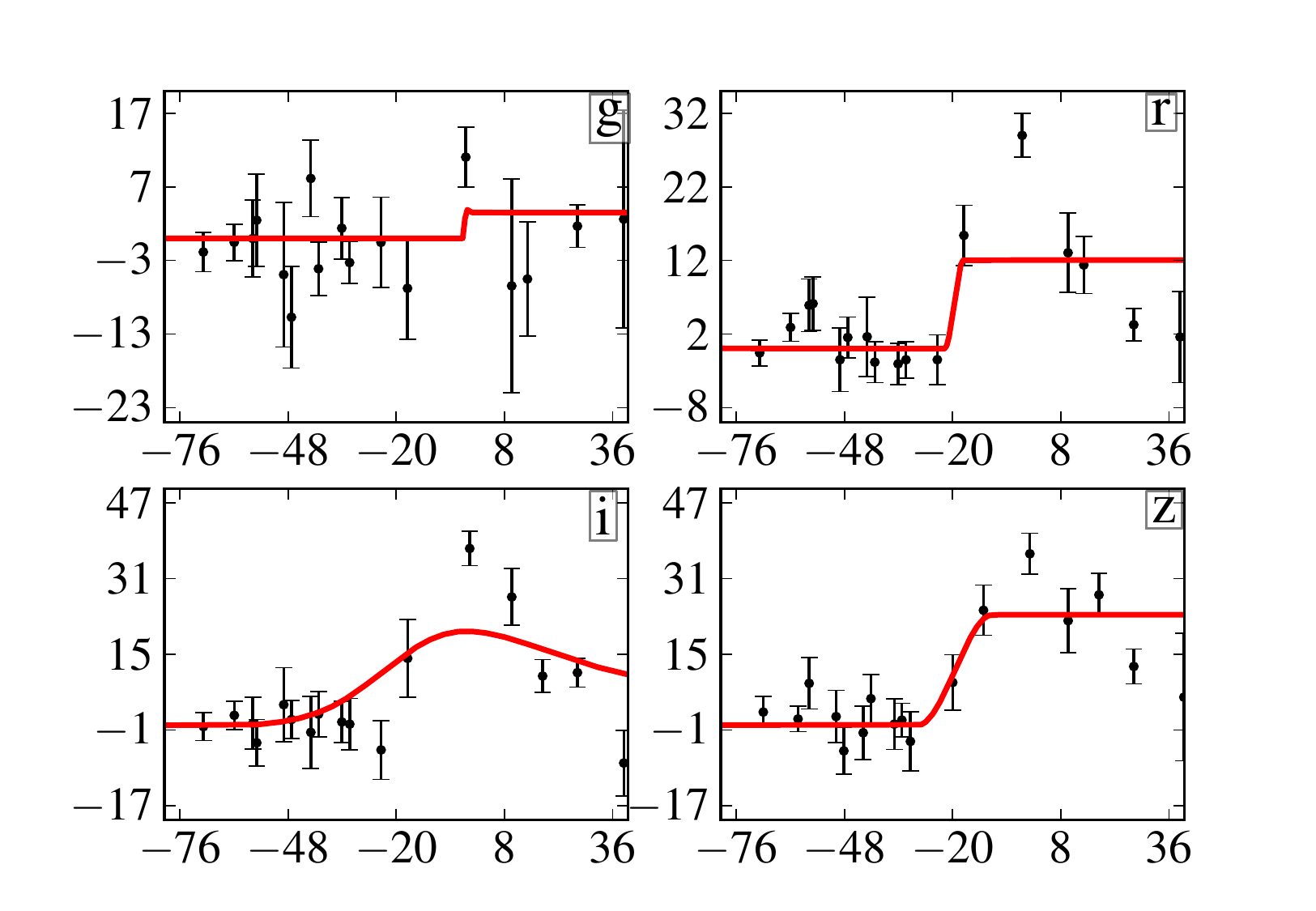}
	\end{center}
\caption{Ia SN at z = 0.692. Boosting value of -0.869. \tickNo}
\label{fig:Ia_5134.pdf}
\end{figure}

\begin{figure}
	\begin{center}
		\includegraphics[width=3.5in]{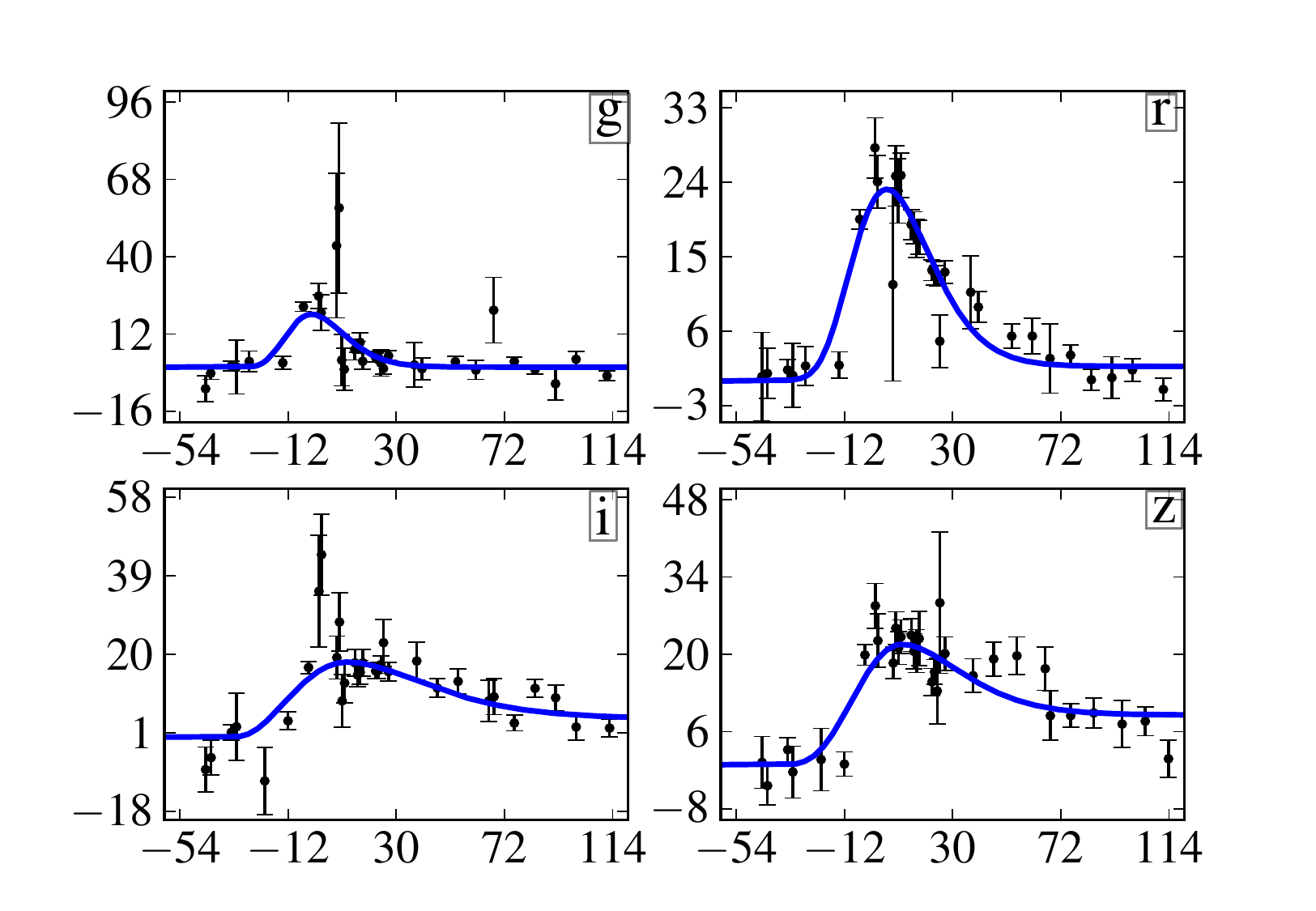}
	\end{center}
\caption{non-Ia SN at z = 0.578. Boosting value of -5.79. \tickYes}
\label{fig:nIa_10871.pdf}
\end{figure}

\begin{figure}
	\begin{center}
		\includegraphics[width=3.5in]{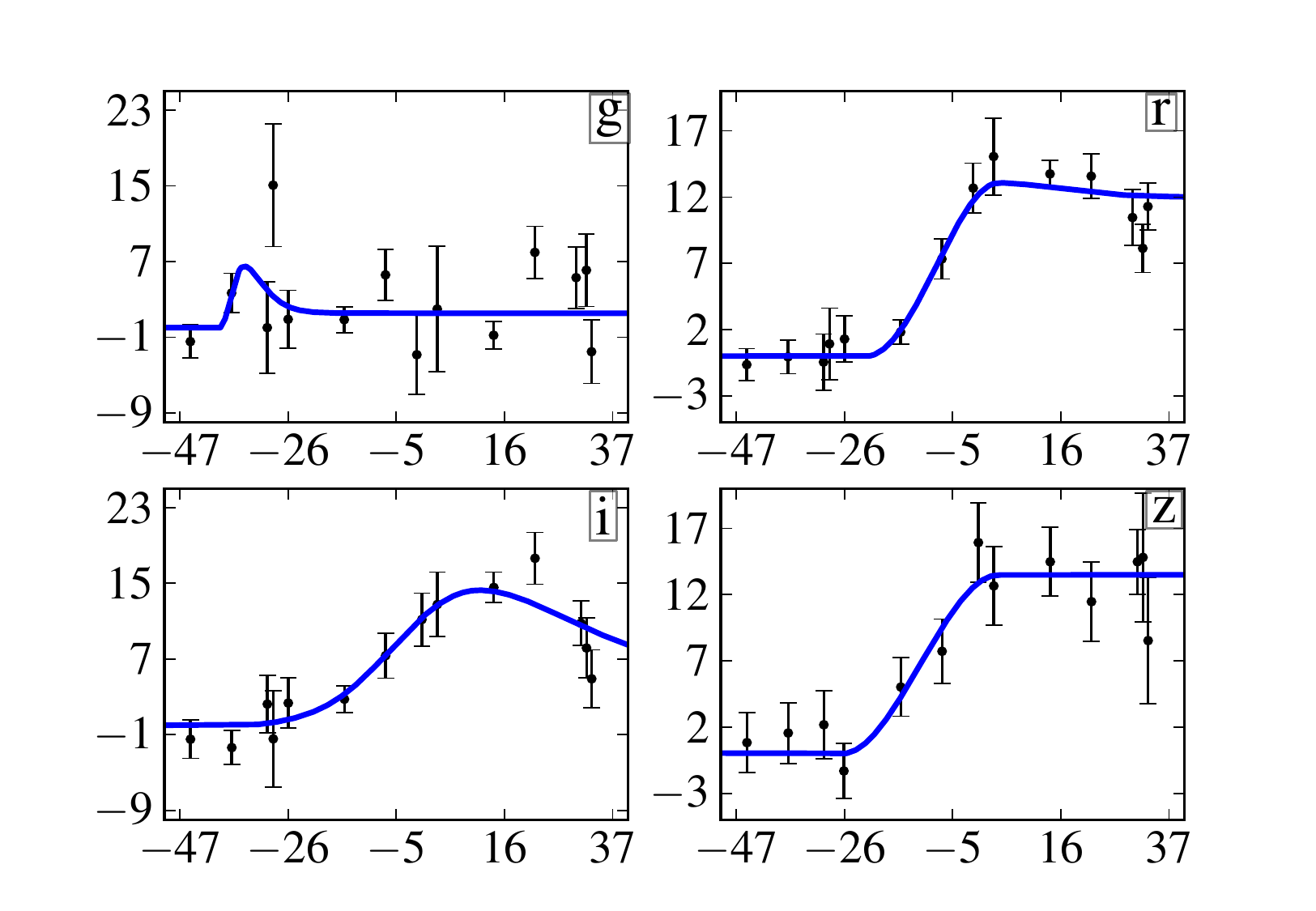}
	\end{center}
\caption{non-Ia SN at z = 0.674. Boosting value of -3.17. \tickYes}
\label{fig:nIa_1188.pdf}
\end{figure}

\newpage

\begin{figure}
	\begin{center}
		\includegraphics[width=3.5in]{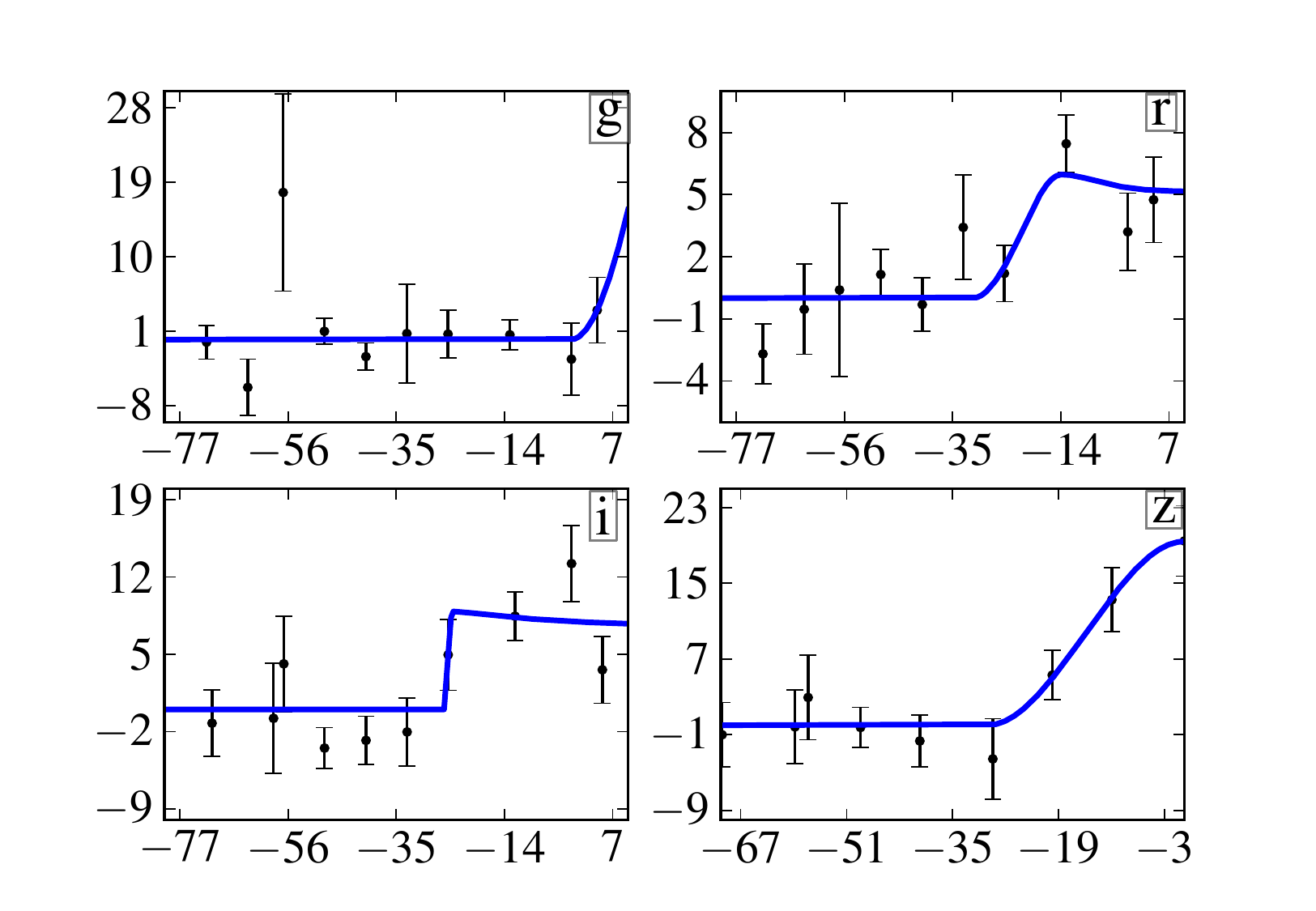}
	\end{center}
\caption{non-Ia SN at z = 1.04. Boosting value of -1.13. \tickYes}
\label{fig:nIa_15225.pdf}
\end{figure}

\begin{figure}
	\begin{center}
		\includegraphics[width=3.5in]{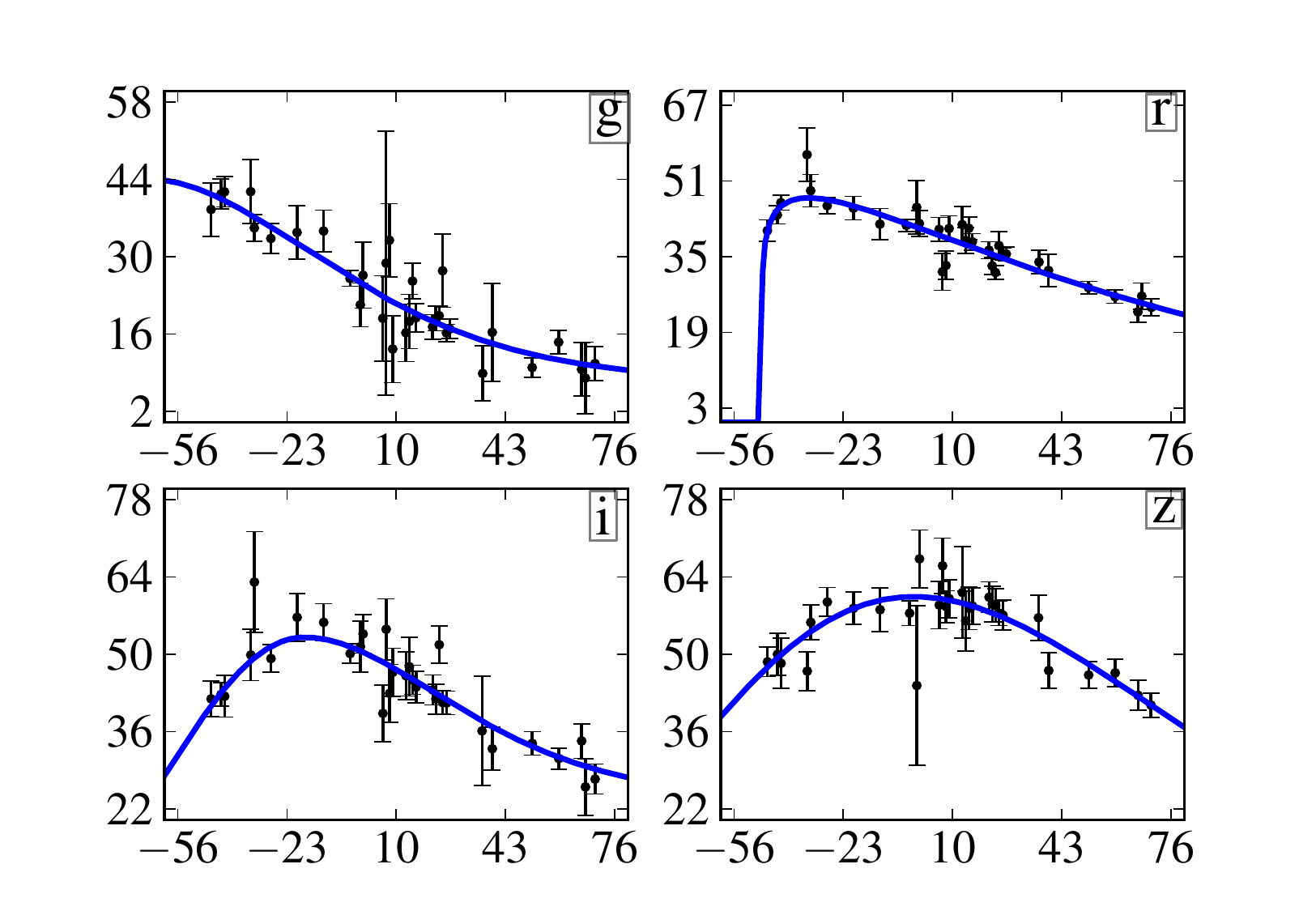}
	\end{center}
\caption{non-Ia SN at z = 0.722. Boosting value of -3.33. \tickYes}
\label{fig:nIa_15117.pdf}
\end{figure}

\begin{figure}
	\begin{center}
		\includegraphics[width=3.5in]{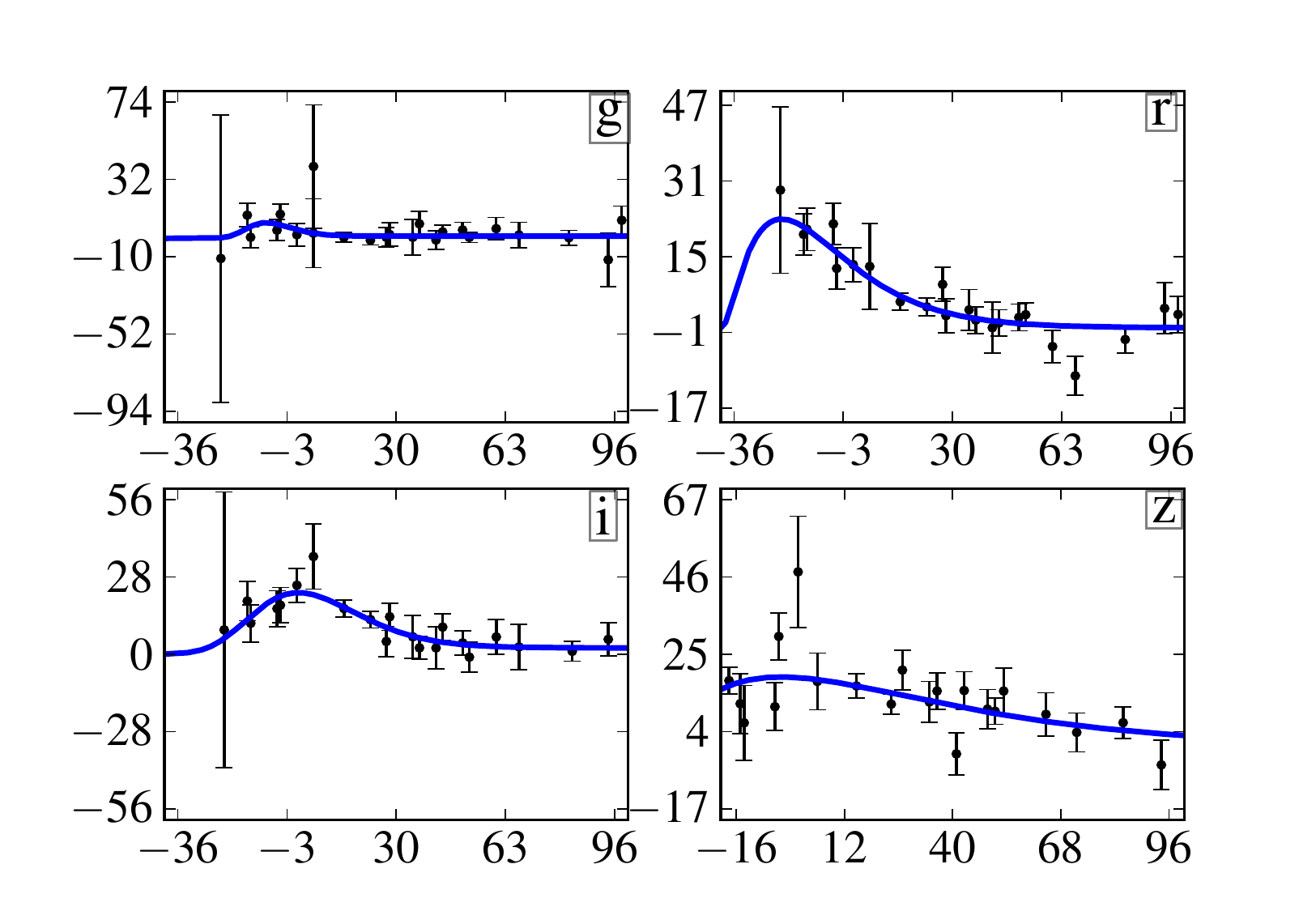}
	\end{center}
\caption{non-Ia SN at z = 0.688. Boosting value of 0.52. \tickNo}
\label{fig:nIa_10567.pdf}
\end{figure}

\bibliography{sn_typing_mat,Data_Bib}{}
\bibliographystyle{mn2e}

\end{document}